\documentclass[a4paper,11pt]{article}
\pdfoutput=1 
\usepackage{jheppub} 
\usepackage[numbers]{natbib}
\usepackage[T1]{fontenc} 
\usepackage{subfig}
\usepackage{hyperref}
\usepackage{float}
\usepackage{xcolor}
\usepackage[export]{adjustbox}
\usepackage{blindtext}
\usepackage{multirow}
\usepackage{verbatim}
\usepackage{bm}
\usepackage[normalem]{ulem}
\usepackage{chngpage}
\usepackage{array}

\newcommand{\oo}{\Delta}

\newcommand{\omm}[1]{\OO\!\left(M^{-#1}\right)}

\newcommand{\sss}{\scriptscriptstyle}

\renewcommand{\phi}{\ensuremath{\varphi}}

\newcommand{\bpm}{\begin{pmatrix}}      
\newcommand{\epm}{\end{pmatrix}} 
\newcommand\mw{m_{\sss W}}
\newcommand\mwhat{\hat{m}_{\sss W}}
\newcommand\mz{m_{\sss Z}}

\newcommand\mh{m_{\sss H}}
\newcommand\mt{m_{\sss t}}
\newcommand{\mphi}{m_{\sss{\Phi}}}

\newcommand{\OO}{\ensuremath{\mathcal{O}}}

\newcommand{\Op}[1]{\OO_{\sss #1}}
\newcommand{\Opp}[2]{\OO_{\sss #1}^{\sss #2}}

\newcommand{\Cp}[1]{C_{\sss #1}}

\newcommand{\Cpp}[2]{C_{\sss #1}^{\sss #2}}
\newcommand{\sw}{s_{\sss w}}
\newcommand{\cw}{c_{\sss w}}
\newcommand{\swh}{s_{\sss \hat{w}}}
\newcommand{\cwh}{c_{\sss \hat{w}}}

\newcommand{\ctwh}{c_{\sss 2\hat{w}}}

\newcommand{\lh}{\hat{\lambda}}

\newcommand{\gpsq}{g^{\prime\,2}}
\newcommand{\gpsqhat}{\hat{g}^{\prime\,2}}
\newcommand{\vT}{v_{\sss T}}
\newcommand{\voL}[1]{\frac{\vT^{#1}}{\Lambda^{#1}}}

\newcommand{\vh}{\hat{v}}
\newcommand{\vhoL}[1]{\frac{\vh^{#1}}{\Lambda^{#1}}}

\newcommand{\de}[2]{\delta^{\sss (#1)}{#2} }
\newcommand{\DE}[2]{\Delta^{\sss (#1)}{#2} }
\newcommand{\OHbx}{\Op{H\Box}}
\newcommand{\OHD}{\Op{HD}}
\newcommand{\OtH}{\Op{tH}}
\newcommand{\OH}{\Op{H}}
\newcommand{\OHso}{\Opp{H^6}{(1)}}
\newcommand{\OHst}{\Opp{H^6}{(2)}}
\newcommand{\OquH}{\Op{quH^5}}
\newcommand{\OHH}{\Op{H^8}}
\newcommand{\CHbx}{\Cp{H\Box}}
\newcommand{\CHD}{\Cp{HD}}
\newcommand{\CtH}{\Cp{tH}}
\newcommand{\CH}{\Cp{H}}
\newcommand{\CHso}{\Cpp{H^6}{(1)}}
\newcommand{\CHst}{\Cpp{H^6}{(2)}}
\newcommand{\CquH}{\Cp{quH^5}}
\newcommand{\CHH}{\Cp{H^8}}
\newcommand{\mx}{M_{\sss \Xi}}
\newcommand{\kx}{\kappa_{\sss \Xi}}
\newcommand{\lx}{\lam{\Xi}}
\newcommand{\ex}{\eta_{\sss \Xi}}
\newcommand{\lam}[1]{\lambda_{\sss #1}}
\newcommand{\ks}{\kappa_{\sss S}}
\newcommand{\ksss}{\kappa_{\sss S^3}}
\newcommand{\kssss}{\kappa_{\sss S^4}}
\newcommand{\ls}{\lam{S}}
\newcommand{\vs}{v_{\sss S}}
\newcommand{\bs}{b_{\sss S}}
\newcommand{\ms}{M_{\sss S}}
\newcommand{\hdh}{H^\dagger H}
\newcommand{\hsdh}{H^\dagger\sigma^a H}

\title{Dimension-8 SMEFT Analysis of Minimal \\
Scalar Field Extensions of the Standard Model}

\author[a,b]{John Ellis,}
\author[a]{Ken Mimasu,}
\author[a]{Francesca Zampedri}

\affiliation[a]{Theoretical Particle Physics and Cosmology Group, Department of Physics, \\
King's~College~London, London WC2R 2LS, UK}
\affiliation[b]{Theoretical Physics Department, CERN, CH-1211 Geneva 23, Switzerland}

\emailAdd{john.ellis@cern.ch}
\emailAdd{ken.mimasu@kcl.ac.uk}
\emailAdd{francesca.zampedri@kcl.ac.uk}

\preprint{
\begin{flushright}
{KCL-PH-TH/2023-18, CERN-TH-2023-038}
\end{flushright}
}

\abstract{We analyze the constraints obtainable from present data using the Standard Model Effective Field Theory (SMEFT) on extensions of the Standard Model with additional electroweak singlet or triplet scalar fields. We compare results obtained using only contributions that are linear in dimension-6 operator coefficients with those obtained including terms quadratic in these coefficients as well as contributions that are linear in dimension-8 operator coefficients. We also implement theoretical constraints arising from the stability of the electroweak vacuum and perturbative unitarity. Analyzing the models at the dimension-8 level constrains scalar couplings that are not bounded at the dimension-6 level. The strongest experimental constraints on the singlet model are provided by Higgs coupling measurements, whereas electroweak precision observables provide the strongest constraints on the triplet model. In the singlet model the present di-Higgs constraints already play a significant role. We find that the current constraints on model parameters are already competitive with those anticipated from future di- and tri-Higgs measurements. We compare our results with calculations in the full model, exhibiting the improvements when higher-order SMEFT terms are included. We also identify regions in parameter space where the SMEFT approximation appears to break down. We find that the combination of current constraints with the theoretical bounds still admits regions where the SMEFT approach is not valid, particularly for lower scalar boson masses.}

\begin{document}
\maketitle
\flushbottom

\section{Introduction}
\label{sec:intro}
The Standard Model Effective Field Theory (SMEFT)~\cite{Buchmuller:1985jz} 
is a convenient tool for assessing the sensitivities of
present and future experimental measurements to possible extensions of the Standard Model (SM) containing
additional massive degrees of freedom that decouple at low energies. In principle, the SMEFT offers a
systematic framework for approximating progressively the low-energy effects of such decoupled physics. In
practice, many SMEFT analyses include only dimension-6 operators (see, e.g.,~\cite{Grzadkowski:2010es,Giudice:2007fh}), often working to linear order and hence
to ${\cal O}(1/\Lambda^2)$ in the new physics scale $\Lambda$ (see, e.g., \cite{Pomarol:2013zra,Berthier:2015oma,Berthier:2015gja,Berthier:2016tkq,Brivio:2017bnu,Biekoetter:2018ypq,deBlas:2019rxi,Ellis:2020unq}), though sometimes also considering some terms of quadratic order that are of ${\cal O}(1/\Lambda^4)$ in the new physics scale (see, e.g.,~\cite{Corbett:2015ksa,Brivio:2019ius,Durieux:2019rbz,Ethier:2021bye}). However, working 
consistently to ${\cal O}(1/\Lambda^4)$ requires in general also including effects that are linear in
dimension-8 operators. An understanding of the relevance of dimension-8 effects is crucial for
establishing the validity of the SMEFT as a model-independent framework to search for new physics at collider experiments. This has been made possible in a general sense with the determination of complete, non-redundant dimension-8 operator bases in recent years~\cite{Murphy:2020rsh,Li:2020gnx}.

A complete treatment of all dimension-8 operators, even to linear order, is beyond
our current reach, but there are several ways to explore the possible importance of dimension-8
operators. Specific phenomenological studies have also been performed quantifying dimension-8 effects in key processes and observables such as Higgs production and decay~\cite{Hays:2018zze,Corbett:2021iob,Asteriadis:2022ras}, Electroweak precision observables (EWPOs)~\cite{Corbett:2021eux}, neutral- and charged-current Drell Yan processes~\cite{Alioli:2020kez,Boughezal:2021tih,Kim:2022amu}, and diboson production~\cite{Degrande:2023iob,Corbett:2023qtg}.
Alternatively, one may identify processes to which there are no dimension-6 contributions, examples of which
include light-by-light scattering~\cite{Ellis:2022uxv}, gluon-gluon scattering to photon pairs~\cite{Ellis:2018cos} and $Z \gamma$~\cite{Ellis:2021dfa}, 
as well as triple neutral-gauge-boson
vertices~\cite{Ellis:2019zex,Ellis:2020ljj,Ellis:2022zdw}. 
Another possibility is to look at specific minimal extensions of the SM and assess the relative
importances of linear dimension-6, quadratic dimension-6 and linear dimension-8 effects in these models,
which is the approach taken in this paper. 

We consider two minimal extensions of the SM that include massive scalar fields, either an electroweak
singlet $S$ or a  hypercharge-zero triplet $\Xi$, and compare the sensitivities to the masses and couplings of these fields
estimated using consistent truncations of SMEFT effects 
to ${\cal O}(1/\Lambda^2)$ and ${\cal O}(1/\Lambda^4)$.
In the former case, we include only the linear contributions of dimension-6
operators due to their interferences with SM amplitudes, and in the second case
we include both the quadratic contributions of dimension-6 operators to
experimental rates and the linear (interference) contributions of
dimension-8 operators. Such a top-down approach provides complementary information about the SMEFT, where by knowing the full result, we can quantify whether the expansion faithfully approximates the heavy new physics model. Several works have studied dimension-8 effects in simple extensions of the SM~\cite{Corbett:2015lfa,Dawson:2021xei,Corbett:2021eux,Dawson:2022cmu,Banerjee:2022thk,Banerjee:2023bzl} including for the two models we study in our paper. Our work continues in this vein, completing the associated tree-level matching at dimension-8 and providing an in-depth exploration of the constraints on the parameter space and the validity of the SMEFT approximation, making use of the \verb|fitmaker| code~\cite{Ellis:2020unq} to combine information from EWPOs, Higgs signal strengths, and di-Higgs production rates.

To set the scene for our analysis, in Section~\ref{sec:Higgs_self} we review the dimension-4, -6 and -8
Higgs operators that are relevant to our analysis (see Tables~\ref{tab:dim4-6} and \ref{tab:dim8}). We then
analyze in Section~\ref{sec:contributions} how the SM expressions for observables such as the
Higgs vacuum expectation value (vev), the Higgs mass, $Z$ and $W$ couplings, and the Yukawa and trilinear and quartic self-couplings of the Higgs boson are modified
in the presence of non-zero coefficients for the dimension-6 and -8 Higgs operators.
The relevant experimental constraints, theoretical predictions and their statistical interpretation are discussed in Section~\ref{sec:exp_constraints}.
Then, in Section~\ref{sec:tree} we derive the tree-level matching conditions for the
relevant dimension-6 and -8 operator coefficients in the two scalar field extensions of the SM that we consider. 

As quantitative applications of these derivations, in Sections~\ref{sec:singlet} and \ref{sec:triplet}, we present a global analysis of EWPOs, Higgs signal strengths and di-higgs rate measurements interpreted via the dimension-8 SMEFT in the contexts of the singlet and triplet scalar
extensions of the SM. On one hand, these results allow us to quantify the impact of extending the analysis to dimension-8, where we find that it brings sensitivity to new model parameters and leads to a richer interplay between the data and model parameters. On the other, it also helps us to uncover the regions in parameter space where the SMEFT expansion does not converge, by comparison to calculations in the full model. For the singlet model, we also calculate the possible magnitude
of the quartic-Higgs coupling that could be generated, given the prospective sensitivities of future measurements of double-
and triple-Higgs production. As an aside, we also quantify the impact of the recent $W$-mass measurement on the triplet model parameter space, focusing on the dimension-8 effects and the EFT validity. Finally in Section~\ref{sec:summary}, we summarise and conclude.

\section{\label{sec:Higgs_self}Setting the Scene}

Experiments have provided many precise verifications of
predictions involving most of the interactions in the SM
Lagrangian. The SMEFT provides a systematic framework that
is suitable for formulating global analyses of
extensions of the SM, such as the single scalar field
extensions that we discuss in this paper. In this Section we 
highlight briefly the SM data sectors that are
particularly relevant for probing these SM extensions.

The scalar field extensions that we consider generate 
SMEFT operators that involve not only the Higgs field but also
other SM fields that could modify processes other than Higgs production. 
The relevant dimension-4 and -6 SMEFT operators
are displayed in Table~\ref{tab:dim4-6}, and those of dimension 8 are
displayed in Table~\ref{tab:dim8}. We denote the high mass scale of BSM physics
by $\Lambda$, and denote the dimensionless Wilson
coefficients of the operators ${\cal O}_i$ by $C_i$.

\begin{table}
\begin{center}
    \begin{tabular}{ ||c||c|c|| } 
    \hline \hline
    \multirow{1}{4em}{Dim - 4} & ${\cal O}_{H4}$ & $(H^{\dagger}H)^2$ \\ \hline \hline \hline
    \multirow{7}{4em}{Dim - 6} & \multicolumn{2}{c||}{$H^6$ and $H^4D^2$} \\
    \cline{2-3}
    & ${\cal O}_H$ & $(H^{\dagger}H)^3$ \\
    & ${\cal O}_{HD}$ & $(H^{\dagger} D^{\mu} H)^*(H^{\dagger} D^{\mu} H)$\\
    & ${\cal O}_{H\Box}$ & $(H^{\dagger}H)\Box (H^{\dagger}H)$\\
    \cline{2-3} 
    & \multicolumn{2}{c||}{$\psi^2 H^3$} \\
    \cline{2-3}
    & ${\cal O}_{eH}$ & $(H^{\dagger}H)(\bar{l}_pe_rH)$ \\
    & ${\cal O}_{uH}$ & $(H^{\dagger}H)(\bar{q}_p u_r \widetilde{H})$\\
    & ${\cal O}_{dH}$ & $(H^{\dagger}H)(\bar{q}_pd_rH)$ \\
    \hline \hline 
    \end{tabular}
\end{center}
\caption{\it Dimension-4 and -6 Higgs operators 
relevant for our analysis (in the Warsaw basis).}
\label{tab:dim4-6}
\end{table}

\begin{table}
\begin{center}
\begin{adjustwidth}{-0.7cm}{-0.5cm}
    \begin{tabular}{ ||c||c|c|c|c|| } 
    \hline \hline
    \multirow{18}{4em}{Dim - 8} & \multicolumn{2}{c|}{$H^8$, $H^6D^2$ and $H^4D^4$} & \multicolumn{2}{c||}{$(\bar{L}R)(\bar{L}R)H^2$ + h.c}\\
    \cline{2-5}
    & ${\cal O}_{H^8}$ & $(H^{\dagger}H)^4$ & ${\cal O}_{lequH^2}^{(1)}$ & $(\bar{l}_p^je_r)\epsilon_{jk}(\bar{q}_s^ku_t)(H^{\dagger}H)$\\
    & ${\cal O}_{H^6}^{(1)}$ & $(H^{\dagger}H)^2(D_{\mu}H^{\dagger}D^{\mu}H)$ & ${\cal O}_{lequH^2}^{(2)}$ & $(\bar{l}_p^je_r)(\sigma^I\epsilon)_{jk}(\bar{q}_s^ku_t)(H^{\dagger}\sigma^I H)$\\
    & ${\cal O}_{H^6}^{(2)}$ & $(H^{\dagger}H)(H^{\dagger}\sigma^I H)(D_{\mu}H^{\dagger}\sigma^ID^{\mu}H)$ & ${\cal O}_{q^2udH^2}^{(1)}$ & $(\bar{q}_p^ju_r)\epsilon_{jk}(\bar{q}_s^kd_t)(H^{\dagger}H)$\\
    & ${\cal O}_{H^4}^{(1)}$ & $(D_{\mu}H^{\dagger}D_{\nu}H)(D^{\nu}H^{\dagger}D^{\mu}H)$ & ${\cal O}_{q^2udH^2}^{(2)}$ & $(\bar{q}_p^ju_r)(\sigma^I \epsilon)_{jk}(\bar{q}_s^kd_t)(H^{\dagger}\sigma^I H)$ \\
    & ${\cal O}_{H^4}^{(3)}$ & $(D_{\mu}H^{\dagger}D^{\mu}H)(D_{\nu}H^{\dagger}D^{\nu}H)$ & ${\cal O}_{leqdH^2}^{(3)}$ & $(\bar{l}_pe_rH)(\bar{q}_sd_tH)$\\
    \cline{2-3}
    &  \multicolumn{2}{c|}{$\psi^2 H^5$} & ${\cal O}_{l^2e^2H^2}^{(3)}$ & $(\bar{l}_pe_rH)(\bar{l}_se_tH)$\\
    \cline{2-3}
    & ${\cal O}_{leH^5}$ & $(H^{\dagger}H)^2(\bar{l}_pe_rH)$ & ${\cal O}_{q^2u^2H^2}^{(5)}$ & $(\bar{q}_pu_r\widetilde{H})(\bar{q}_su_t\widetilde{H})$ \\
    & ${\cal O}_{quH^5}$ & $(H^{\dagger}H)^2(\bar{q}_p u_r \widetilde{H})$ & ${\cal O}_{q^2d^2H^2}^{(5)}$ & $(\bar{q}_pd_rH)(\bar{q}_su_tH)$\\
    & ${\cal O}_{qdH^5}$ & $(H^{\dagger}H)^2(\bar{q}_pd_rH)$ &  &  \\
    \cline{2-5}
    & \multicolumn{2}{c|}{$\psi^2 H^3 D^2$ + h.c.} & \multicolumn{2}{c||}{$(\bar{L}L)(\bar{R}R) H^2$}\\
     \cline{2-5}
    & ${\cal O}_{leH^3D^2}^{(1)}$ & $(D_{\mu}H^{\dagger} D^{\mu}H)(\bar{l}_pe_rH)$  & ${\cal O}_{l^2e^2H^2}^{(1)}$ & $(\bar{l}_p \gamma^{\mu}l_r)(\bar{e}_s \gamma_{\mu}e_t)(H^{\dagger}H)$ \\
    & ${\cal O}_{leH^3D^2}^{(2)}$ & $(D_{\mu}H^{\dagger} \sigma^I D^{\mu}H)(\bar{l}_pe_r\sigma^I H)$  & ${\cal O}_{l^2e^2H^2}^{(2)}$ & $(\bar{l}_p \gamma^{\mu}\sigma^Il_r)(\bar{e}_s \gamma_{\mu}e_t)(H^{\dagger}\sigma^I H)$ \\
    & ${\cal O}_{quH^3D^2}^{(1)}$ & $(D_{\mu}H^{\dagger} D^{\mu}H)(\bar{q}_pu_rH)$ & ${\cal O}_{q^2u^2H^2}^{(1)}$ & $(\bar{q}_p \gamma^{\mu}q_r)(\bar{u}_s \gamma_{\mu}u_t)(H^{\dagger}H)$\\
    & ${\cal O}_{quH^3D^2}^{(2)}$ & $(D_{\mu}H^{\dagger} \sigma^I D^{\mu}H)(\bar{q}_pu_r\sigma^I H)$ & ${\cal O}_{q^2u^2H^2}^{(2)}$ & $(\bar{q}_p \gamma^{\mu}\sigma^I q_r)(\bar{u}_s \gamma_{\mu}u_t)(H^{\dagger}\sigma^I H)$\\
    & ${\cal O}_{qdH^3D^2}^{(1)}$ & $(D_{\mu}H^{\dagger} D^{\mu}H)(\bar{q}_pd_rH)$ & ${\cal O}_{q^2d^2H^2}^{(1)}$ & $(\bar{q}_p \gamma^{\mu}q_r)(\bar{d}_s \gamma_{\mu}d_t)(H^{\dagger}H)$ \\
    & ${\cal O}_{qdH^3D^2}^{(2)}$ & $(D_{\mu}H^{\dagger} \sigma^I D^{\mu}H)(\bar{q}_pd_r \sigma^I H)$ & ${\cal O}_{q^2d^2H^2}^{(2)}$ & $(\bar{q}_p \gamma^{\mu}\sigma^I q_r)(\bar{d}_s \gamma_{\mu}d_t)(H^{\dagger}\sigma^I H)$ \\
    \cline{2-5}
     & \multicolumn{4}{c||}{$(\bar{L}R)(\bar{R}L) H^2$ + h.c.} \\
     \cline{2-5}
     & ${\cal O}_{leqdH^2}^{(1)}$ & $(\bar{l}_p^je_r)(\bar{d}_sq_{tj})(H^{\dagger}H)$  & ${\cal O}_{lequH^2}^{(5)}$ & $(\bar{l}_pe_rH)(\widetilde{H}^{\dagger}\bar{u}_sq_t)$ \\
     & ${\cal O}_{leqdH^2}^{(2)}$ & $(\bar{l}_pe_r)\sigma^I(\bar{d}_sq_{t})(H^{\dagger}\sigma^I H)$ & ${\cal O}_{q^2u^2H^2}^{(5)}$ & $(\bar{q}_pd_rH)(\widetilde{H}^{\dagger}\bar{u}_sq_t)$\\
     \hline \hline
    \end{tabular}
\end{adjustwidth}
\end{center}
\caption{\it Dimension-8 Higgs operators relevant for our analysis.}
\label{tab:dim8}
\end{table}

\subsection{\label{sec:Higgs_strengths}
Higgs Coupling Strengths}

As we discuss in Section~\ref{sec:singlet}, the most relevant couplings
for the singlet scalar extension of the SM are those of the Higgs boson.
The most important impact of this model is via mixing of the singlet scalar
with the SM Higgs field through a universal mixing angle $\alpha$, which
has the effect of suppressing the couplings of the Higgs boson to other SM
particles by a universal factor $\cos \alpha$:
\begin{equation}
h^{SM}_{HXX} \; \to \; \cos \alpha \times h^{SM}_{HXX} \, .
\end{equation}
The ATLAS and CMS Collaborations have recently published legacy papers
summarizing their measurements of the couplings of the Higgs boson 10 years after its
discovery~\cite{ATLAS:2022vkf,CMS:2022dwd}. Analysing the constraints from these measurements in the SMEFT framework and interpreting the results
in the singlet model, we find an upper limit
\begin{equation}
    \sin^2 \alpha \; \le \; 0.114 \,.
    \label{alphaconstraint}
\end{equation}
We use this bound in Section~\ref{sec:singlet} in our analyses of the singlet scalar
extension of the SM at the dimension-6 and -8 levels, comparing these results
to check the convergence of the SMEFT expansion and towards an analysis of the full model.

\subsection{\label{sec:EWPOs}
Electroweak Precision Observables}

As we discuss in Section~\ref{sec:triplet}, the most relevant SM sector
for constraining the triplet scalar extension of the SM is that of
electroweak precision observables (EWPOs). These include measurements
of electroweak couplings at the $Z$ peak as well as measurements of the
$W$ mass. As is well known, the current status of $W$ mass measurements is
somewhat unsettled. Measurements from LEP and the LHC are consistent with
the SM prediction, but a recent measurement from the CDF Collaboration~\cite{CDF:2022hxs}
is in significant disagreement with the SM. A global dimension-6 SMEFT 
analysis~\cite{Bagnaschi:2022whn} identified several single-field extensions 
of the SM that could potentially alleviate this discrepancy, including the triplet
scalar field that we analyze here in Section~\ref{sec:triplet}. We also analyse this
extension of the SM at the dimension-6 and -8 levels along
with an analysis of the full model, considering two possibilities for the $W$ mass:
the pre-CDF world average and the new CDF value~\cite{CDF:2022hxs}.

\subsection{\label{sec:Higgs_selfcouplings}
Higgs Self-Interactions}

In contrast to the two categories of observables summarized above,
one category of SM interactions that has not yet been constrained significantly
is that of the self-interactions of the Higgs boson~\cite{Papaefstathiou:2015paa}. 
These Higgs self-couplings, which include terms proportional to $h^n$ 
in the Higgs potential as well as possible derivative interactions, could provide a 
deeper understanding of the nature of  the electroweak symmetry breaking
(EWSB) mechanism. They are relevant for the stability of the electroweak
vacuum, and determine the shape of the Higgs potential,
which has implications for the nature of the electroweak phase transition (EWPT) and baryogenesis \cite{Roloff:2019crr}, as
well as the possible generation of a stochastic background of 
gravitational waves (see, e.g.,~\cite{Kamionkowski:1993fg}). For all these reasons, probing the Higgs
self-couplings is one of the main objectives of current and future
particle colliders.

Within the SM, the Higgs cubic and quartic self-couplings are completely
determined by the Higgs boson mass $\mh \simeq 125$ GeV and its vev, $v \simeq 246$ GeV:
\begin{equation}
    V_{\text{self,SM}} \; = \; \frac{\mh^2}{2v} h^3 + \frac{\mh^2}{8v^2} h^4\,.
    \label{eq:V_self_SM}
\end{equation}
However, physics beyond the Standard Model (BSM) may enter the Higgs sector and
induce deviations from the SM predictions for the cubic and quartic Higgs self-couplings,
which can be parametrized as:
\begin{equation}
    V_{\text{self}} = \frac{\mh^2}{2v}(1+c_3) h^3 + \frac{\mh^2}{8v^2}(1+d_4) h^4\,,
    \label{eq:V_self_dev}
\end{equation}
where $c_3$ and $d_4$ are model-dependent parameters. As we discuss in more detail below,
the BSM physics would, in general, modify the SM expressions for $\mh$ and $v$ in terms of 
underlying parameters in the Lagrangian, as well as the Higgs field normalisation. 
These modifications are taken into account in
specific models when calculating $c_3$ and $d_4$. Significant deviations from the
SM predictions would provide indirect evidence that there exists BSM
physics that couples to the EWSB 
sector~\cite{Barger:2003rs}.

One of the outputs of our analyses of the SMEFT framework including dimension-6 and
dimension-8 Higgs operators is to study how specific models of high-scale BSM physics 
involving singlet and triplet scalar fields can induce deviations from the SM predictions 
for the Higgs cubic and quartic self-couplings, and compare them with present and
prospective experimental constraints on these couplings. Interestingly, we find that
the present experimental limit on di-Higgs production already plays a role in our
analysis by excluding regions of parameter space where a second minimum of the likelihood appears in the SMEFT analysis using single-Higgs data alone.

\section{SMEFT contributions to experimental measurements}
\label{sec:contributions}

In this section we detail the calculations needed to determine the impact of the singlet and triplet scalar extensions of the SM on EWPOs, Higgs signal strengths and di-Higgs production. We do this in the SMEFT formalism up to operator dimension-8, calculating the relevant shifts in couplings that lead to modifications of the observables of interest. We do not present a complete calculation considering the effects of all possible dimension-8 operators, but rather focus on the following operators, which are those generated by the scalar field models studied in this work:
\begin{align}
    \text{Dimension-6}&: \OHbx, \OHD, \OtH, \OH \, ;\\
    \text{Dimension-8}&: \OHso, \OHst, \OquH, \OHH \, .
\end{align}
As shown later in Tables~\ref{tab:singlet_results} and~\ref{tab:triplet_results}, all other operators involving light quarks that could potentially affect the aforementioned observables are suppressed by at least one power of a non-top quark Yukawa coupling. We neglect such Yukawa couplings, keeping only that of the top quark, so only modifications to its coupling to the Higgs boson are relevant for our purposes. The remaining operators of relevance only involve the Higgs field and the EW gauge bosons via its covariant derivative.  

\subsection{Input scheme\label{sec:input_scheme}}
We extract the values of the SM input parameters in the $\{\alpha,\mz,G_F\}$ scheme, including effects from the operators of interest. These are the hypercharge and weak gauge couplings, the Higgs vev and the Higgs quartic coupling, denoted by $\{g^{\prime 2}, g^2, \vT^2, \lambda \}$, respectively.
As reviewed in Appendix~\ref{app:input scheme}, this corresponds to obtaining expressions for the input observables, $\{\alpha_{\sss EM}, m_{\sss Z}^2, G_F, \mh^2\}$ (where $G_F$ is extracted from the measurement of the muon decay lifetime), in terms of the input parameters and the Wilson coefficients and inverting the system to second order in $\Lambda^{-2}$.

The Higgs potential receives corrections from the sextic and octic self-interaction terms:
\begin{equation}
    V_{\text{SMEFT}} = -\mu^2 (H^{\dagger}H) +\lambda (H^{\dagger}H)^2 - \frac{1}{\Lambda^2}\Cp{H} (H^{\dagger}H)^3 - \frac{1}{\Lambda^4} \Cp{H^8} (H^{\dagger}H)^4 \,.
    \label{eq:V_SMEFT}
\end{equation}
Extremising with respect to $(H^{\dagger}H)$ results in the minimisation condition
\begin{align}
    \label{eq:mincond}
    \mu^2 - \lambda \vT^2 
    + \frac{\vT^4}{\Lambda^2}\frac{3\CH}{4}
    + \frac{\vT^6}{\Lambda^4}\frac{\CHH }{2}= 0 \, ,
\end{align}
which implies a correction to the vev with respect to the SM:
\begin{align}
    \label{eq:vT}    
    \langle H^{\dagger}H \rangle = \frac{v^2}{2} \left(
    1 + \frac{v^2}{\Lambda^2}\frac{3\Cp{H}}{4\lambda}
    + \frac{v^4}{\Lambda^4}\frac{ 9\Cp{H}^2 + 4\lambda  \Cp{H^8}}{8\lambda^2}
    \right) + \mathcal{O}\!\left(\Lambda^{-6}\right)
    \equiv \frac{\vT^2}{2}\,,
\end{align}
where $v^2 \equiv  \mu^2/\lambda$ is the expression for the Higgs vev in the SM and $\vT\simeq 246$ GeV 
is the physical value extracted from data, including the BSM effects. We then expand the Higgs field around its vev as follows:
\begin{equation}
    H = \frac{1}{\sqrt{2}}
    \begin{pmatrix}
    0\\ h+\vT
    \end{pmatrix}\,.
\end{equation} 

The relevant effects for our purposes are the shifts in the $W$, $Z$ and Higgs boson mass terms. The latter two are direct input observables and the former affects the expression for the muon decay amplitude, which is used to define the Fermi constant, another input quantity. None of the operators of interest affect the diagonalisation of the EW gauge boson kinetic terms, which is affected by other operators such as $\Op{HWB}=(H^\dagger \tau^I H) W^I_{\mu\nu}B^{\mu\nu}$ at dimension-6, and $\Opp{(1)}{WBH^4}=(H^\dagger  H)(H^\dagger \tau^I H) W^I_{\mu\nu}B^{\mu\nu}$ and $\Opp{(3)}{W^2H^4}=(H^\dagger \tau^I H)(H^\dagger \tau^J H) W^I_{\mu\nu}W^{J,\mu\nu}$ at dimension-8. This simplifies the exercise at hand, since the compositions of the photon and $Z$-boson mass eigenstates in terms of the hypercharge and neutral SU(2) fields are the same as in the SM, there is no induced shift of the fine structure constant, and the $Z$-boson mass term is much simpler. We point interested readers to Ref~\cite{Helset:2020yio} for a treatment of these effects in a compact, geometric formalism.

Following EW symmetry breaking, the $Z$ and $W$ masses are given by
\begin{align}
    \mz^2 &= \frac{(g^2+g^{\prime2}) \vT^2}{4}\bigg[
    1 + \voL{2}\bigg(\frac{\CHD}{2}\bigg)
    + \voL{4}\bigg(\frac{\CHso}{4} + \frac{\CHst}{4}\bigg)\bigg]\,,
    \label{eq:mzEW}\\
    \mw^2 &= \frac{g^2 \vT^2}{4} \bigg[
    1 + \voL{4}\bigg(\frac{\CHso}{4} - \frac{\CHst}{4}\bigg)
    \bigg]=\frac{g^2 \vT^2}{4} \bigg[
    1 + \DE{8}{\mw}
    \bigg]\,,
    \label{eq:mwEW}
\end{align}
where the shifts in the latter, arising purely at dimension-8, feed into the muon decay amplitude that defines the Fermi constant: 
\begin{align}
    \label{eq:GF}    
    G_F =  \frac{1}{\sqrt{2}\vT^2}\bigg[
    1 - \voL{4}\bigg(
    \frac{\CHso}{4} - \frac{\CHst}{4}
    \bigg)
    \bigg]\,.
\end{align}
The dynamical Higgs field receives corrections to its kinetic term, and requires canonical normalisation by a field redefinition:
\begin{align}
    \label{eq:hredef}
    \begin{split}
          h&\to h \left[
          1 - 2\voL{2} \left(\CHbx -\frac{\CHD}{4}\right) 
          + \voL{4}\left(\frac{\CHso}{4} + \frac{\CHst}{4} \right) 
          \right]^{-\frac{1}{2}}\,,\\
         &\approx h\left(1 
        +\Delta_h
        \right)\equiv h\left(1 
        + \voL{2}\Delta^{\sss (6)}_h 
        + \voL{4}\Delta^{\sss (8)}_h
        \right),
    \end{split}\\
\Delta^{\sss (6)}_h &=\CHbx-\frac{\CHD}{4}\,,\quad
\Delta^{\sss (8)}_h = \frac{1}{2}\left(3(\Delta^{\sss (6)}_h )^2 - \frac{\CHso}{4}-\frac{\CHst}{4}\right) \, ,
\end{align}
after which we can read off the Higgs mass, having additionally made use of Eq.~\eqref{eq:mincond} to eliminate $\mu$:
\begin{align}
    \label{eq:mhEW}
    \mh^2=2\lambda \vT^2\left[
    1 
    + \voL{2}\left(
      2\Delta^{\sss (6)}_h-\frac{3\CH}{2\lambda}
    \right)
    + \voL{4}\left(
      4(\Delta^{\sss (6)}_h )^2 -\frac{3\CH}{\lambda}\Delta^{\sss (6)}_h
      - \frac{\CHso}{4}-\frac{\CHso}{4}
      -\frac{3\CHH}{2\lambda}
    \right)
    \right] \, .
\end{align}
We can now make use of Eqs.~\eqref{eq:mzEW}, \eqref{eq:GF}, \eqref{eq:mhEW} and the fact that the fine structure constant receives no corrections in our models to identify the various $\DE{i}{\mathcal{O}_n}$ shifts as in Eqs.~\eqref{eqn:D_aEM}--~\eqref{eqn:D_mh} and extract the derived SM parameters, $ g_i =\{g^{\prime 2}, g^2, \vT^2, \lambda \}$. We follow the notation of~\cite{Brivio:2020onw}, defining as follows the relative shifts in the parameters:
\begin{align}
    g_i = \hat{g}_i\left(
      1 + \de{6}{g_i}\,\vhoL{2} + \de{8}{g_i}\,\vhoL{4}
    \right) \, ,
\end{align}
where the hat notation denotes the corresponding SM function of the input parameters, as in Eq.~\eqref{eq:SMsol}. Since we are computing to dimension-8, we have to account for the explicit dependence of the $\Delta^{(6)}\mathcal{O}_n$ on the derived parameters, including the $\voL{2}$ factors that appear throughout. The shifts in the derived parameters are found to be:
\begin{align}
    \label{eq:dvt}
    \de{6}{\vT^2} &= 0\,; &
    \de{8}{\vT^2} &=\frac{\CHst}{4} - \frac{\CHso}{4} \,;\\
    \label{eq:dlam}
    \de{6}{\lambda} &= \frac{3\CH}{2\lh} - 2\CHbx + \frac{\CHD}{2}\,; &
    \de{8}{\lambda} &= \frac{3\CHH}{2\lh} + \frac{\CHso}{2} \,; \\  
    \label{eq:dgpsq}
    \de{6}{\gpsq} &= \frac{\swh^2}{\ctwh}\frac{\CHD}{2}\,; &
    \de{8}{\gpsq} &=\frac{\swh^2}{2\ctwh}\Bigg(
        \frac{\swh^2(\cwh^2+\ctwh)}{\ctwh^2}\frac{\CHD^2}{2}
        +\CHst
            \Bigg) \,;\\
    \label{eq:dgwsq}
    \de{6}{g^2} &= -\frac{\cwh^2}{\ctwh}\frac{\CHD}{2}\,; &
    \de{8}{g^2} &=\frac{\cwh^2}{2\ctwh}\Bigg(
        \frac{\cwh^2(\ctwh-\swh^2)}{\ctwh^2}\frac{\CHD^2}{2}
        -\CHst
    \Bigg) \, .
\end{align}
These can be used to calculate the shift in the $\mw^2$ prediction, which includes both the direct contribution from Eq.~\eqref{eq:mwEW} and the indirect ones from the derived parameters:
\begin{align}
    \begin{split}\label{eq:mw}
    \mw^2=&\,\frac{\hat{g}^2\vh^2}{4}\left(
      1
    + \de{6}{\mw^2}\,\frac{\vh^2}{\Lambda^2}
    + \de{8}{\mw^2}\,\frac{\vh^4}{\Lambda^4}
    \right) \, , \\
    \de{6}{\mw^2} =& \,\de{6}{g^2} + \de{6}{\vT^2}=\de{6}{g^2} \, , \\
    \de{8}{\mw^2} =& \,\de{6}{g^2} \de{6}{\vT^2} 
    + \de{8}{g^2}  +  \de{8}{\vT^2} + \DE{8}{\mw^2}=\de{8}{g^2} \, ,
    \end{split}
\end{align}
  where in the second equalities we have used the fact that $\de{6}{\vT^2}=0$ and $\de{8}{\vT^2} =-\DE{8}{\mw^2}$ in our case 
  (\emph{cf.} Eqs.~\eqref{eq:mwEW} and~\eqref{eq:dvt}).
\subsection{$Z$ and $W$ boson couplings\label{sec:ZW_couplings}}
Since we can neglect direct contributions from two-fermion operators that are not relevant for the extended scalar models considered here, only the indirect effects arising from the derived parameters affect the weak boson couplings to fermions. The generic $Z$ coupling to a pair of fermions $\psi$ of chirality $\chi$ is
\begin{align}
    G^{\sss Z,\psi}_{\sss\chi}=&g_{\sss Z}\left(T^{\sss \chi}_{\sss 3,\psi} - Q_{\sss \psi} s_{\sss Z}^2\right),\quad\left[ T^{\sss L}_{\sss 3,\psi}=T^{\sss \psi}_{\sss 3},\,T^{\sss R}_{\sss 3,\psi}=0\right] \, ,
\end{align}
where $T^{\chi}_{3,\psi}$ and $Q_\psi$ denote the third component of hypercharge and electric charge, respectively, and in the SM, $g_Z\to \hat{g}_Z=\sqrt{\hat{g}^\prime +\hat{g}^2}$ and $s_{\sss Z}^2\to \swh^2$. We note that, beyond dimension 6, $s_{\sss Z}^2$ does not correspond to the corrected sine of the Weinberg angle, $\sw^2$~\cite{Helset:2020yio,Corbett:2021eux}. However, this is not relevant for our models, which do not induce direct corrections to the weak mixing angle, in the sense that the $Z,\gamma$ eigenvalues are the same as for the SM mass matrix. Nevertheless, the input parameter shifts translate into corrections to the weak mixing angle, which, in turn, affect the chiral couplings of the Z boson:
\begin{align}
\begin{split}
\label{eq:dsw}
    \sw^2 =& \frac{\gpsqhat}{\hat{g}^2+\gpsqhat}
      + \de{6}{\sw^2}\vhoL{2} 
      + \de{8}{\sw^2}\vhoL{4} \,;\\
   \de{6}{\sw^2}=&-\frac{\hat{g}^2\gpsqhat}{(\hat{g}^2 + \gpsqhat)^2}\left(
      \de{6}{g^2}-\de{6}{\gpsq}
    \right)=\frac{\swh^2\cwh^2}{\ctwh}\frac{\Cp{HD}}{2} \,;\\
    \de{8}{\sw^2}=&
    -\frac{\hat{g}^2\gpsqhat}{(\hat{g}^2 + \gpsqhat)^2}\left(
      \de{8}{g^2}-\de{8}{\gpsq}
      \right)
    -\frac{\de{6}{\sw^2}}{\hat{g}^2 + \gpsqhat}\left(
    \hat{g}^2\de{6}{g^2}+\gpsqhat\de{6}{\gpsq}
    \right)\\
    =&\frac{\swh^2\cwh^2}{2\ctwh}\left(\Cpp{H^6}{(2)}+
    \frac{\swh^2\cwh^2}{\ctwh}\frac{\Cp{HD}^2}{2}\right).
\end{split}
\end{align}
The corrections to the overall $Z$ coupling prefactor, $g_Z$, read:
\begin{align}
  \begin{split}
  g_{\sss Z} =&\frac{\hat{g}}{\cwh}\left(
    1 +\de{6}{g_{\sss Z}}\vhoL{2} + \de{8}{g_{\sss Z}}\vhoL{4}
  \right) \, , \\
  \de{6}{g_{\sss Z}}=&\frac{1}{2}\frac{1}{\hat{g}^2+\gpsqhat}\left(
  \hat{g}^2\de{6}{g^2} + \gpsqhat\de{6}{\gpsq}
  \right)=-\frac{\CHD}{4} \, , \\
  \de{8}{g_{\sss Z}}
  =&\frac{1}{2}\Bigg(
    \frac{1}{\hat{g}^2+\gpsqhat}\left(
    \hat{g}^2\de{8}{g^2} + \gpsqhat\de{8}{\gpsq}
    \right) -  (\delta^{(6)}g_Z)^2
  \Bigg)=\frac{1}{4}\left(\frac{3\CHD^2}{8}-\CHst\right) \, . \\
  \end{split}
\end{align}
\noindent The effective interaction term can be expressed as: 
\begin{equation}
    \begin{gathered}
    \mathcal{L}_{Z\bar{\psi}\psi} = \hat{g}_Z\,\bar{\psi}\gamma^\mu( 
    G^{\sss Z,\psi}_{\sss V} -  G^{ \sss Z,\psi}_{ \sss A}\gamma_5
    )\psi\, Z_\mu \, : \\
    G^{ \sss Z,\psi}_{ \sss V} = 
     \frac{1}{2}(G^{ \sss Z,\psi}_{ \sss L} + G^{ \sss Z,\psi}_{ \sss R}) 
    =\frac{g_{\sss Z}}{\hat{g}_{\sss Z}}\left(\frac{T^{\psi}_{3}}{2} - Q_\psi s_{\sss Z}^2\right)\,;\quad
    G^{ \sss Z,\psi}_{ \sss A}= \frac{1}{2}(G^{ \sss Z,\psi}_{ \sss L} - G^{ \sss Z,\psi}_{ \sss R}) = \frac{g_{\sss Z}}{\hat{g}_{\sss Z}}\frac{T^{\psi}_{3} }{2}   \, .     
    \end{gathered}
\end{equation}
Defining the shifts as the difference between the SMEFT coupling and its SM version, the vector and axial-vector $Z$-boson couplings are corrected as follows:
\begin{align}
\begin{split}
\label{eq:Zcoup}
    \delta g^{\sss \psi}_{\sss V,A} =& G^{\sss Z}_{\sss V,A} - G^{\sss Z,\mathrm{SM}}_{\sss V,A}
    =\de{6}{g^{\sss \psi}_{\sss V,A}}\vhoL{2}
    +\de{8}{g^{\sss \psi}_{\sss V,A}}\vhoL{4} \, , \\
    \de{6}{g^{\sss \psi}_{\sss V}} =& \de{6}{g_{\sss Z}}\, G^{\sss Z,\mathrm{SM}}_{\sss V} 
    - Q_{\sss \psi}\de{6}{\sw^2}\, ,\\
   \de{8}{g^{\sss \psi}_{\sss V}} =& \de{8}{g_{\sss Z}}\, G^{\sss Z,\mathrm{SM}}_{\sss V} 
    - Q_{\sss \psi}(\de{8}{\sw^2} +  \de{6}{g_{\sss Z}}\de{6}{\sw^2} ) \, ,\\
   \de{6,8}{g^{\sss \psi}_{\sss A}} =& \de{6,8}{g_{\sss Z}}\, G^{\sss Z,\mathrm{SM}}_{\sss A} \, .
\end{split}
\end{align}
On our scenario, the $W$ boson only receives corrections to its couplings stemming from the overall factor of the weak coupling, $g$. It therefore maintains its left-handed coupling structure, proportional to the CKM and PMNS matrices for the quarks and leptons, respectively, \emph{i.e.}:
\begin{align}
\begin{gathered}
    \mathcal{L}_{W_\pm\bar{\psi}^\prime\psi} = \frac{\hat{g}}{\sqrt{2}}\bar{\psi}^\prime\gamma^\mu( 
    G^{\sss W_\pm,\psi}_{\sss V} -  G^{ \sss W_\pm,\psi}_{ \sss A}\gamma_5
    )\psi\, W^\pm_\mu ;  \\
        G^{ \sss W_\pm,\psi}_{ \sss V} =  G^{ \sss W_\pm,\psi}_{ \sss A}
    =\frac{g}{\hat{g}}\frac{\mathbf{V}}{2} \, ,\quad \mathbf{V}=V_{\sss CKM}\text{ or }V_{\sss PMNS}.
\end{gathered}
\end{align}
The coupling corrections are therefore given by:
\begin{align}
\begin{gathered}
\label{eq:Wcoup}
     \delta g^{\sss W_\pm,  \psi}_{\sss V,A} =\frac{\mathbf{V}}{2}\left(\sqrt{\frac{g}{\hat{g}}}-1\right)\approx
     \de{6}{g^{\sss W_\pm, \psi}_{\sss V,A}}\vhoL{2}
    +\de{8}{g^{\sss W_\pm, \psi}_{\sss V,A}}\vhoL{4} \,;\\
   \de{6}{g^{\sss W_\pm, \psi}_{\sss V,A}}= \frac{1}{4}\de{6}{g^2};\quad
   \de{8}{g^{\sss W_\pm, \psi}_{\sss V,A}} = \frac{1}{4}\left(\de{8}{g^2} - \frac{1}{4}(\de{6}{g^2})^2\right).
 \end{gathered}
\end{align}
The above coupling shifts, along with the $W$ mass prediction, are sufficient to compute the impact of our scalar extensions of the SM on the EWPOs. We note, in particular, that they only depend on the custodial symmetry-violating operators, $\Op{HD}$ and $\Opp{H^6}{(2)}$, which are only generated in the triplet scalar model, as
we show later. This reflects the fact that, as is well known, this model receives strong constraints from EWPOs, whereas the singlet model does not. The latter model is most strongly bounded by its modifications of Higgs boson couplings, which we discuss in the next Subsection.

\subsection{Higgs boson couplings\label{sec:higgs_couplings}}
\subsubsection{Yukawa couplings}
The SM Yukawa interactions generally receive corrections from higher-dimensional operators of the form:
\begin{align}
\frac{C_{fH}^{(n)}}{\Lambda^{n-4}}(H^\dagger H)^{\frac{n-4}{2}}\left(H \bar{F}_L f_r\right) + \mathrm{h.c.} \, ,
\end{align}
with $n=6,8,\dots$.
The top quark Yukawa coupling is the only relevant interaction for our purposes, and is modified by $\OtH$ and $\OquH$. After EWSB, the top quark mass term is
\begin{align}
\mathcal{L}_{\mt}=\left(-\frac{y_t \vT}{\sqrt{2}}  + \Delta \mt \right)\bar{t}_Lt_R + \mathrm{h.c.} \, ,
\end{align}
and the shift, $\Delta \mt$, can be absorbed by a redefinition of the Yukawa coupling, which then defines the mass parameter,
\begin{align}
    y_t^\prime &= y_t - \frac{\sqrt{2}}{\vT}\Delta m_t;\quad
    m_t=\frac{y_t^\prime \vT}{\sqrt{2}} \, .
\end{align}
The coupling to the Higgs boson generically receives different direct shifts, $\Delta y_t$, and can be written in terms of the input parameter, $m_t$,
\begin{align}
    \begin{split}
        \label{eq:Lhtt}
    \mathcal{L}_{h\bar{t}t}&=\left(-\frac{y_t}{\sqrt{2}} + \Delta y_t\right)h\bar{t}_Lt_R + \mathrm{h.c.}\,,\\
    &=\left(-\frac{\mt+\Delta m_t}{\vT} +\Delta y_t\right)
     h\bar{t}_Lt_R + \mathrm{h.c.} \, .
    \end{split}
\end{align}
To dimension-8, the shifts are:
\begin{align}
    \frac{\Delta m_t}{\vT}=\frac{\CtH}{2 \sqrt{2}}\voL{2}
    +\frac{\CquH}{4 \sqrt{2}}\voL{4} \, ,\quad
    \Delta y_t=\frac{3 \CtH}{2 \sqrt{2}}\voL{2}+
    \frac{5 \CquH}{4 \sqrt{2}}\voL{4} \, .
\end{align}
Including also the Higgs field redefinition and the extraction of $\vT$ (cf, Eqs.~\eqref{eq:hredef} and~\eqref{eq:dvt}) gives the following top Yukawa shifts in terms of the input parameters:
\begin{align}
    \begin{split}
  \mathcal{L}_{h\bar{t}t}\simeq&-\frac{\mt}{\vh} \left(
    1 
    + \de{6}{y_t}\vhoL{2}
    + \de{8}{y_t}\vhoL{4}
  \right)h\,\bar{t}_Lt_R+ \mathrm{h.c.}  \, :    
    \end{split}\\
    \begin{split}
    \label{eq:dyt}
  \de{6}{y_t} =& -\CtH\frac{\vh}{\sqrt{2}\mt}
  + \Delta^{\sss (6)}_h-\frac{1}{2} \de{6}{\vT^2}
  = \CHbx-\frac{\CHD}{4}-\CtH\frac{\vh}{\sqrt{2}\mt} \, ,
  \\
  \de{8}{y_t} =& 
    -\CquH\frac{\vh}{\sqrt{2}\mt}
  + \Delta^{\sss (8)}_h-\frac{1}{2}  \de{8}{\vT^2}
  -\CtH\frac{\vh}{\sqrt{2}\mt}\left(
    \Delta^{\sss (6)}_h+ \de{6}{\vT^2}
  \right)\\ 
  &+\frac{1}{2}\de{6}{\vT^2}\left(
    \Delta^{\sss (6)}_h+ \frac{3}{4}\de{6}{\vT^2}
  \right)\\
  &=-\CquH\frac{\vh}{\sqrt{2}\mt} - \frac{\CHst}{4}
  + \frac{3}{2}\left( \CHbx-\frac{\CHD}{4}\right)\left(
  \CHbx-\frac{\CHD}{4}-\CtH\frac{\sqrt{2}\vh}{3\mt} 
  \right) \, .
    \end{split}
\end{align}

\subsubsection{Gauge couplings}
The SM-like gauge couplings of the Higgs come from the mass generation mechanism, \emph{i.e.} its kinetic term. The $hVV$ couplings are therefore correlated with the particle masses, and in the SM the two are proportional. The relevant operators have exactly two covariant derivatives and a certain number of Higgs fields: $D^2H^{(d-2)}$, where $d$ is the operator canonical dimension. In the SM, we have $H^2\propto(v+h)^2$, hence a factor of two between the mass term and the Higgs couplings. At dimensions 6 \& 8, however, these factors are 4 and 6, respectively. This means that, as with the top quark above, the effect cannot be fully absorbed into the definition of the gauge boson masses. Furthermore, it leads to an increasing relative impact of the effective operators on the $hVV$ coupling with respect to the SM as the operator dimension increases.

As mentioned above, the operators in question do not affect the diagonalisation of the neutral gauge sector, so the $hZZ$ coupling is proportional to the $Z$ mass eigenvalue of Eq.~\eqref{eq:mzEW}, with an overall correction factor that accounts for the field redefinition for the Higgs boson:
\begin{align}
    \begin{split}
    \mathcal{L}_{hZZ}&=
    \frac{(g^2+g^{\prime 2})\vT}{4}
          \left(1+\Delta h_Z\right)
          h(1+\Delta h) Z^\mu Z_\mu \, , \\      
    \Delta h_Z &= \CHD\voL{2} + \frac{3}{4}\left(\CHso+\CHst\right)\voL{4} \, .
    \end{split}
\end{align}
Using Eqs.~\eqref{eq:mzEW},~\eqref{eq:hredef} and~\eqref{eq:dvt} yields the coupling shifts in terms of the input parameters:
\begin{align}
   \mathcal{L}_{hZZ}\simeq&\frac{\mz^2}{\vh}\left(
          1 
          + \de{6}{h_{\sss Z}}\vhoL{2}
          + \de{8}{h_{\sss Z}}\vhoL{4}
          \right)\,h Z^\mu Z_\mu \, ,\\
          \begin{split}
          \de{6}{h_{\sss Z}} =& \frac{\CHD}{2} + \Delta^{\sss (6)}_h
                               -\frac{1}{2}\de{6}{\vT^2}
                               =\CHbx+\frac{\CHD}{4} \, ,\\
          \de{8}{h_{\sss Z}} 
          =&\frac{1}{2}\left(
          \CHso+\frac{\CHst}{2}+3\CHbx^2-\frac{9}{16}\CHD^2-\frac{1}{2}\CHbx\CHD
          \right) \, .
         \end{split}
\end{align}
The analogous expression for the $hWW$ coupling in terms of the inputs, accounting for the fact that $\mw$ is not an input parameter, is:
\begin{align}
    \begin{split}
    \mathcal{L}_{hWW} &=
    \frac{g^2\vT}{2} \left(1+\Delta h_{\sss W}\right) 
                \left(1+\Delta_h\right)\,hW_+^\mu W^-_\mu \, ,\\
\Delta h_{\sss W} &= \frac{3}{4}(\CHso - \CHst)\voL{4} \, .
    \end{split}
\end{align}
In terms of the input parameters, we have:
\begin{align}
    \begin{split}
    \mathcal{L}_{hWW} = & \frac{2\mwhat^2}{\vh}\left(
   1
   + \de{6}{h_{\sss W}}\vhoL{2}
   + \de{8}{h_{\sss W}}\vhoL{4}
   \right)\,hW_+^\mu W^-_\mu \, ,\\
  \de{6}{h_{\sss W}} =&
   \de{6}{\mw^2}  + \Delta^{\sss (6)}_h 
   - \frac{1}{2}\de{6}{\vT^2}=
   \CHbx - \left(1+\frac{2\cwh^2}{\ctwh}\right)\frac{\CHD}{4} \, ,\\
   \de{8}{h_{\sss W}} =& \frac{1}{2}\bigg(\CHso
   -\left( \frac{3}{2} + \frac{\cwh^2}{\ctwh}\right)\left(\CHst+\CHbx \CHD\right)+3\CHbx^2\\
   &\quad\,\,+\left(
   \frac{\cwh^2}{\ctwh^3}(10\cw^4-8\cw^2+1)
   +\frac{3}{8}
   \right)\frac{\CHD^2}{2} \bigg) \, .  
    \end{split}
\end{align}

\subsubsection{Di-Higgs production\label{subsubsec:diHiggs}}

The gluon-fusion Di-Higgs production rate can be obtained from the dimension-8 Lagrangian using existing results that employ a general EFT formalism. Inclusive and differential production rates have been computed as a function of the following anomalous coupling Lagrangian~\cite{Goertz:2014qta,Azatov:2015oxa}, which contains all relevant, derivative-free interactions that can contribute to the $gg\to hh$ amplitude up to one loop:
\begin{align}
    \label{eq:Lgghh}
    \mathcal{L}_{gg\to hh} = 
    -m_t t\bar{t}\left(1 + c_t \frac{h}{\vh}+ c_{2t} \frac{h^2}{\vh^2}\right)
    -c_3\frac{\mh^2}{2\vh}h^3
    +\frac{g_s^2}{4\pi^2}\left(
      c_g\frac{h}{\vh}+ c_{2g}\frac{h^2}{\vh^2}
    \right)G^a_{\mu\nu}G^{a\,\mu\nu} \, .
\end{align}
In keeping with the rest of our assumptions, the only fermion-Higgs couplings we consider are those of the top quark. Besides these, dipole operators $\sim\bar{t}\,T^A\sigma_{\mu\nu}t\, G^{\mu\nu}_A$ can contribute to di-Higgs production starting from dimension-6, and their contributions to $gg\to hh$ at leading order are known. However, along with the gluon contact interactions, these are not generated in our models so we can safely neglect them, as well as $c_g$ and $c_{2g}$.
A generic EFT predicts in addition trilinear derivative interactions of the Higgs field $\sim h(\partial h)^2$ that can affect the process of interest, but are not present in Eq.~\eqref{eq:Lgghh}. However, it turns out that they can be removed by a suitable non-linear Higgs field redefinition, $h\to h + a h^2$, such that their effect is moved into a redefinition of the above couplings, as well as higher-point Higgs interactions that are not relevant for $gg\to hh$~\footnote{We note that this field redefinition does not respect the spontaneously broken $SU(2)$ symmetry of the SM. Nevertheless, it leads to an action that is equivalent for the leading-order predictions of interest to us.}. The appropriate field redefinition up to dimension-8 is
\begin{align}
    \label{eq:hredef_gghh}
    h\to h + \frac{h^2}{v}\bigg[
      \left(\CHbx-\frac{\CHD}{4}\right)\voL{2}
      - \frac{1}{4}\left(
        \CHso +  \CHst 
        - 12\left(\CHbx-\frac{\CHD}{4}\right)^2
      \right)\voL{4}
    \bigg] \, .
\end{align}
Defining the usual decomposition in terms of dimension-6 and -8 components
\begin{align}
    c_x = 1 + \de{6}{c_x} \vhoL{2}+ \de{8}{c_x} \vhoL{4}\,,
\end{align}
we find that:
\begin{align}
    \begin{split}
    \label{eq:c3}
    \de{6}{c_3}=& 3\left(\CHbx-\frac{\CHD}{4}\right) 
    - \frac{2 \vh^2}{\mh^2}\CH\,,\\
    \de{8}{c_3}=&
      -\frac{\CHso}{2}-\frac{3\CHst}{4}-\frac{4\vh^2}{\mh^2} \CHH
      +\frac{3}{2}\left(\CHbx-\frac{\CHD}{4}\right)\left(
        5\left(\CHbx-\frac{\CHD}{4}\right)-\frac{4\vh^2}{\mh^2} \CH
      \right) \, , 
    \end{split}
\end{align}
where $\de{6}{c_t}$ and $\de{8}{c_t}$ can be identified with the top Yukawa coupling shifts $\de{6}{y_t}$ and $\de{8}{y_t}$ from Eq.~\eqref{eq:Lhtt}, and 
\begin{align}
 \begin{split}\label{eq:c2t}
     c_{2t} =& 
      c_{2t}^{\sss (6)}\vhoL{2}
     + c_{2t}^{\sss (8)}\vhoL{2}\, ,\\
     c_{2t}^{\sss(6)}=&-\frac{3}{2}\frac{\CtH}{\hat{y}_t} + \CHbx-\frac{\CHD}{4}\,,\quad
    \hat{y}_t\equiv\frac{\sqrt{2}\hat{m}_t}{\vh} \, ,\\
     c_{2t}^{\sss (8)}=&-\frac{5}{2}\frac{\CquH}{\hat{y}_t} 
     -4\frac{\CtH}{\hat{y}_t}\left(
     \CHbx-\frac{\CHD}{4} 
     \right)
     -\frac{1}{4}\left( \CHso+\CHst\right)+4\left(\CHbx-\frac{\CHD}{4}\right)^2 \, ,
    \end{split}
\end{align}
where we haved used the fact that $\de{6}{\vT^2}=0$ in our case. The dimension-6 parts are consistent with previous results in the literature~\cite{Goertz:2014qta,Azatov:2015oxa}.

Beyond dimension-6, the possibility of contact terms involving higher derivatives of Higgs fields also arises. For example, among the $t\bar{t}hh$ interactions there are terms like $t\bar{t}(\partial h)^2$ starting from dimension-8, and similarly for the $gghh$ contact interaction. Such effects would induce genuinely new contributions to $gg\to hh$, beyond the effective Lagrangian of Eq.~\eqref{eq:Lgghh} that, to our knowledge, have not been calculated explicitly. This is particularly relevant in our case, since our scalar models generate $\OO^{\sss (1)}_{\sss qu H^3 D^2}=(D^\mu H)^\dagger(D^\mu H)\left(\bar{Q}t\tilde{H}\right)+\mathrm{h.c.}$ and $\OO^{\sss (2)}_{\sss qu H^3 D^2}=(D^\mu H)^\dagger\sigma^a(D^\mu H)\left(\bar{Q}t\sigma_a\tilde{H}\right)+\mathrm{h.c.}$, which predict such an interaction with top quarks proportional to $|y_t|$. 
We note that their contribution should be straightforwardly calculable, since the effective vertex that they introduce has an identical spinor structure to the SM Yukawa, apart from an additional dependence on the momenta of the external Higgs bosons. Adding such an effective interaction term to the top quark part of our effective Lagrangian, 
\begin{align}
    \mathcal{L}_{hhtt} = -m_t t\bar{t}\left(1 + c_t \frac{h}{\vh}+ c_{2t} \frac{h^2}{\vh^2}+ c_{2t,\partial} \frac{(\partial h)^2}{\vh^4}\right) \, ,
\end{align}
one can check that the ratio of the $t\bar{t}hh$ Feynman rules induced by $c_{2t,\partial}$ and $c_{2t}$ is
\begin{align}
    \frac{\Gamma_{2t,\partial}}{\Gamma_{2t}}=-\frac{c_{2t,\partial}}{c_{2t}}\frac{p_3\cdot p_4}{\vh^2} \, ,
\end{align}
where we label the momenta of the two external Higgs bosons by $p_3$ and $p_4$. The form factors entering  the $gg\to hh$ amplitudes should therefore be unchanged, and one should be able to obtain the $c_{2t,\partial}$ contribution from that of $c_{2t}$, simply rescaling by the above factor, where $p_3\cdot p_4=\frac{s}{2}-\mh^2$. An explicit calculation of such a contribution is beyond the scope of our current investigation, and we leave it for further work. 

This neglect is justified in our study of the singlet and triplet scalar models for the following reason. In both cases, the relevant Wilson coefficient $C^{\sss (1)}_{\sss qu H^3 D^2}$ is proportional to the trilinear coupling of the heavy scalar with a pair of Higgs bosons, $\ks$ or $\kappa_\Xi$, for the singlet and triplet, respectively. These are significantly constrained already at dimension-6 by Higgs signal strength measurements and EWPOs, respectively. 
Indeed, we find that the impact of di-Higgs cross section measurements is limited to providing additional sensitivity to other parameters of the scalar potential for the singlet, and is completely negligible in the triplet case.
Assuming that the inclusive di-Higgs rate is dominated by the threshold region, where $s\approx4\mh^2\to p_3\cdot p_4\approx\mh^2$, we can roughly estimate the contribution of $c_{2t,\partial}$ by shifting $c_{2t}\to c_{2t}- c_{2t,\partial} (\mh^2/\vh^2)$. In parameter regions of the singlet model where an interplay between single and di-Higgs data is observed, we found that this approximation amounts to a 1--2\% effect on the di-Higgs production rate.
 We therefore do not expect such contributions to have significant impact on the results of the present study, although they merit future investigation, since they should yield much larger effects in the high-energy tails of $gg \to hh$.

\subsection{Quartic Higgs self-coupling\label{subsec:higgs_quartic}}
Further field redefinitions involving higher powers of $h$: $h\to h+ b h^3 + c h^4\dots$ can be performed to remove successively higher-point, 2-derivative Higgs self-interactions in favour of even higher-point, derivative-free contact interactions, without affecting the lower-point self-interactions. For example, in order to remove the 2-derivative, 4-point self-couplings $\sim h^2(\partial h)^2$, one can set 
\begin{align}
    b=\frac{1}{\vT^2}\bigg[
    \frac{1}{3}\left(\CHbx-\frac{\CHD}{4}\right)\voL{2}
    - \left(\frac{\CHso+\CHst}{4}-4\left(\CHbx-\frac{\CHD}{4}\right)^2
    \right)\voL{4}
    \bigg] \, .
\end{align} 
In this basis, the corrections to the quartic Higgs self-coupling, $d_4$, defined as:
\begin{align}
    \mathcal{L}\supset-\frac{\mh^2}{8\vh^2}\left(
      1+d_4^{\sss (6)}\vhoL{2}+d_4^{\sss (8)}\vhoL{4}
    \right)h^4 \, ,
\end{align}
are
\begin{align}
    \label{eq:d4}
    d_4^{\sss (6)}=& \frac{50}{3}\left(\CHbx-\frac{\CHD}{4}\right) 
    - \frac{12 \vh^2}{\mh^2}\CH \,,\\
    d_4^{\sss (8)}=&
      -5\CHso-\frac{11\CHst}{2}-\frac{32\vh^2}{\mh^2} \CHH
      +8\left(\CHbx-\frac{\CHD}{4}\right)\left(
        11\left(\CHbx-\frac{\CHD}{4}\right)-\frac{9\vh^2}{\mh^2} \CH
      \right) \, .
\end{align}
However, at dimension-8 and beyond there are four-derivative self-interactions $\sim(\partial h)^4$ that cannot be removed in this way. These would be relevant for triple-Higgs production at colliders and are generated by operators like $C_{\sss H^4}^{\sss(1)}$ and $C_{\sss H^4}^{\sss(3)}$. Such operators also contribute to longitudinal vector boson scattering and their coefficients are bounded by recent LHC measurements at the level of 2--4 TeV$^{-4}$~\cite{CMS:2019qfk}. Finally, they are also subject to positivity bounds~\cite{Zhang:2018shp} arising from basic properties of the $S$-matrix, if one assumes a UV completion that can be described by a QFT.
\section{Experimental constraints\label{sec:exp_constraints}}
\subsection{Input data}
The coupling shifts induced by the set of operators considered can be constrained by the measurements of EWPOs, Higgs boson signal strengths and multi-Higgs production processes. We take these constraints into account via global fits to the underlying model parameters using the \verb|fitmaker| framework~\cite{Ellis:2020unq}. We make use of the EWPOs present in the public version of the code, which include the pseudo-observables measured on the $Z$ resonance by LEP and SLD~\cite{ALEPH:2005ab}, together with a combination of pre-2022 $W$ boson mass measurements by CDF and D0 at the Tevatron and ATLAS at the LHC~\cite{CDF:2013dpa}, which we naively combine in a uncorrelated way with the recent LHCb measurement~\cite{LHCb:2021bjt}:~\footnote{We discuss later some potential implications of the recent CDF measurement~\cite{CDF:2022hxs}.}
\begin{align}
   \quad \{\Gamma_Z, \sigma^0_\text{had.}, R_l^0, A_{FB}^l, A_l, R_b^0, R_c^0, A_{FB}^b, A_{FB}^c, A_b, A_c, M_W \} \, .
\end{align}
Additionally, we implement a private version of the \verb|fitmaker| code that incorporates the latest Higgs signal strength measurements by ATLAS~\cite{ATLAS:2022vkf,hepdata.130266} and CMS~\cite{CMS:2022dwd,hepdata.127765}, taken from the respective \verb|HEPData| records. Finally, for di-Higgs production, we interpret recently published upper bounds on the total production rate at the LHC as a measurement, assuming that the SM value is observed and that the upper bound corresponds to a 95\% Confidence Level (C.L.) one-sided upper bound derived from a Gaussian PDF. The ATLAS and CMS experiments quote upper bounds of 2.4~\cite{ATLAS:2022kbf} and 3.4~\cite{CMS:2022dwd} times the SM, from which we extract our approximate signal strengths and 1-$\sigma$ uncertainties as $\mu^{hh}_{\mathrm{ATLAS}}=1\pm0.907$ and $\mu^{hh}_{\mathrm{CMS}}=1\pm1.29$.

\subsection{Theoretical predictions}
Our analysis is simplified by the fact that the operators of interest only lead to shifts of SM couplings, rather than introducing new Lorentz structures. The predictions for the EW precision observables in terms of the $Z$ boson coupling shifts of Eq.~\eqref{eq:Zcoup} are derived from known, tree-level expressions in terms of generic vector and axial-vector $Z$ boson couplings to leptons and quarks, assuming the narrow width approximation for the $Z$ (see, \emph{e.g.}, Ref.~\cite{Brivio:2017vri}), and expanding out to dimension 8. 

For the Higgs boson signal strengths, we similarly assume the narrow-width approximation, decomposing the rates into the production cross-section multiplied by the decay branching fraction into a given final state. As already mentioned, in our restricted scenario, $gg\to h$ and $h\to gg/\gamma\gamma/\gamma Z$ are only modified by top quark Yukawa coupling shifts (\emph{cf}. Eq.~\eqref{eq:dyt}). For Higgs production in association with a vector boson, we can account for the modifications by combining the relative shifts of the Higgs coupling to the corresponding gauge boson, and of its partial width to a given quark-anti-quark channel, which has the same coupling dependence $\sim ((G^{q}_{\sss V})^2+(G^{q}_{\sss A})^2)^2$ as the $q\bar{q}\to Vh$ amplitude. For $Zh$ production, the contributions from coupling shifts of up and down quarks were weighted by the relative importance of the $u\bar{u}$ and $d\bar{d}$ initial states, computed using \verb|Madgraph5_aMC@NLO|~\cite{Alwall:2014hca}, assuming the SM hypothesis and using the \verb|NNPDF3.1| PDF sets~\cite{NNPDF:2017mvq}. For $Wh$ production, this approach neglects the effect of the $W$ mass shift. This should not have any impact on the sensitivity of Higgs data, since that shift is strongly constrained by the EWPO dataset. We perform a similar computation for Vector Boson Fusion (VBF) Higgs production, which is complicated by the fact that the process is mediated by both $W$ and $Z$ intermediate bosons. We approximate the modification of this process by splitting the cross-section into $W$- and $Z$-mediated parts, again using the integrated channel information obtained with \verb|Madgraph5_aMC@NLO|. The overall modification is then a weighted rescaling according to the respective $W$ and $Z$ coupling shifts. This neglects the effect of couplings shifts on the interference term  between the two amplitudes. Furthermore, shifts of the $W$ mass and $W/Z$ widths are also neglected. Finally, the $t\bar{t}h$ signal strength modifier is obtained from the top quark Yukawa coupling shift. 

Since we neglect all Yukawa couplings besides that of the top, the only other, tree-level decay rates that we need to compute are those to four fermions via intermediate $W$ and $Z$ bosons. For these modes, we assume that they proceed via one on-shell and one off-shell state, \emph{i.e.}, $h\to W/Z f^\prime\bar{f}\to 4f$. We can therefore combine the rescalings of the relevant Higgs-$W/Z$ coupling and the fermionic coupling of the off shell-state, as we did for the $Vh$ production processes. The shift of the $W$ mass is taken into account via numerically determined dependences derived from the phase space integral for the 3-body decay quoted in Ref.~\cite{Rizzo:1980gz}. Finally, we multiply by the branching fraction correction to the relevant final state for the on-shell $W/Z$ to obtain the net correction. This approach neglects width effects in the off-shell leg, as well as interference between crossed diagrams mediated by the $W$ and $Z$ present in certain channels. Photon-mediated diagrams are also not taken into account and, although they are generically present at tree level in the SMEFT, the relevant operators are not generated by the models of interest to us, and are therefore not relevant for our study.

Finally, the inclusive di-Higgs production rates are obtained as functions of the effective couplings of Eq.~\eqref{eq:Lgghh}. Specifically, we make use of the analytic parametrisation published in Ref.~\cite{Carvalho:2015ttv}, in the form of a quartic polynomial in the effective couplings. We expand this polynomial in order to isolate the dimension-6 and -8 parts as functions of the parameters in Eqs.~\eqref{eq:dyt}, \eqref{eq:c3} and~\eqref{eq:c2t}, truncating the remaining terms. We use predictions for the inclusive signal strengths as inputs to the statistical analysis, neglecting the impact of the models on the decay branching fractions that may differ for individual di-Higgs channels. Branching fraction effects are confirmed to be sub-leading with respect to the total cross section in the relevant regions of parameter space, given the relatively tight constraints on the former from single-Higgs data. 

\subsection{Statistical interpretation}
Combining the input measurements and the theoretical predictions, we construct a $\chi^2$ function:
\begin{align}
    \chi^2 \big(\vec{\theta}\,\big)= \left(\vec{\mu}_{\mathrm{obs.}} - \vec{\mu}_{\mathrm{th.}}\!\big(\vec{\theta}\,\big)\right)^\top \!\!
    \cdot \mathbf{V}^{-1}\cdot
    \left(\vec{\mu}_{\mathrm{obs.}} - \vec{\mu}_{\mathrm{th.}}\!\big(\vec{\theta}\,\big)\right),
\end{align}
which is used as the log-likelihood for the analysis. Here $\vec{\mu}_{\mathrm{obs.}}$ represents the experimentally-determined values of the observables, $ \mathbf{V}^{-1}$ is the associated inverse covariance matrix for the dataset, and $\vec{\mu}_{\mathrm{th.}}$ denotes the theoretical predictions for the observables, as functions of the parameters of interest, $\vec{\theta}$, which in our case are the model parameters appearing via the Wilson coefficients of the SMEFT. We extract confidence intervals as regions of parameter space where $\Delta \chi^2$, defined with respect to the global minimum of the $\chi^2$ function, $\chi^2_\mathrm{min.}$, is below a critical value, $\chi^2_c$, which depends on the number of degrees of freedom, $n_p$ in the model:
\begin{align}
    \Delta\chi^2\big(\vec{\theta}\,\big)\equiv \chi^2\big(\vec{\theta}\,\big) - \chi^2_{\mathrm{min.}}\leq\chi^2_c;\quad \chi^2_c=3.84,5.99,\dots\text{ for }n_p=1,2,
\dots\,\text{at 95\% CL.}
\end{align}
We also define a profiled log-likelihood for particular parameters of interest as the resulting $\chi^2$ function, having minimised over all other degrees of freedom. We use the same criteria as above to determine the profiled confidence intervals. 
\section{Tree-level Matching}
\label{sec:tree}

Having determined the impacts of the relevant Wilson coefficients on our observables of interest up to dimension-8, the next step is to derive their values in the single-field extensions of the SM considered in this work. To this end, we match the model parameters to the SMEFT coefficients at tree level, integrate out the scalar fields
in a gauge-covariant way using the Covariant Derivative Expansion (CDE) method~\cite{Zuk:1985sw,Cheyette:1985ue,Gaillard:1985uh} (see, \emph{e.g.},~\cite{Henning:2014wua,Drozd:2015rsp} for more recent reviews and applications to the SMEFT). We briefly review the method before applying it at tree level to our scalar extensions of the SM.

Consider a UV model with a heavy scalar field $\Phi$ of mass $m_{\Phi}$ that we would like to integrate out and match to the SMEFT at some high-energy scale, $\Lambda$. The action containing $\Phi$ and its interactions with the SM
fields $\phi$ is $S[\Phi,\phi]$, so that the action describing the EFT at
$\Lambda \sim m_{{\Phi}}$ is given by:
\begin{equation}
    e^{iS_{\text{eff}}[\phi]} = \int \mathcal{D}\Phi e^{iS[\Phi,\phi]}\,,
\label{eq:S_EFT}
\end{equation}
where $S_{\text{eff}}[\phi]$ contains only the SM fields. The effective
action can be computed in the standard way by expanding $\Phi$ around its
minimum, $\Phi = \Phi_c + \eta$, where $\Phi_c$ is determined by solving
the classical equation of motion. Expanding the action $S[\Phi,\phi]$
around this minimum and computing the integral
gives~\cite{Henning:2014wua}:
\begin{equation}
    S_{\text{eff}}[\phi] \approx S[\Phi_c] + \frac{i}{2} \text{Tr} \log \left(-\frac{\delta^2 S}{\delta \Phi^2}\Bigr| _{\Phi=\Phi_c}\right)
\end{equation}
up to one-loop order. We focus here on the tree-level effective action,
which corresponds to the first term in the equation above, which is given
by replacing $\Phi$ by the classical field $\Phi_c$ in $S[\Phi,\phi]$. 

A tree-level contribution to the effective action arises only when the UV Lagrangian $\mathcal{L}[\Phi,\phi]$ contains a term that is linear in the heavy field $\Phi$~\cite{Henning:2014wua}. Hence we consider a Lagrangian 
\begin{equation}
   \mathcal{L}[\Phi,\phi] \supset \eta[\Phi^{\dagger}(-D^2 - \mphi^2-U(x))\Phi] + [\Phi^{\dagger}B(x) +\text{h.c.}] + \mathcal{O}(\Phi^3) 
  \label{eq:L_UV}
\end{equation}
for a UV model containing a real ($\eta = 1/2$) or complex ($\eta = 1$)
scalar field, where $B(x)$ and $U(x)$ are functions of the SM fields
$\phi(x)$. The classical field, $\Phi_c$,
is found by solving the corresponding equation of motion:
\begin{equation}
    \frac{\delta \mathcal{L}[\Phi, \phi]}{\delta \Phi} = 0 \; \Rightarrow  \; (-D^2-\mphi^2-U(x))\Phi_c = - B(x) +\mathcal{O}(\Phi^2) \, .
\label{eq:EOM}
\end{equation}
Taking the linear approximation yields the solution
\begin{align}
\label{eq:linsol}
     \Phi^{\sss (1)}_c \approx -\dfrac{1}{P^2-\mphi^2-U(x)}B(x)\,,
\end{align}
where $P_{\mu} \equiv iD_{\mu}$ is the covariant derivative. Note, however, that mass scales other than $\mphi$ can also arise in the UV Lagrangian, \emph{i.e.}, within $B(x)$ or in front of the $\Phi^3$ interaction terms.
Denoting a generic new physics mass scale by $M$, one can see that, in the absence of tadpole terms for $\Phi$, $B(x)$ can be at most $\mathcal{O}(M)$ so that, in the linearised solution of Eq.~\eqref{eq:EOM}, $\Phi^{\sss (1)}_c$ is of order $1/M$. Substituting $\Phi^{\sss (1)}_{c}$ back into~Eq.~\eqref{eq:L_UV} yields the 
tree-level effective Lagrangian:
\begin{equation}
    \mathcal{L}_{\text{eff,tree}} = - \eta B^{\dagger}(x) \dfrac{1}{P^2-m_{\sss{\Phi}}^2-U(x)}B(x) + \mathcal{O}\!\left(({\Phi_c^{\sss (1)}})^3\right)\,,
\end{equation}
where the substitution should also be performed in terms with higher powers of $\Phi_c^{\sss (1)}$, which we have omitted here for brevity.
One can then perform a CDE to obtain the local operators that are generated by integrating out the heavy field:
\begin{equation}
\begin{split}
    &\dfrac{1}{P^2-m_{\sss{\Phi}}^2-U(x)} = - \left[1 - \frac{1}{m_{\sss{\Phi}}^2}(P^2-U)\right]^{-1}\frac{1}{m_{\sss{\Phi}}^2} = \\
    &= - \frac{1}{m_{\sss{\Phi}}^2} - \frac{1}{m_{\sss{\Phi}}^2}{(P^2-U)}\frac{1}{m_{\sss{\Phi}}^2} -  \frac{1}{m_{\sss{\Phi}}^2}{(P^2-U)}\frac{1}{m_{\sss{\Phi}}^2}{(P^2-U)}\frac{1}{m_{\sss{\Phi}}^2} - ...\,.
\end{split}
\end{equation}
The factor $1/m_{{\Phi}}^2$ is the inverse of the mass-squared matrix, 
and may not commute with $U$, so one needs to be careful about the
ordering when performing the CDE. It leads
to the following effective Lagrangian in the inverse mass expansion:
\begin{equation}
    \mathcal{L}_{\text{eff,tree}} = \eta \bigg( B^{\dagger}\frac{1}{m_{\sss{\Phi}}^2}B+B^{\dagger}\frac{1}{m_{\sss{\Phi}}^2}(P^2-U)\frac{1}{m_{\sss{\Phi}}^2}B+...\bigg)+\mathcal{O}(\Phi_c^3)\,.
\label{eq:L_eff}
\end{equation}
As $B$ and $U$ depend only on the SM fields, the heavy field $\Phi$
has been integrated out and the UV model can be matched to the local operators of a particular SMEFT basis, in
which the effects of the heavy field are encoded in terms scaled by 
inverse powers of $\Lambda = m_{{\Phi}}$.

It turns out that this level of approximation is sufficient to perform the tree-level matching up to order $1/M^2$, {i.e.}, dimension-6. This is not immediately obvious, since potential subleading terms in the solution to the classical equation of motion of order $1/M^2$ or $1/M^3$ look like they might generate operators suppressed by $1/M^2$. Going beyond the linearised solution relies on additional details of the model not specified in Eq.~\eqref{eq:L_UV}, namely the nature of the self-interactions. If there are none, the linearised solution of Eq.~\eqref{eq:linsol} is exact. If a cubic or quartic self-interaction is present, the equation of motion will include a $\Phi^2$ or $\Phi^3$ term. A solution can be found iteratively, starting from $\Phi^{\sss (1)}_c$, which generates subleading terms of order $1/M^3$ or $1/M^5$ from the cubic and quartic interactions respectively. This indicates that a higher-order solution to the classical equation of motion is needed to obtain the correct matching conditions up to dimension-8, as has also been shown to be the case for the 1-loop effective action at dimension-6~\cite{Haisch:2020ahr}. In Section~\ref{sec:singlet} we go through the tree-level exercise explicitly.

Performing calculations in the full UV theory will in general
yield predictions that are more accurate than those obtained using the
SMEFT. However, we assume that their difference can be neglected 
at the level of the current experimental precision~\cite{Brivio:2017vri}. 
The accuracy of the SMEFT depends on the order of the inverse mass expansion
in $\mathcal{L}_{\text{eff,tree}}$ (\ref{eq:L_eff}), and keeping
higher-order terms systematically increases the accuracy of the
predictions. The tree-level matching to dimension-6 has already been
considered extensively in the literature~\cite{Henning:2014wua,deBlas:2017xtg,Jenkins:2017jig,
Corbett:2017ieo}. However, the possibility of reaching a higher level 
of precision in the SMEFT predictions provides motivation to perform 
the matching to include higher-dimensional terms. This will allow us to quantify the validity of the EFT expansion given the data at hand and more accurately reflect how it translates into bounds on the underlying model parameters.

In the following Sections we perform the tree-level matching up to dimension-8 for the singlet and triplet scalar extensions of 
the SM, deriving the coefficients of the dimension-6
and -8 operators relevant for our analysis displayed in 
Tables~\ref{tab:dim4-6} and \ref{tab:dim8}. We have checked that the coefficients resulting from matching to dimension-6 are consistent with those obtained in Ref.~\cite{deBlas:2017xtg}. We also perform a validation of a subset of the coefficients by explicit computation of the $hh\to hh$ scattering amplitude detailed in Appendix~\ref{app:hhhh} as well as explicit computations of the shifts to Higgs boson couplings and the $W$-boson mass prediction in the singlet and triplet models, respectively.

\section{\label{sec:singlet}Singlet scalar model}
In this Section we consider the SM extended by a single real
singlet scalar field $S$ with hypercharge $Y_{\sss S}=0$, performing the tree-level matching to the SMEFT up to dimension-8. A subset of these matching calculations were performed in Ref.~\cite{Corbett:2015lfa}. We do not consider one-loop corrections to the tree-level dimension-6 matching, which have been considered in~\cite{Jiang:2018pbd,Haisch:2020ahr}, nor the associated constraints taking into account renormalisation-group evolution effects~\cite{Dawson:2021jcl}. 

The only possible interactions between $S$ and the SM fields consistent 
with the SM gauge symmetry group are the so-called portal interactions between $S$ and a pair of SM Higgs
SU(2) doublets, $H^\dagger H$. Following the conventions of Ref.~\cite{deBlas:2017xtg}, the corresponding
model Lagrangian terms are:
\begin{equation}
\begin{split}
    \mathcal{L}_s =& \, \frac{1}{2}(D_{\mu}S)(D^{\mu}S) -  \frac{1}{2}\ms^2 SS - (\kappa_{s}) SH^{\dagger}H \\
    &- (\lambda_{s}) SSH^{\dagger}H - \ksss SSS - (\kssss) SSSS\,,
\end{split}
\label{eq:L_S}
\end{equation}
where we have included terms describing the triple and quartic self-interactions
of the $S$, and the $S$ couplings have mass dimensions
$[\ks]=[\ksss]=1$ and $[\lam{S}]=[\kssss]=0$. In general, a tadpole term linear in $S$ is also permitted by the symmetries of the model. However, it can be removed by shifting the $S$ field by a constant and redefining the remaining parameters. Since the field shift does not affect $S$-matrix elements, physical observables are unchanged and the two theories are equivalent. We note, however, that in the above representation the $S$-field generically obtains a vev and refer the reader to Appendix~\ref{app:hhhh} for more details. 

For future reference, we note that the
following constraints must be satisfied in order
for the scalar potential to be bounded from below~\cite{Chen:2014ask}:
\begin{equation}
\lambda, \kssss \ge 0, \; |\lam{S}| \ge -4 \sqrt{\lambda \kssss} \, .
\label{bounded}
\end{equation}
Furthermore, the following constraints can be derived by imposing partial-wave unitarity of the $2\to2$ bosonic scattering matrix~\cite{Dawson:2021jcl}:
\begin{equation}
|\lambda| \le \frac{8\pi}{3}, \; |\lam{S}| \le 4\pi, \; |\kssss| \le \frac{2\pi}{3} \, .
\label{perturbation}
\end{equation}
Moreover, the scalar potential generally admits several minima as well as the EW minimum. We therefore ensure that, for a given point in parameter space, the EW minimum is the global minimum, which we do by a numerical procedure informed by the discussions in Refs.~\cite{Espinosa:2011ax,Chen:2014ask}.

As the Lagrangian contains an interaction term that is linear in $S$, there are
tree-level contributions to the effective action.
We evaluate these using the CDE method described in the previous Section.
Writing Eq.~\eqref{eq:L_S} in a form similar to Eq.~\eqref{eq:L_UV}:
\begin{equation}
\label{eq:L_S_B_and_U}
    \mathcal{L}_{s} = \frac{1}{2} S(-D^2 - \ms^2-U)S - SB - \ksss SSS - (\kssss) SSSS\,,
\end{equation}
we have
\begin{equation}
    U = 2(\lambda_{s})H^{\dagger}H \hspace{0.7cm} \text{and} \hspace{0.7cm} B = (\kappa_{s})H^{\dagger}H\,.
\end{equation}
The equation of motion may be written as:
\begin{equation}
    \Delta^{-1}S = B + 3(\kappa_{\sss{S^3}})SS + 4(\kappa_{\sss{S^4}})SSS\,,
\end{equation}
where we have defined the inverse propagator
\begin{align}
    \oo^{-1} \equiv -D^2-\ms^2-U \, .
\end{align}
As discussed in Section~\ref{sec:tree}, to linear order the solution is
\begin{equation}
\label{eq:linearsol}
  S_c^{\sss(1)} = \oo B +\omm{3} \, ,
\end{equation}
where $M$ denotes a generic UV mass parameter in the set $\{\ms, \ks,\ksss\}$.
The higher-order solution needed to obtain the matching through dimension-8 can be found iteratively from Eq.~\eqref{eq:linearsol}, and is given by
\begin{align}
  \label{eq:sol}
  S_c &= \oo \bigg[B + 3\,\ksss\,(\oo B)^2 
  + 18\,\ksss^2\,\oo \Big[(\oo B)^2\Big]\oo B 
  + 4\,\kssss\,(\oo B)^3\bigg]+ \omm{6}\,,
\end{align}
which introduces a dependence on the heavy scalar self-interactions.
The square brackets emphasise that the differential operator $\oo$ is acting on the entire expression within. The first additional term is $\omm{3}$ while the second and third are $\omm{5}$.
Performing an inverse mass expansion for $S_c$ and plugging it back in (\ref{eq:L_S}), we find the following tree-level effective Lagrangian 
up to dimension-8:
\begin{equation}
\begin{split}
    \begin{split}
    \mathcal{L}^{(8)}_{\sss \text{eff.,tree},S} &= 
     \frac{1}{2\ms^2}BB
    +\frac{1}{2\ms^4}B(P^2-U)B
    +\frac{1}{2\ms^6}B(P^2-U)(P^2-U)B+\frac{\ksss}{\ms^6}BBB \\
    &+ \frac{3\ksss}{\ms^8}BB(P^2-U)B
    - \frac{\kssss}{\ms^8} BBBB + \frac{9\ksss^2}{2\ms^{10}} BBBB + \omm{5}\,.
    \end{split}
    \label{eq:L_eff_S}
\end{split}    
\end{equation}
The last term comes from the higher-order solution in Eq.~\eqref{eq:sol}.
Evaluating each term in Eq.~(\ref{eq:L_eff_S}) 
and using the SM Higgs equation of motion: 
\begin{equation}
    D^2H = \mu^2H - 2\lambda(H^{\dagger}H)H - y^{\dagger}_e (\bar{e}l) - y_u \varepsilon_{ik}(\bar{q}^k u) - y^{\dagger}_d (\bar{d}q)
    \label{eq:Higgs_EOM}
\end{equation}
to remove higher-order covariant derivatives in order to match the Warsaw basis~\cite{Buchmuller:1985jz,Grzadkowski:2010es} of dimension-6 operators and the dimension-8 basis of Ref.~\cite{Murphy:2020rsh} yields the Wilson coefficients reported in Table~\ref{tab:singlet_results}. Since our basis reduction only involves eliminating dimension-8 operators, we can safely use the SM equation of motion of Eq.~\eqref{eq:Higgs_EOM} to this order, which is not the case for the triplet model considered in the next Section.

Besides operators of dimension-6 and beyond, integrating out the $S$ field also generates a dimension-4 $(H^\dagger H)^2$ term that shifts the coupling of the quartic potential term, $\lambda$. This effect can be absorbed in the definition of $\lambda$ after integrating out the heavy field, and is therefore unobservable. However, if $\lambda$ appears in the matching at dimension-6, the redefinition will induce a shift of the matching in terms of the new $\lambda$ at dimension-8. This does not occur for the singlet model, but does for the triplet model. In the expressions for the dimension-6 Wilson coefficients of Table~\ref{tab:singlet_results}, we have retained higher-order terms, suppressed by $\mu^2/\ms^2$ with respect to the leading contributions, that arise from the application of the Higgs equation of motion, Eq.~\eqref{eq:EOM}. The dimensionful parameter, $\mu^2$, lowers the effective dimension of an operator by two units, generating dimension-6 operators from dimension-8 ones involving $D^2H$. These pieces are genuine dimension-8 contributions and form a part of the complete $\mathcal{O}(1/M^4)$ corrections induced by integrating out the $S$ field. Extending the analysis to dimension-8 therefore involves not only calculating the coefficients of the dimension-8 operators but also corrections to the matching conditions for those of dimension-6.
In an analysis truncated up to $\mathcal{O}(1/\Lambda^2)$, one would not typically include these terms, however the possibility of including these types of partial higher-order corrections has been previously considered as part of the so-called ``$v$-improved matching'' procedure~\cite{Brehmer:2015rna}. 

In Appendix~\ref{app:hhhh}, we validate our results obtained using the CDE by calculating the matched SMEFT amplitude for $hh\to hh$ scattering and comparing it with that of the full theory, expanded to $\OO(1/M^4)$. The EFT reproduces the full theory to dimension-8 only when the higher-order solution to the equation of motion is used and the dimension-8 corrections to the dimension-6 coefficients are taken into account. Modulo the additional contribution from the higher-order equation of motion, our results agree with a matching calculation of Higgs operators up to dimension-8 reported in Ref.~\cite{Corbett:2015lfa}.
\begin{table}[h!]
    \begin{center}  
    \renewcommand{\arraystretch}{1.5}
    \resizebox{\textwidth}{!}{
    \begin{tabular}{|c||c|c|}
   \hline 
     \multirow{2}{4em}{Dim - 6} & $C_H$ & $-\frac{\ks^2 }{\ms^2}\left(\ls\left(1-\frac{4\mu^2}{\ms^2}\right) - \frac{\ks \ksss}{\ms^2}\left(1-\frac{6\mu^2}{\ms^2}\right)\right)$ \\ [0.6ex]
     \cline{2-3}
      & $C_{H\Box}$ & $ -\frac{\ks^2}{2\ms^2}\left(1- \frac{4\mu^2}{\ms^2}\right)$ \\[0.6ex]
     \hline \hline 
     \multirow{18}{4em}{Dim - 8} & $C_{H^8}$ & $\frac{\ks^2}{\ms^2}\Bigl(2(\ls - 2\lambda)^2 - \frac{6\ks \ksss}{\ms^2}(\ls - 2\lambda) + \frac{\ks^2}{\ms^2}\Bigl(\frac{9\ksss^2}{2\ms^2} - \kssss \Bigr)\Bigr)$ \\[0.6ex]
     \cline{2-3}
     & $C_{H^6}^{(1)}$ & $\frac{2   \ks^2}{\ms^2}\Bigl(2(\ls - 2\lambda) - \frac{3\ks \ksss}{\ms^2}\Bigr)$ \\[0.6ex]
     \cline{2-3}
     & $C_{H^4}^{(3)}$ & $\frac{2\ks^2}{\ms^2}$ \\[0.6ex]
     \cline{2-3}
     & $[C_{l\psi H^5}/C_{q\psi H^5}]_{\text{\tiny wx}}$ & $-[y_\psi]_{\text{\tiny wx}}\frac{\ks^2}{\ms^2}\Bigl(2( \ls-2\lambda ) - \frac{3\ks \ksss}{\ms^2}\Bigr)$; $\psi=u,d,e$ \\[0.6ex]
     \cline{2-3}
     & $[C_{l^2\psi^2H^2}^{(1)}/C_{q^2\psi^2H^2}^{(1)}]_{\text{\tiny wxyz}}$ & $-[y_\psi]_{\text{\tiny wz}}[y_\psi^\dagger]_{\text{\tiny yx}} \frac{\ks^2}{4\ms^2}$; $\psi=u,d,e$ \\[0.6ex]
     \cline{2-3}
     & $[C_{l^2e^2H^2}^{(2)}/C_{q^2d^2H^2}^{(2)}]_{\text{\tiny wxyz}}$ & $-[y_\psi]_{\text{\tiny wz}}[y_\psi^\dagger]_{\text{\tiny yx}} \frac{\ks^2}{4\ms^2}$; $\psi=d,e$ \\[0.6ex]
     \cline{2-3}
     & $[C_{q^2u^2H^2}^{(2)}]_{\text{\tiny wxyz}}$ & $[y_u]_{\text{\tiny wz}}[y_u^\dagger]_{\text{\tiny yx}} \frac{\ks^2}{4\ms^2}$ \\[0.6ex]
     \cline{2-3}
     & $[C_{l^2\psi^2H^2}^{(3)}/C_{q^2\psi^2H^2}^{(5)}]_{\text{\tiny wxyz}}$ & $[y_\psi]_{\text{\tiny wx}}[y_\psi]_{\text{\tiny yz}}\frac{\ks^2 }{2\ms^2}$; $\psi=u,d,e$ \\[0.6ex]
     \cline{2-3}
     & $[C_{lequH^2}^{(1)}]_{\text{\tiny wxyz}}$ & $-[y_e]_{\text{\tiny wx}}[y_u]_{\text{\tiny yz}}\frac{\ks^2}{2\ms^2}$ \\[0.6ex]
     \cline{2-3} 
     & $[C_{leqdH^2}^{(1)}]_{\text{\tiny wxyz}}$ & $[y_e]_{\text{\tiny wx}}[y_d^\dagger]_{\text{\tiny yz}}\frac{\ks^2}{2\ms^2}$ \\[0.6ex]
     \cline{2-3}
     & $[C_{q^2udH^2}^{(1)}]_{\text{\tiny wxyz}}$ & $[y_u]_{\text{\tiny wx}}[y_d]_{\text{\tiny yz}}\frac{\ks^2}{2\ms^2}$ \\[0.6ex]
     \cline{2-3}
     & $[C_{lequH^2}^{(2)}]_{\text{\tiny wxyz}}$ & $-[y_e]_{\text{\tiny wx}}[y_u]_{\text{\tiny yz}}\frac{\ks^2}{2\ms^2}$ \\[0.6ex]
     \cline{2-3} 
     & $[C_{leqdH^2}^{(2)}]_{\text{\tiny wxyz}}$ & $[y_e]_{\text{\tiny wx}}[y_d^\dagger]_{\text{\tiny yz}}\frac{\ks^2}{2\ms^2}$ \\[0.6ex]
     \cline{2-3}
     & $[C_{q^2udH^2}^{(2)}]_{\text{\tiny wxyz}}$ & $-[y_u]_{\text{\tiny wx}}[y_d]_{\text{\tiny yz}}\frac{\ks^2}{2\ms^2}$ \\[0.6ex]
     \cline{2-3}
     & $[C_{lequH^2}^{(5)}]_{\text{\tiny wxyz}}$ & $[y_e]_{\text{\tiny wx}}[y_u^\dagger]_{\text{\tiny yz}}\frac{\ks^2}{\ms^2}$ \\[0.6ex]
     \cline{2-3} 
     & $[C_{leqdH^2}^{(3)}]_{\text{\tiny wxyz}}$ & $[y_e]_{\text{\tiny wx}}[y_d]_{\text{\tiny yz}}\frac{\ks^2}{\ms^2}$ \\[0.6ex]
     \cline{2-3}
     & $[C_{q^2udH^2}^{(5)}]_{\text{\tiny wxyz}}$ & $[y_d]_{\text{\tiny wx}}[y_u^\dagger]_{\text{\tiny yz}}\frac{\ks^2}{\ms^2}$ \\[0.6ex]
     \cline{2-3}
     & $[C_{l\psi H^3D^2}^{(1)}/C_{q\psi H^3D^2}^{(1)}]_{\text{\tiny wx}}$ & $-[y_\psi]_{\text{\tiny wx}} \frac{2\ks^2}{\ms^2}$; $\psi=u,d,e$ \\[0.6ex]
     \hline 
    \end{tabular}}
    \end{center}
    \renewcommand{\arraystretch}{1}
    \caption{\it Dimension-6 and -8 Wilson coefficients resulting
    from the tree-level matching of the singlet scalar model to the SMEFT. Flavour indices are denoted by Roman letters 
    $\{\mathrm{w,x,y,z}\}$. The parameters $\mu^2$ and $\lambda$ are the quadratic and quartic coefficients of the Higgs potential at the EW scale, respectively}
    \label{tab:singlet_results}
\end{table}

Using these results, we can evaluate the dimension-6 and -8 contributions to the normalization of the physical Higgs field (\ref{eq:hredef})
in the singlet scalar extension of the SM:
\begin{equation}
\begin{split}
    \Delta_h &= -\frac{\ks^2\vh^2}{2\ms^4}\bigg[1 - \frac{1}{\ms^2}(3\mh^2 - \ls\vh^2) - \frac{3\ks\vh^2}{4\ms^4}(\ks+2\ksss)\bigg]\\ 
    &\approx -\frac{\ks^2 \vh^2}{2\ms^4}\left(1 - \frac{3\mh^2}{\ms^2} - \frac{3\ks^2\vh^2}{4\ms^4}\right) \, ,
    \label{eq:Deltah_S}
\end{split}
\end{equation}
and the trilinear and quadrilinear Higgs self-couplings \eqref{eq:c3}, \eqref{eq:d4}:
\begin{equation}
\begin{split}
    c_3 =& \, \frac{\ks^2\vh^2}{\ms^4}\bigg[-\frac{3}{2} + \frac{\vh^2}{\mh^2}\left(2\ls - \frac{2\ks\ksss}{\ms^2}\right) + \frac{\vh^2}{\ms^2}\bigg(-\frac{3\mh^2}{\vh^2} + 2\ls\left(5 - \frac{4\ls\vh^2}{\mh^2}\right)\\&\, - \frac{3 \ks\ksss}{\ms^2}\left(5 - \frac{8\ls\vh^2}{\mh^2}\right) + \frac{\ks^2}{\ms^2}\bigg(\frac{15}{8} - \frac{\vh^2}{\mh^2}\bigg(3\ls - 4\kssss + \frac{3\ksss}{\ms^2}(6\ksss -\ks)\bigg) \bigg) \bigg]\\
    \approx& \, -\frac{3\ks^2\vh^2}{2\ms^4} \left(1 + \frac{2\mh^2}{\ms^2} - \frac{5\ks^2\vh^2}{4\ms^4}\right) \, ,
    \label{eq:c3_S}
\end{split}
\end{equation}
\begin{equation}
\begin{split}
    d_4 =&\, \frac{\ks^2\vh^2}{\ms^4}\bigg[-\frac{25}{3} + \frac{12\vh^2}{\mh^2}\left(\ls - \frac{\ks\ksss}{\ms^2}\right) + \frac{\vh^2}{\ms^2}\bigg(-\frac{82\mh^2}{3\vh^2} + 4\ls\left(21 - \frac{16\ls\vh^2}{\mh^2}\right)\\&\, - \frac{6 \ks\ksss}{\ms^2}\left(21 - \frac{32\ls\vh^2}{\mh^2}\right) + \frac{\ks^2}{\ms^2}\bigg(22 - \frac{4\vh^2}{\mh^2}\bigg(9\ls - 8\kssss + \frac{3\ksss}{\ms^2}(12\ksss -3\ks)\bigg) \bigg) \bigg]\\
    \approx& \, -\frac{\ks^2\vh^2}{\ms^4} \left(\frac{25}{3} + \frac{82\mh^2}{3\ms^2} - \frac{22\ks^2\vh^2}{\ms^4}\right) \, ,
    \label{eq:d4_S}
\end{split}
\end{equation}
In each of the above equations, the final line, indicated by ``$\approx$'', corresponds to the case where all the couplings except $\ks$ are assumed to be negligible. 

\subsection{Constraints on Model Parameters}

We see in Table~\ref{tab:singlet_results} that the only dimension-6 operator
coefficients that receive contributions in the singlet scalar model are
$C_H$ and $C_{H \Box}$. The former can only be constrained by a measurement of the Higgs self-coupling via, \emph{e.g.}, di-Higgs production, while the latter is relatively much better constrained via Higgs signal strength measurements. To quantify this sensitvity from the Higgs datasets, we find at the individual, linear dimension-6 level, that the coefficients are constrained at 95\% Confidence Level (CL) to lie within the following ranges:
\begin{align}
\label{eq:CHbox_CH_bounds}
    \Cp{H\Box}\subset[-1.22,-0.055]
    \quad\text{and}\quad
    \Cp{H}\subset[-3.82,3.82]
    \,\,\bigg[\text{TeV}^{-2}\bigg].
\end{align}
This highlights the hierarchy in sensitivity between single and double Higgs production as well as a preference for negative $\Cp{H\Box}$ emerging from the most recent Higgs signal strength data: the variance-weighted average of all of the Higgs signal strengths input into the fit is slightly below 1. Since all of the Higgs couplings are shifted proportional to $\Cp{H\Box}$ through the Higgs field redefinition of Eq.~\eqref{eq:hredef}, this is best explained by a negative $\Cp{H\Box}$. The symmetric limits in $\Cp{H}$ are an artefact of our assumption that the SM rate was observed when translating the upper limits into quasi signal strengths for the purposes of our analysis. Although $\Cp{H}$ is less well constrained, unlike $\Cp{H\Box}$ that only depends on $\ks$, it also depends on two additional parameters of the scalar potential, namely $\lam{S}$ and $\ksss$. More than that, it actually requires the presence of at least one of these additional parameters in order to be generated. We therefore expect di-Higgs data to provide crucial information in pinning down the singlet model parameters beyond what is possible with only single Higgs measurements.

At dimension-8, a much larger set of Wilson coefficients are generated that could potentially affect a wider class of scattering processes. Focusing first on Higgs-related processes, the dimension-8 matching yields operators that modify the Yukawa couplings, \emph{e.g.}, $\Op{q\psi H^5}$ as well as $\Opp{H^6}{(1)}$, an operator similar to $\Op{H\Box}$ that modifies the Higgs boson kinetic term. A new feature with respect to the dimension-6 matching is that these Wilson coefficients introduce a dependence of single Higgs data on $\lam{S}$ and $\ksss$. We therefore expect this next order in the SMEFT expansion to contribute non-trivially and lead to a richer structure of the bounds on the parameter space. 

The octic Higgs operator $\Op{H^8}$ is also generated, and introduces for the first time a dependence on the singlet quartic self-coupling $\kssss$. This could eventually be probed by future measurements of triple Higgs production. The remainder of the operators generated involve four fermion fields and are not relevant for the datasets included in our study, particularly since we neglect light-quark Yukawa couplings, thereby suppressing all light-quark operators in this model. The only exceptions are the class of two-derivative Yukawa-like operators, \emph{e.g.}, $\Opp{q\psi H^3D^2}{(1)}$, which we have previously discussed in Section~\ref{subsubsec:diHiggs}, arguing that their contribution to inclusive di-Higgs production is expected to be suppressed. 

Moving now to fits in the singlet model parameter space, we begin by examining the dependence on the trilinear $S H^\dagger H$ coupling, $\ks$. Since it is linear in the heavy field, this coupling is essential for generating non-zero tree-level Wilson coefficients, as can be seen from Table~\ref{tab:singlet_results}, where all coefficients are proportional to $\ks^2/\ms^2$. In general, the results in Table~\ref{tab:singlet_results} are almost completely independent of the sign of $\ks$. The only exception is in terms proportional to $\ks^3\ksss$ that also involve the singlet self-coupling. The entire set of results up to dimension-8 are invariant under the simultaneous sign flips, $\ks\to-\ks$ and $\ksss\to-\ksss$. In our subsequent analysis, we therefore only show half of the accessible parameter space, represented by $\ks$. In cases where both $\ks$ and $\ksss$ are probed, it should be understood that constraints on the other half of the parameter space can be inferred from the aforementioned symmetry.

\begin{figure}[h!]
\centering
\includegraphics[width=0.46\textwidth,valign=t]{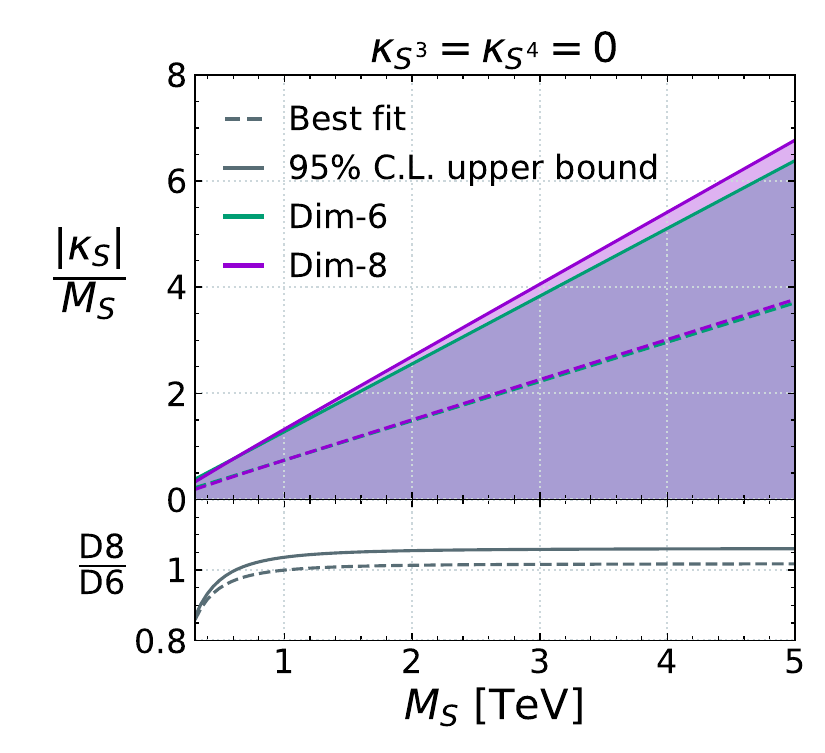}
\includegraphics[width=0.48\textwidth,valign=t,trim=0 0 0 2.5mm]{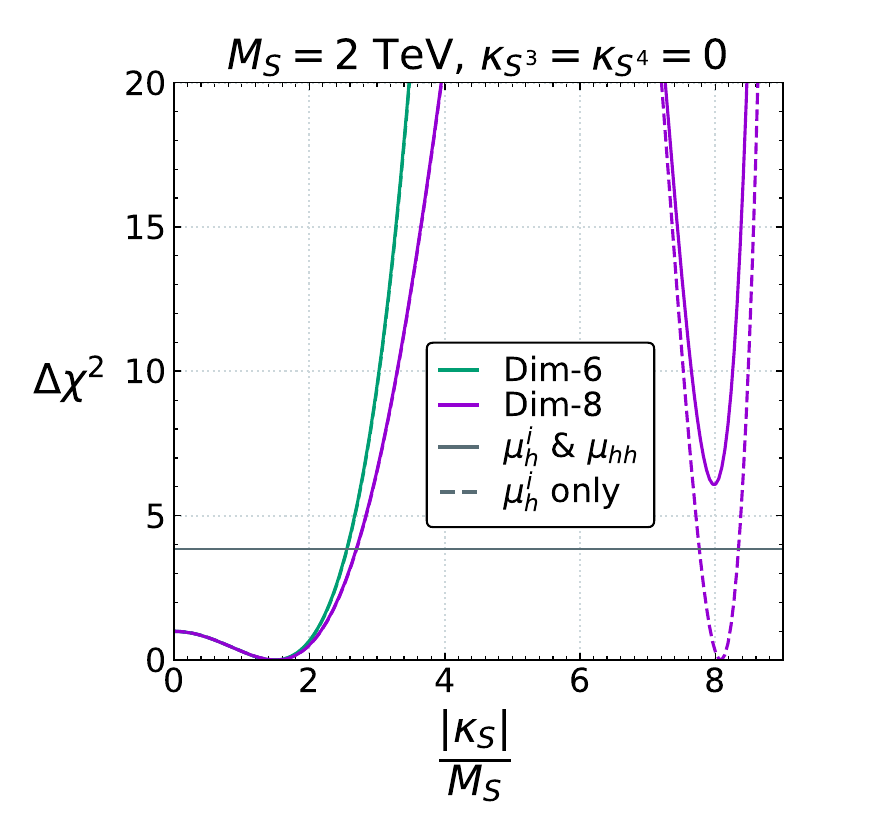}
\caption{\label{fig:singlet_ks}
\it (Left plot) The range of $|\ks|/\ms$ allowed as a function of $\ms$ by the current experimental data.
The green shaded area represents the allowed interval at 95\% Confidence Level from a dimension-6 analysis including only linear (interference) effects, and the purple shaded area is from an analysis including also linear dimension-8 contributions and quadratic dimension-6 contributions. The best fit value is represented by the correspondingly coloured dashed line. The subplot shows the ratio of the dimension-8 and dimension-6 determinations of the upper bounds and best fit points in solid and dashed, respectively.\\
(Right plot) $\Delta\chi^2$ as a function of $|\ks|/\ms$ for $\ms=2$ TeV. The green (purple) lines indicate the results of a dimension-6 (dimension-8) analysis, while the dashed lines show the result of excluding the di-Higgs cross-section measurements from the dataset.}

\end{figure}

Since the dimension-6 contribution to $\Cp{H\Box}$ is negative definite in this model, the single Higgs data translate into non-zero best-fit values for $\ks$. The constraints in the simplest case, where only this parameter is assumed to be non-zero, are shown in the left plot of Fig.~\ref{fig:singlet_ks}.
In the upper panel, the 95\% CL allowed regions for $|\ks|/\ms$ are shown in green for the linear dimension-6 analysis, and the dimension-8 counterpart is shown in purple. When considering only $\ks$ at dimension-6, there is a one-to-one mapping between $\ks/\ms$ and $\Cp{H\Box}$, so one can reconstruct the expected bounds at dimension-6 by translating the observed interval on $\Cp{H\Box}$ from Eq.~\eqref{eq:CHbox_CH_bounds} into a bound on the Higgs field-redefinition parameter $\Delta_h$, given for our special case in the last line of Eq.~\eqref{eq:Deltah_S}.

The relative impact of including dimension-8 effects is shown in the lower panel, which plots the ratio of the dimension-8 and dimension-6 determinations of the upper bound as well as the best-fit point. 
As expected, the dimension-8 effects are found to be most important at low masses around and below 1 TeV. Below 500 GeV, they lead to a tighter constraint on $\ks$. Interestingly, the dimension-8 bound asymptotes to a constant value slightly above 1, which can be understood as due to the relatively large values of $|\ks|/\ms$ being probed in that region. As shown in Eq.~\eqref{eq:Deltah_S}, at dimension-8 $\Delta_h$ receives corrections of order $\vh^2/\ms^2$ and $\vh^2\ks^2/\ms^4$, and the latter contribution comes to dominate as $\ks/\ms\to\ms/\vh$. Fortunately, the upper bounds obtained are relatively far from that limit, remaining within $\ks/\ms<\frac{1}{2}\ms/\vh$. Nevertheless, from the presence of this term one can show that the asymptotic ratio of the dimension-8 and -6 upper bounds is expected to be $\ne 1$ in the large-$\ms$ limit, with an exact value that depends on the sensitivity of the Higgs data. 

In general, including $\mathcal{O}(\Lambda^{-4})$ effects can lead to the appearance of additional minima in the $\chi^2$ function. One can immediately see from the expression for $\Delta_h$ that this will occur for some large $\ks$ value, around where the quantity flips sign. The right plot of Fig.~\ref{fig:singlet_ks} highlights the appearance of such a second minimum in the $\Delta\chi^2$ for $\ms=2$ TeV. The green and purple lines plot the quantity for the dimension-6 and -8 hypotheses respectively. We also show the impact of the di-Higgs measurements by plotting the $\Delta\chi^2$ with and without these data as solid and dashed lines, respectively. On the one hand, this confirms that current di-Higgs data have no impact in determining the primary allowed region in this simplified scenario where only $\ks$ is switched on. On the other hand, we see that taking Higgs data alone would lead to the appearance of a second allowed region in the dimension-8 analysis, at large $\ks$ values. The di-Higgs data bring sufficient additional information to lift that second minimum and exclude the associated region of $\ks$, strengthening the validity of our EFT analysis.

So far, we have assumed that the other singlet scalar couplings, namely the second Higgs portal coupling $\lam{S}$ and the two singlet self-couplings, $\ksss$ and $\kssss$,
all vanish. As we have discussed, the interplay between single Higgs and di-Higgs data at dimension-6 and dimension-8 is expected to become more interesting as this assumption is relaxed. We begin by looking at 2-dimensional subspaces, switching on one additional parameter at a time alongside $\ks$. Due to the fact that the quartic singlet self-coupling, $\kssss$ appears only in di-Higgs rates and at dimension-8, we find that the data included in our analysis is almost insensitive to its value. We will therefore focus on $\lam{S}$ and $\ksss$, only varying $\kssss$ in the last analysis in which we derive profiled bounds on $\ks$.

\begin{figure}[h]
\centering
\includegraphics[width=\textwidth]{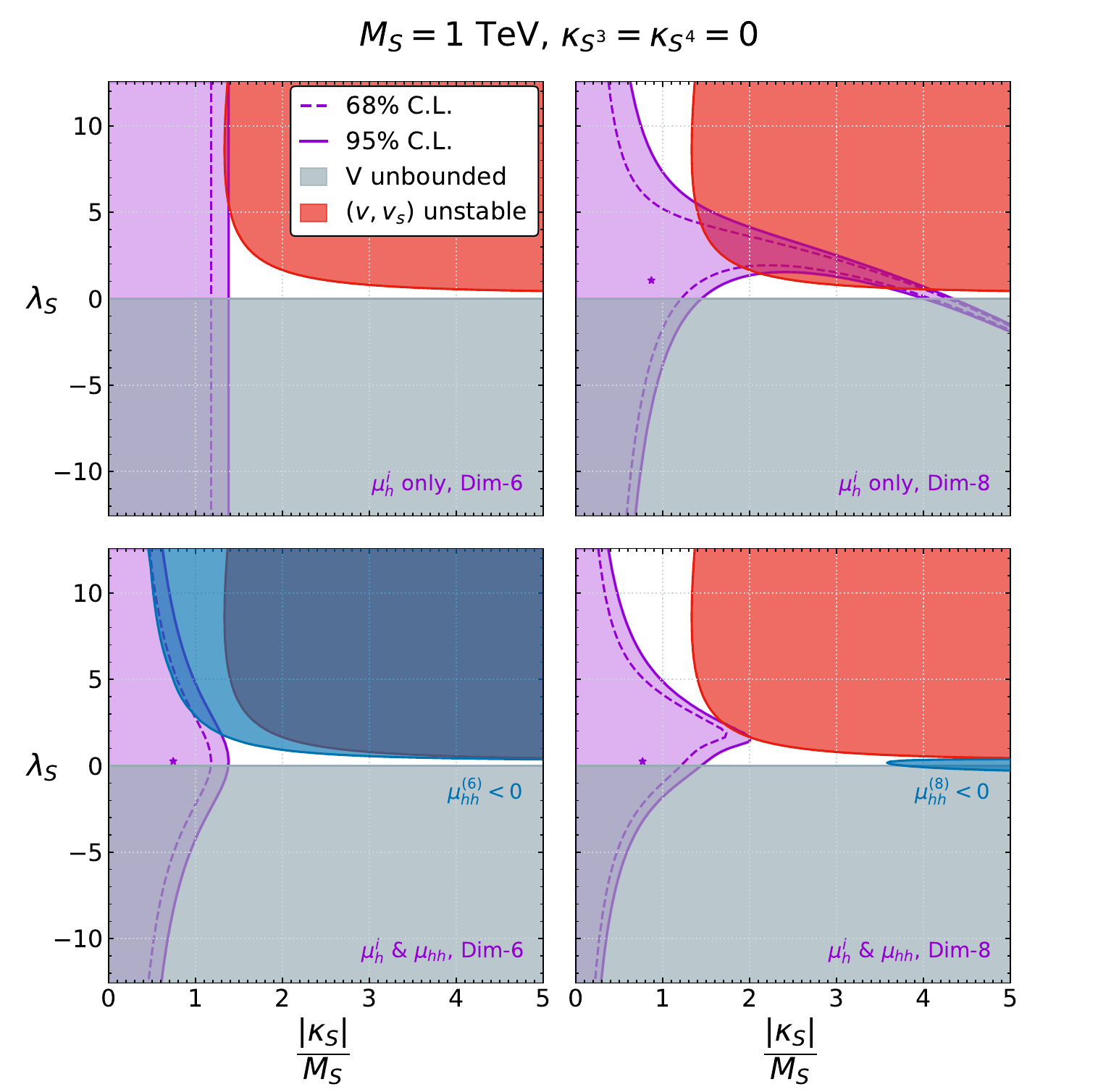}
\caption{\it Values of $(|\ks|/\ms, \lam{S}$) for $\ms = 1$~TeV and $\ksss = \kssss = 0$
that are allowed at the 68\% CL (dashed lines) and 95\% CL (solid lines)
by the present Higgs coupling measurements (upper panels)
and also including the present constraint in double-Higgs production (lower panels), as functions of
$|\ks|/\ms$ including only the linear effects of dimension-6 operators (left panels) and
including both the quadratic effects of dimension-6 operators and the linear effects of dimension-8
operators (right panels). The grey shaded regions with $\lam{S} < 0$ are excluded because the
potential is unbounded below, and the blue shaded regions in the upper part of the lower left panel,
\emph{i.e.}, in the linear dimension-6 case, and at large $|\ks|/\ms$ in the lower right panel
highlight regions where the di-Higgs cross section becomes negative in that approximation.}
\label{fig:singlet_kS_lS}
\end{figure}

In Fig.~\ref{fig:singlet_kS_lS} we analyze the constraints on $|\ks|$ and $\lam{S}$
for $\ms = 1$~TeV and $\ksss = \kssss = 0$ at the 68\% CL (dashed lines) and 95\% CL (solid lines),
due to Higgs coupling strengths (upper panels) and also including double-Higgs production (lower panels).
In all the panels we shade in grey the region with
$\lam{S} < 0$, which is excluded because there the effective scalar potential is unbounded from below when $\kssss=0$, see
Eq.~\eqref{bounded}. We also shade in red the region where the EW vacuum is not the global minimum of the scalar potential,
denoted for simplicity by the legend ``$(v, v_s)$ unstable'', though
we have not checked whether the electroweak vacuum is truly unstable or merely metastable.

We see in the upper left panel of Fig.~\ref{fig:singlet_kS_lS} that in the linear approximation for dimension-6 
operators the constraint on $|\ks|$ due to Higgs coupling strengths is independent of $\lam{S}$, whereas 
this constraint on $|\ks|$ depends strongly on $\lam{S}$ when the quadratic effects of 
dimension-6 operators and the linear effects of dimension-8 operators are also included (upper right panel).
On the other hand, we see in the lower left panel of Fig.~\ref{fig:singlet_kS_lS}
that the di-Higgs data also introduce a dependence on $\lam{S}$
into the constraint on $|\ks|$ even in the linear dimension-6 approximation, which is accentuated
when the quadratic effects of dimension-6 operators and the linear effects of dimension-8 operators 
are also included (lower right panel). We note also in the lower left panel the appearance of a blue shaded region at large 
$|\ks|$ and positive $\lam{S}$ that we disregard in the linear dimension-6 approximation because
the di-Higgs cross section becomes negative in this approximation~\footnote{This region overlaps
extensively with the region where the electroweak vacuum is not the lowest-energy state, and the region
where both conditions apply is shaded dark blue.}. However, 
the positivity of the di-Higgs cross section is much less of an issue
when the quadratic effects of dimension-6 operators and the linear effects of dimension-8 operators 
are also included, as seen in the lower right panel where the problematic region lies far away from the allowed parameter space. 

Overall, we see that the `complete' analysis including both single and di-Higgs at dimension-8 yields the strongest constraints, and that they also overlap the least with the unphysical regions of parameter space. The inclusion of dimension-8 and di-Higgs information drastically affects the conclusions that can be drawn about the singlet model compared to the dimension-6 interpretation. In particular, they allow for potentially larger values of $\ks$, due to the possibility of cancellations with $\lam{S}$.

\begin{figure}[h]
\centering
\includegraphics[width=\textwidth]{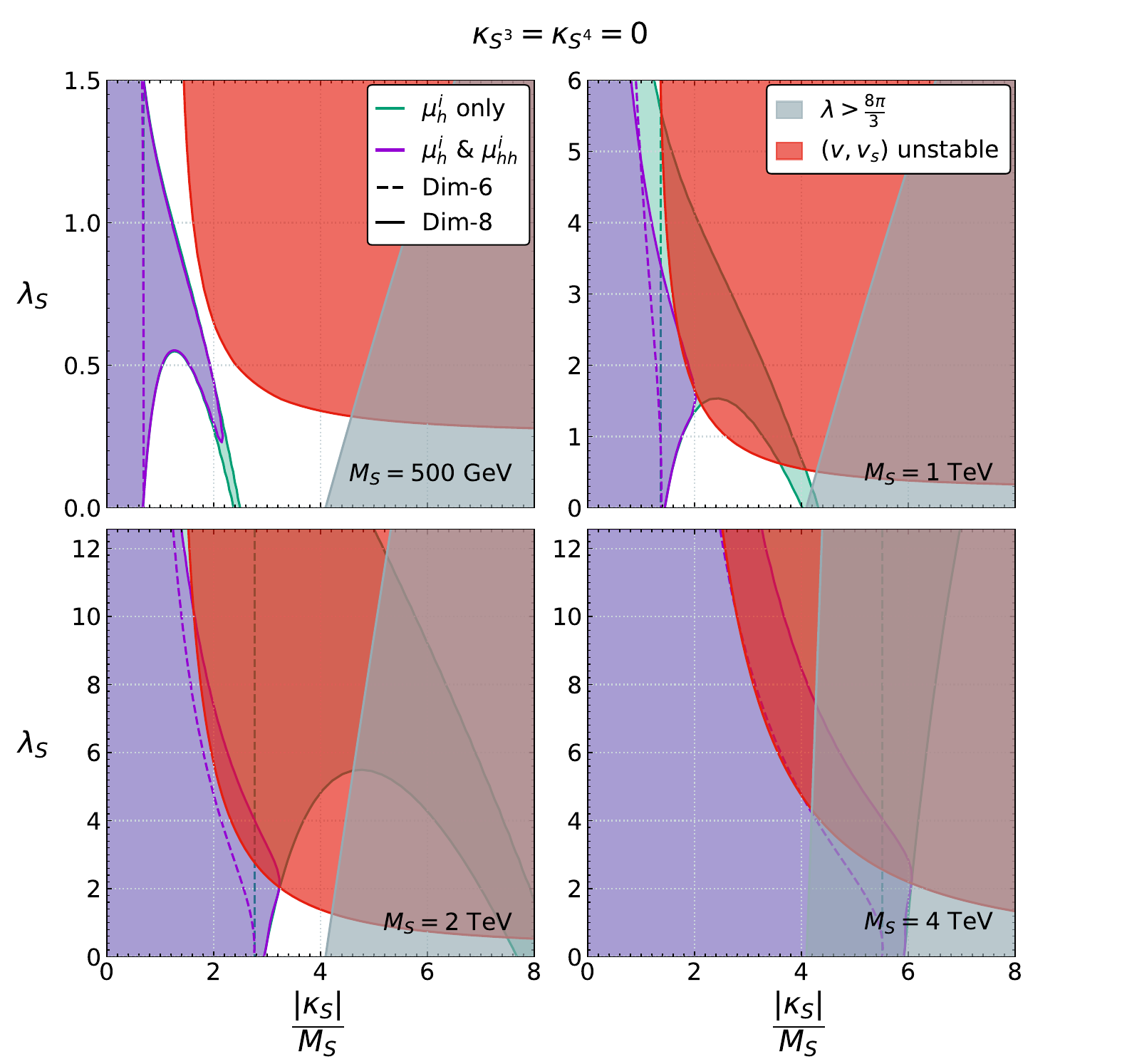}
\caption{\it Values of $(|\ks|/\ms, \lam{S}$) for the indicated values of $\ms$ and $\ksss = \kssss = 0$
that are allowed at the 68\% CL (dashed lines) and 95\% CL (solid lines)
by the present Higgs coupling measurements (green shading)
and also including the present constraint in double-Higgs production (purple shading) as functions of
$|\ks|/\ms$ including only the linear effects of dimension-6 operators (dashed lines) and
including both the quadratic effects of dimension-6 operators and the linear effects of dimension-8
operators (solid lines). The grey shaded regions at large $|\ks|/\ms$ are excluded by the 
perturbative unitarity constraint $\lambda \le 8 \pi/3$, and the electroweak vacuum is not stable in
regions at large $|\ks|/\ms$ that are shaded red.}
\label{fig:singlet_kS_lS_MS}
\end{figure}

Fig.~\ref{fig:singlet_kS_lS_MS} shows how the
constraints on $|\ks|/\ms$ as functions of
$\lam{S}$ vary with $\ms$, again for 
$\ksss = \kssss = 0$. The constraints
found at the linear dimension-6 level are shown as 
dashed lines and those including quadratic
dimension-6 contributions and linear dimension-8
contributions by solid lines. The allowed regions
found including only Higgs signal strengths are
shaded green and those found including di-Higgs
constraints are shaded purple. The region shaded grey
is excluded by 
partial-wave unitarity constraints on the Higgs self-coupling parameter $\lambda$:
$\lambda \le 8 \pi/3$, see Eq.~\eqref{perturbation} and Appendix~\ref{app:hhhh}.
As in Fig.~\ref{fig:singlet_kS_lS}, the region where the electroweak vacuum is not
stable is shaded red. We note in all the panels of Fig.~\ref{fig:singlet_kS_lS_MS} significant differences between
the regions allowed at the linear dimension-6 and quadratic dimension-6/linear dimension-8 levels, and concentrate below on the latter, which are shaded and have solid boundaries.

Comparing the green and purple shaded regions in Fig.~\ref{fig:singlet_kS_lS_MS}, we see that
they are very similar for $\ms = 500$~GeV, and are unaffected by the perturbativity and 
electroweak vacuum stability constraints. The di-Higgs data serve to rule out the tip of the cancellation region. There is a greater difference between the green and
purple regions when $M = 1$~TeV (upper right panel of Fig.~\ref{fig:singlet_kS_lS_MS}), but
most of the (green) region that would have been allowed if the di-Higgs data were neglected
is excluded by the electroweak vacuum stability requirement.
Starting from this mass, a significant portion of the boundary of the `complete' analysis for this mass now overlaps with the theory bounds. As expected, the sensitivity to $|\ks|/\ms$ degrades with increasing $\ms$. We can see that the di-Higgs data rule out a second minimum in the $|\ks|/\ms$ direction for all values of $\ls$ where it exists, except for the $\ms=500$ GeV case, where some values of $\ls$ still allow it.

Finally we note the trend that, at dimension-8, the inclusion of di-Higgs data has an increasing impact as $\ms$ increases. This can be explained by going back to the matching results of Table~\ref{tab:singlet_results}. At dimension-6, the $\lam{S}$ dependence only enters though di-Higgs data and the relative importance of including these constraints is independent of $\ms$, which can be seen by observing the similarity between the dashed purple and green curves in each panel. The impact of di-Higgs constraints is only significant for large values of $\lam{S}>1$. Going to dimension-8 introduces a linear $\lam{S}$ dependence in single Higgs rates, while the di-Higgs dependence can be quadratic, from contributions to $\Cp{H}^2\sim \ks^4\lam{S}^2$ and $\Cp{H^8}\sim \ks^2\lam{S}^2$. At lower masses, the bounds on $\ks,\lam{S}$ from single-Higgs data are strong enough that the enhanced quadratic dependence does not play a big role. Increasing $\ms$ yields larger possible values in this space, allowing for the quadratic $\lam{S}$ dependence from di-Higgs to dominate in certain regions, leading to the significant differences in the bounds. 

\begin{figure}[h]
\centering
\includegraphics[width=\textwidth]{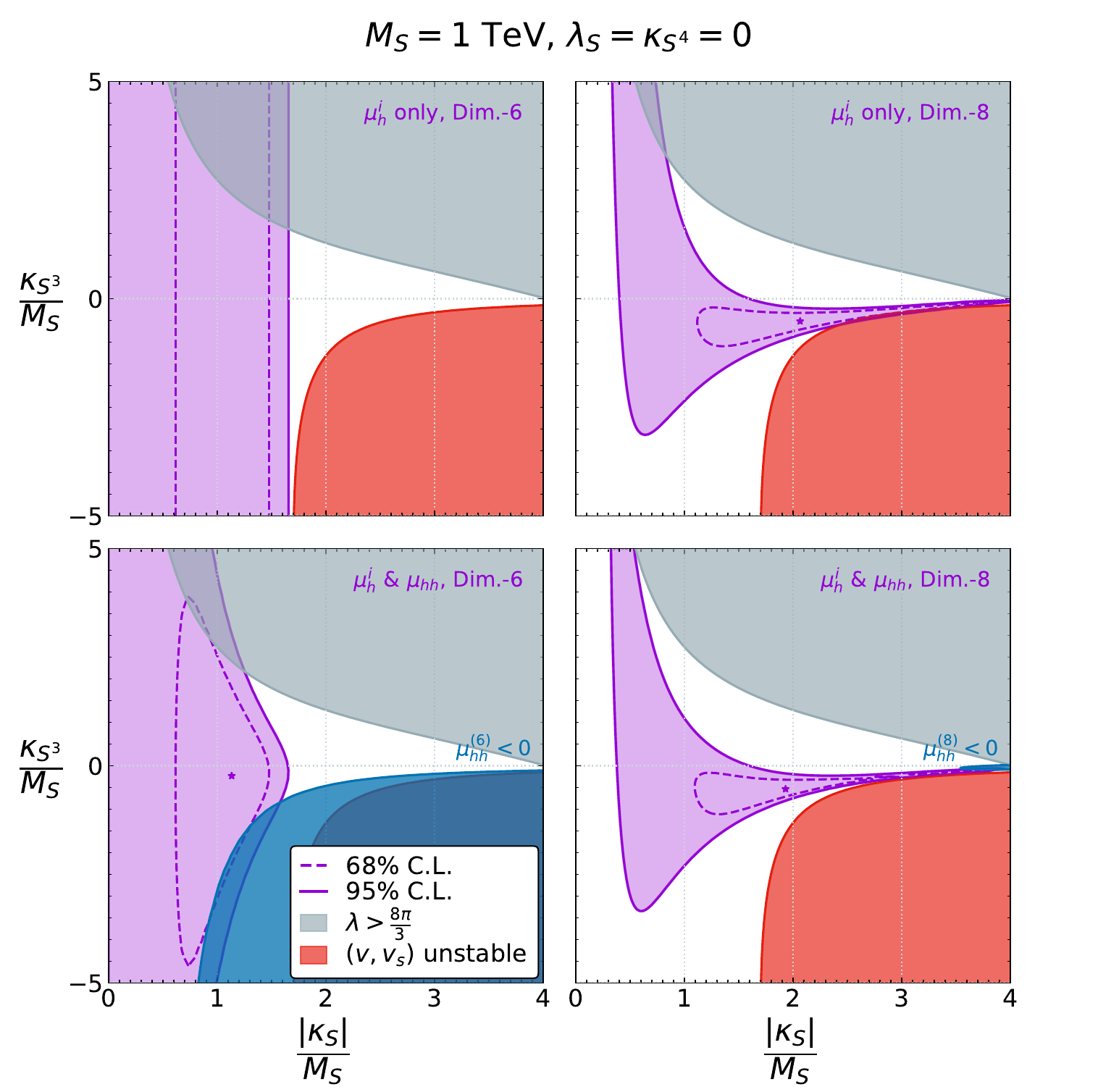}
\caption{\it Values of $(|\ks|/\ms, \ksss/\ms)$ for $\ms = 1$~TeV and $\lam{S} = \kssss = 0$
that are allowed at the 68\% CL (dashed lines) and 95\% CL (solid lines)
by the present Higgs coupling measurements (upper panels)
and also including the present constraint in double-Higgs production (lower panels) as functions of
$|\ks|/\ms$ including only the linear effects of dimension-6 operators (left panels) and
including both the quadratic effects of dimension-6 operators and the linear effects of dimension-8
operators (right panels). The grey shaded regions are excluded because $\lambda$ is
nonperturbative. The di-Higgs cross section is negative in the blue region in the lower left panel.}
\label{fig:singlet_kS_kS3}
\end{figure}

Fig.~\ref{fig:singlet_kS_kS3} shows a corresponding analysis of the Higgs constraints on
$|\ks|/\ms$ as functions of $\ksss/\ms$, now assuming $\lam{S}=\kssss=0$. Since $\ksss$ is a dimensionful coupling, we choose to scale it with $\ms$ in the same way as we did for $\ks$. The upper and lower bounds in $\ksss/\ms$ are chosen at the arbitrary, large value of $\pm 5$. Since it is a trilinear coupling, it is subject to neither stability nor perturbative unitarity bounds derived in the high-energy limit~\footnote{It is, in general, possible to use partial-wave unitarity to bound trilinear couplings~\cite{Goodsell:2018fex,Krauss:2018orw,Goodsell:2018tti} by avoiding the high-energy limit. Since the associated bounds are more involved and depend on the full details of the parameter space, for simplicity we do not include them in our study, although it would be interesting to determine their impact in future work.}.  

As in Fig.~\ref{fig:singlet_kS_lS}, we see the
impact of the constraints on $\lambda$ from vacuum stability and perturbativity (grey shading)
and electroweak stability (red shading). We also note in the lower left panel that large 
$|\ks|/\ms$ and negative $\ksss/\ms$ are disallowed in the linear dimension-6 
case because the di-Higgs cross section becomes negative in this approximation (blue shading).
As in the case of $\lam{S}$, the constraint on $|\ks|/\ms$ from Higgs data
is independent of $\ksss$ in the linear dimension-6 approximation (upper left panel), 
but strongly dependent on $\ksss$ when the quadratic effects of dimension-6 operators 
and the linear effects of dimension-8 operators are also included (upper right panel).
As in Fig.~\ref{fig:singlet_kS_lS}, the double-Higgs constraint introduces $\ksss$
dependence into the constraint on $|\ks|/\ms$, which is accentuated when 
the quadratic effects of dimension-6 operators 
and the linear effects of dimension-8 operators are also included (lower panels). Narrow regions exist in which the effects of both parameters approximately cancel, leading to much larger potentially allowed values of $|\ks|$ or large, positive values of $\ksss$. These regions start to overlap with the theoretically forbidden regions due to EW vacuum stability and perturbative unitarity, respectively. 

\begin{figure}[h]
\centering
\includegraphics[width=\textwidth]{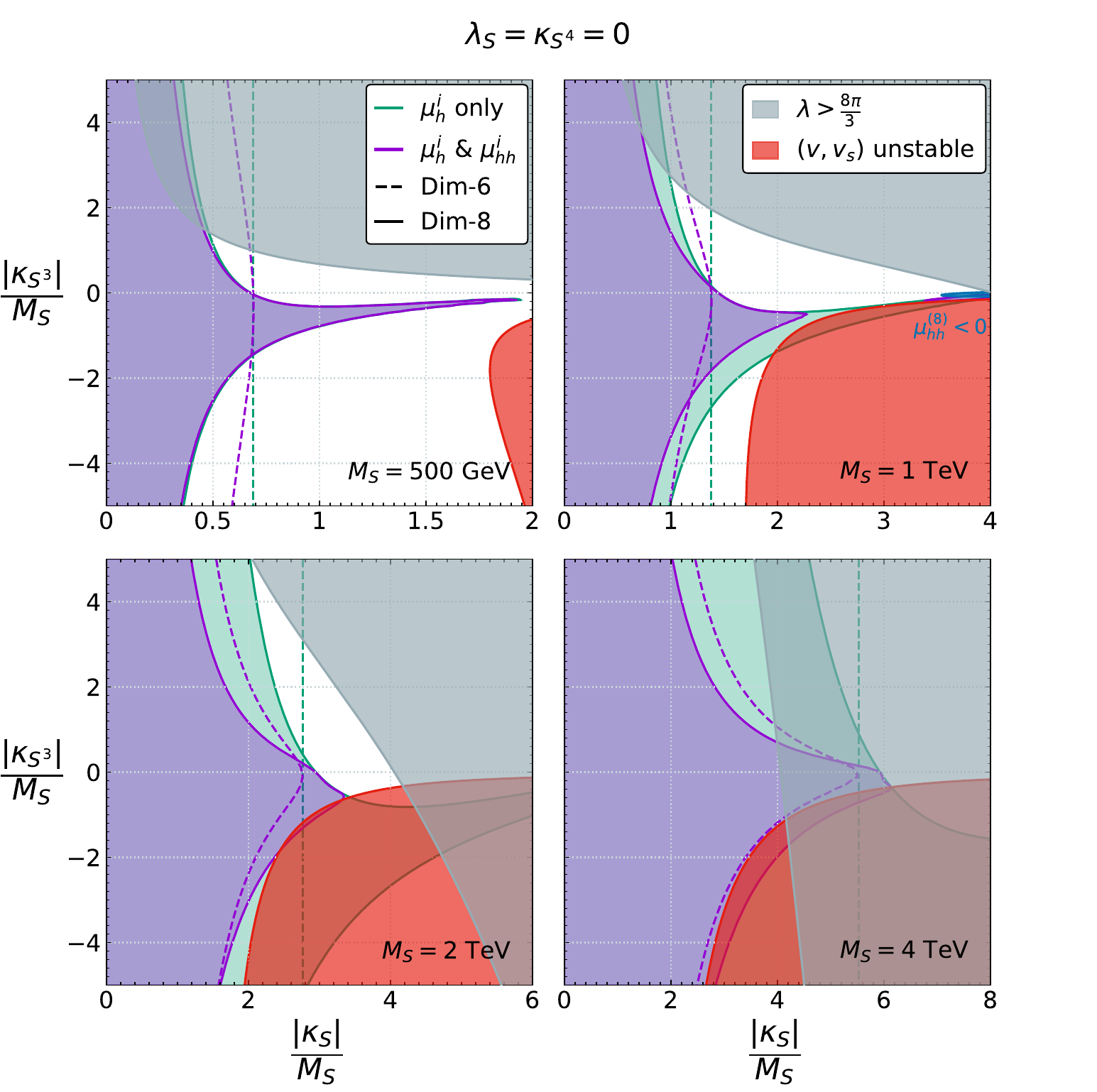}
\caption{\it  Values of $(|\ks|/\ms, \ksss/\ms)$ for the indicated values of $\ms$ and $\lam{S} = \kssss = 0$
allowed at the at the linear dimension-6 level (dashed lines) and including quadratic
dimension-6 contributions and linear dimension-8 contributions
(solid lines), taking into account Higgs coupling measurements (green shading)
and also including also the present constraint on double-Higgs production (purple shading).
As previously, the grey regions are excluded by perturbativity constraints on $\lam{S}$,
and the red regions are excluded by requiring the stability of the electroweak vacuum.}
\label{fig:singlet_kS_kS3_MS}
\end{figure}

Fig.~\ref{fig:singlet_kS_kS3_MS} illustrates how the features seen in
Fig.~\ref{fig:singlet_kS_kS3} vary as functions of $\ms$.
We see in Fig.~\ref{fig:singlet_kS_kS3} how the cancellation region in the single Higgs data gradually shifts downward in the $(|\ks|/\ms,\ksss/\ms)$ plane with increasing $\ms$, where it also overlaps more and more with the region in which the EW vacuum is not the global minimum.
Furthermore, the absolute impact of the di-Higgs constraint is minor in the upper panels of Fig.~\ref{fig:singlet_kS_kS3_MS},
but more important in the lower panels, for the same reasons as discussed before involving $\lam{S}$, \emph{i.e.} a quadratic dependence of di-Higgs rates on $\ksss$ at dimension-8 that becomes increasingly relevant as larger values are allowed by single-Higgs data. Altogether, this results in the di-Higgs data being able to rule out a larger and larger part of the cancellation region as $\ms$ increases. 
Both cases reflect interesting examples of regions in model space where quadratic dimension-6 effects are dominant over their linear dimension-8 counterparts.

Finally, we consider a more general scenario where all of the singlet scalar parameters are allowed to vary. Since $\ks$ is the crucial parameter without which no tree-level effects would be predicted, we consider the impact of varying the three other parameters on the derived bounds on $\ks$. From a statistical point of view, such a bound can be seen as a more robust limit, which is independent of the remaining details of the extended scalar potential. As we have demonstrated in this Section, there is a significant interplay between the experimental and theoretical bounds on the parameter space. In particular, the relevance of the theoretical bounds increases as one increases $\ms$. We therefore also include the  constraints from boundedness, perturbativity, and the requirement that the EW vacuum be the global one.
Specifically, for a given $\ms$ we evaluate the full likelihood as a function of the other parameters and derive a profiled likelihood that depends only on $\ks$, where $\chi^2$ has been minimised over all other directions, subject to the theoretical constraints. 

\begin{figure}[h!]
\centering
\includegraphics[width=0.5\textwidth]{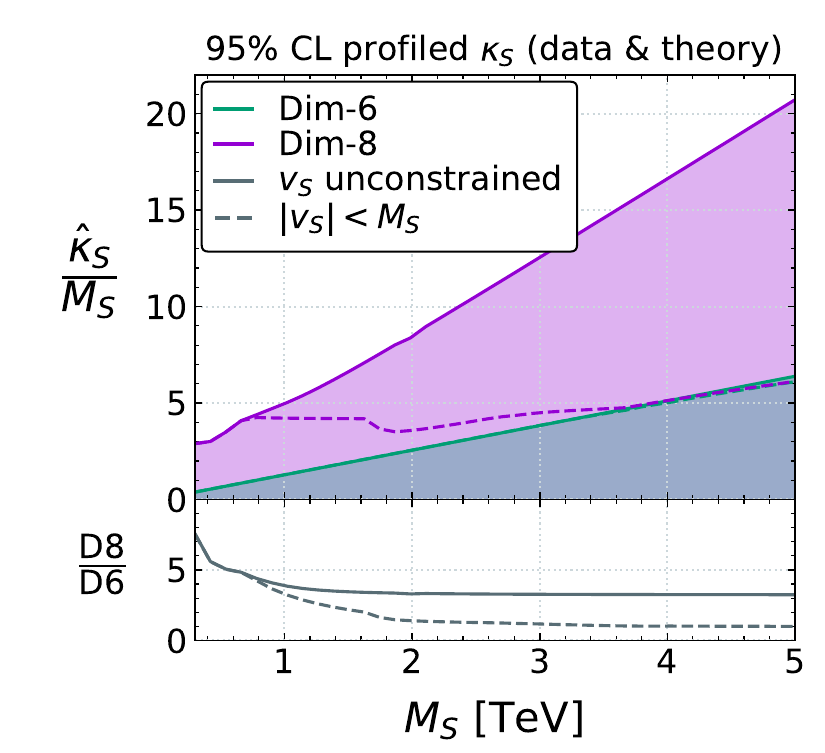}
\caption{\label{fig:singlet_kS_profiled}
{\it Profiled 95 \% confidence level bounds on $\ks/\ms$ as functions of $\ms$
at the linear dimension-6 level (solid green lines \& shaded area) and including quadratic dimension-6 and linear dimension-8 contributions (solid purple lines \& shaded area). The bounds are obtained by minimising the $\Delta\chi^2$ over all other singlet model parameters, subject to theoretical constraints from boundedness, perturbativity and stability of the EW vaccuum. The dashed lines reflect the upper bound obtained when enforcing the additional constrain that $\vs\leq\ms$.
}
}
\label{fig:singlet_kS_prof}
\end{figure}

Fig.~\ref{fig:singlet_kS_profiled} shows the resulting 95 \% CL profiled bounds on $|\hat{\kappa}_{\sss S}|/\ms$ as functions of $\ms$, where the hat notation denotes our use of a profile likelihood in the remaining parameter space. Dimension-6 intervals are shaded in green, while those determined including linear dimension-8/quadratic dimension-6 predictions are shaded in purple. The regions are bounded by solid lines labelled ``$\vs$ unconstrained'' for reasons that will be explained shortly. The subplot shows the ratio of the dimension-8 and dimension-6 bounds.
The first observation to make is that the profiled bounds obtained at dimension-6 level are identical to the individual ones reported in Fig.~\ref{fig:singlet_ks}. This can be understood from the fact that in this approximation, for a given value of $\ks$, the maximum likelihood point is the one in which all other parameters are 0. This can be seen in, \emph{e.g.}, Figs.~\ref{fig:singlet_kS_lS} and~\ref{fig:singlet_kS_kS3}, where the dimension-6 bound on $\ks$ is maximal when the accompanying parameter is exactly 0, and only tightens for non-zero values. Physically, this is because $\Cp{H}$ is the most relevant coefficient for di-Higgs constraints, since it induces significant deviations in this process while being unconstrained by single-Higgs rates. However, as previously mentioned, this operator is only switched on when at least one of $\lam{S},\ksss$ are non-zero, and we have already shown that di-Higgs bounds are not relevant in determining the primary allowed region when the other couplings are assumed to be 0. Hence, one is always in the least-constrained case when only switching on $\ks$ by itself. The theoretical bounds are not found to play any role in the profiling, as this minimally-constrained point is not subject to any of them. 

The story is much changed at the dimension-8 level, since the parametric dependence of the observables is richer than that we have discussed so far. We begin by noting the significant changes in the profiled bounds in this approximation, showing a much greater impact than for the individual case of Fig.~\ref{fig:singlet_ks}. The upper bound is increased by nearly a factor 8 at low masses, going down to a factor of around 3 at large masses. Taken at face value, these results imply that the seemingly robust conclusions that one can draw about $\ks$ at the dimension-6 level are subject to large uncertainties from missing higher orders in the EFT expansion. One could have guessed that the impact would be large from the appearance of cancellation regions in the various figures shown previously in this Section. Additionally, the lack of sensitivity to $\kssss$ from the data has an impact by allowing a wider theoretically-viable range for a given set of the other parameters. 
We find that, without the inclusion of the theoretical bounds, we are not able to find a profiled upper bound on $\ks$. In other words, using constraints from the data alone leaves flat directions that are only closed by the theoretical consistency requirements. 
 Even though extreme parameter values are required in the aforementioned cancellation regions in order to cancel the potentially large effects from extreme $\ks$ values, there is always some narrow space in which the theory constraints can be satisfied. In many cases this appeared to be quite a fine-tuned region where $\kssss\lesssim10^{-3}$, clearly on the edge of the boundary of the theoretically-allowed parameter space. The true bounds were consequently difficult to determine numerically and our results were obtained using \verb|scipy|'s built-in differential evolution algorithm. Since this is a gradient-free sampling method, there is no absolute guarantee that the upper bounds obtained correspond to the true global minimum of the profile-likelihood. Nevertheless, the smoothness of our result as a function of $\ms$ gives us some confidence that we have successfully minimised.

We found that the majority of the space in which the cancellations were achieved for extreme parameter values also corresponded to regions of large singlet vev, $\vs$. As discussed in Appendix~\ref{app:hhhh}, the singlet field minimisation condition is a cubic equation in $\vs$ that can admit additional real solutions far from the origin. When present, these often provide the global minimum of the scalar potential, see Eq.~\eqref{eq:mincondS}. In this limit the correction to the Higgs self-coupling parameter in Eq.~\eqref{eqn:lambda_singlet} becomes independent of $\ks$, scaling like $\lam{S}^2/\kssss$, making it more likely to satisfy perturbative unitarity, even for very large $\ks$, especially since $\kssss$ is essentially unconstrained by our analysis. However, one of the main assumptions of the EFT expansion is that $\vs$ is a suppressed quantity with respect to the EW scale $\sim v^2\ks/\ms^2$, see Eq.~\eqref{eqn:vs_EFT}. The $\vs$ solutions far from the origin therefore cannot be approximated by an expansion in the heavy mass scale, $1/M$, which would select the incorrect $\vs$ in these cases. The EFT expansion and therefore our experimental bounds are likely to be invalid in these regions of model space. In these regions of parameter space, the physical singlet mass receives significant contributions from EW symmetry breaking, and is consequently quite different from $\ms$. In these scenarios, it is known that the SMEFT expansion may not converge and a HEFT expansion is more appropriate~\cite{Corbett:2015lfa,Buchalla:2016bse,Falkowski:2019tft,Cohen:2020xca}.

In order to assess the amount of invalid parameter space, we performed a second constrained profiling, with the additional requirement that $|\vs|\leq\ms$. The upper bounds on $\hat{\kappa}_{\sss S}$ determined in this case are plotted with dashed lines in Fig.~\ref{fig:singlet_kS_prof}. They show that an increasingly large portion of the parameter space seemingly allowed at dimension-8 above $\ms\approx1$ TeV corresponds to cases where $\vs > \ms$. Even in the dimension-6 analysis, a small slice of the upper region above $\ms\approx 4$ TeV is not likely to be well approximated by the EFT.
Nevertheless, interpreting this constraint as an attempt to restrict the parameter space to regions in which we can rely on the EFT approximation, we can still draw significantly different conclusions between the dimension-6 and -8 analyses. The observed upper bound at dimension-8 is significantly modified, and coincides with that obtained at dimension-6 around $\ms=3.5$  TeV. We therefore conclude that the dimension-8 contributions represent important corrections in the singlet model interpretation of LHC Higgs data in the SMEFT framework, even for $\ms\lesssim1$ TeV, as seen in Fig.~\ref{fig:singlet_kS_prof}.

\subsection{Comparison between the Dimension-6 and -8 Results and the Full Model}
\label{sec:comparisonsinglet}
Since the goal of going to higher orders in the SMEFT expansion is to provide a better approximation to the full model, it is instructive to compare dimension-6 and -8 SMEFT predictions with those of the full model. We use the mixing angle, $\alpha$, between the singlet field and the neutral component of the Higgs doublet (See Appendix~\ref{app:hhhh} for a precise definition) as a proxy for testing the SMEFT approximation~\footnote{It would also be interesting to use a comparison of di-Higgs production rates to assess further the SMEFT approximation to the full model, since this process probes higher energies than single Higgs measurements and may also have significant resonant contributions from on-shell scalar production. However, a complete calculation of the di-Higgs rate in the singlet model parameter space is beyond the scope of this work and we leave it for future investigations.}. It is well known that the signal strengths for single Higgs production and decay scale like $\cos^2\alpha$. Since mixing with a gauge-singlet field results in a  universal reduction of Higgs couplings, as long as the singlet mass is greater than half of the Higgs mass (so that the decay channel to a pair of singlets is kinematically closed) the branching fractions of the Higgs-like scalar remain the same as in the SM. This means that, in the narrow width approximation, all modifications to the signal strengths come from the modification of the Higgs production rates, and hence scale with the square of the global coupling modifier, $\cos\alpha$. As shown in Section~\ref{sec:higgs_couplings}, the coupling modifications to fermions and gauge bosons predicted by the SMEFT depend on a number of different Wilson coefficients. The fact that these all coincide with the Taylor-expanded expression for $\cos\alpha$ in the limit of large $\ms,\ks$  and $\ksss$: 
\begin{align}
    \cos\alpha\approx
    1 
    - \frac{\ks^2 v^2}{\ms^4}\left(
\frac{1}{2}
- \frac{(2\lam{S} v^2-\mh^2)}{\ms^2}
- \frac{3\ks v^2(\ks-8\ksss)}{8\ms^4}
   \right),
\end{align}
provides an additional validation of our matching results up to dimension-8. 

\begin{figure}[h!]
\centering
\includegraphics[width=\textwidth]{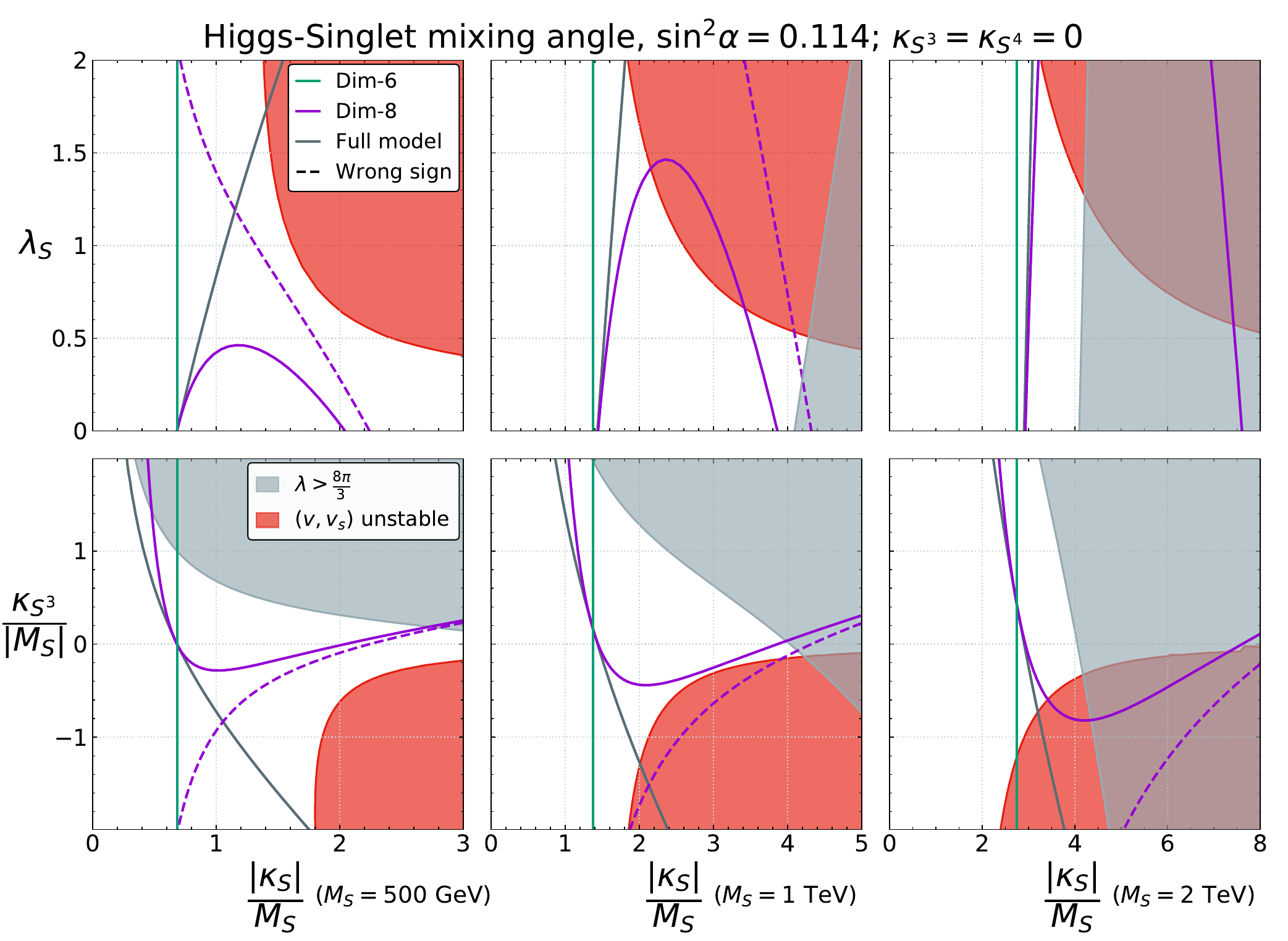}
\caption{\it
Isocontours of the Higgs-singlet mixing parameter $\sin^2 \alpha = 0.114$
in the singlet scalar extension of the SM obtained at the dimension-6 (green line) and -8 (purple line) levels compared with those
predicted in the full model (grey line), for $\ms = 500$~GeV, 1 and 2 TeV. The upper panels show the prediction in the $(\ks/\ms,\lam{S})$ plane, assuming $\ksss,\kssss=0$, while the lower panels show the prediction in the $(\ks/\ms,\ksss/\ms)$ plane, assuming $\lam{S},\kssss=0$. Dashed lines indicate regions where the Higgs couplings are modified by an equal magnitude but with the wrong sign (see text for details).
\label{fig:singlet_comparison}
}
\end{figure}

Fig.~\ref{fig:singlet_comparison} compares the SMEFT predictions for this global signal strength modifier with the true values predicted by the full model. Isocontours of a critical value of the mixing angle, $\sin^2\alpha=0.114$ are shown for a selected values of the singlet mass parameter, $\ms$. This critical value corresponds to the maximum value allowed by the Higgs data, \emph{i.e.}, the value that the mixing angle takes along the 95\% C.L. contour of allowed parameter values in Figs.~\ref{fig:singlet_kS_lS}--\ref{fig:singlet_kS_kS3_MS}.
The upper panels plot the contours in $(\ks/\ms,\lam{S})$ plane with $\ksss,\kssss=0$ for $\ms = 500$~GeV,
1 and 2 TeV, and the lower panels show the corresponding
contours in the  $(\ks/\ms,\ksss/\ms)$ plane taking $\lam{S},\kssss=0$. In each panel the
dimension-6 predictions are shown as green lines, the
dimension-8 predictions as purple lines, and the
full model predictions are shown as grey lines. Dashed lines indicate isocontours of $\cos^2\alpha=1.114$, where the coupling modifier is shifted by the same amount from the SM value of 1, but in the wrong direction, reflecting an increase in Higgs couplings. Such a coupling shift is unphysical from the point of view of Higgs-singlet mixing, which can only reduce Higgs boson couplings, and is therefore labelled `wrong sign'. As in the previous Figures, the
EW vacuum is unstable in the regions shaded red, and
the Higgs quartic coupling, $\lambda$, becomes nonperturbative in the regions shaded
grey in the lower panels.

Fig.~\ref{fig:singlet_comparison} shows how the independence of the dimension-6 prediction for the mixing angle on $\lam{S}$ and $\ksss$ is a bad representation of the full model prediction, and how the dimension-8 approximation brings in a dependence on both of these parameters. In the upper panels 
the dimension-6 prediction as a function of
$\lam{S}$ diverges from the full model constraint by amounts
that increase with $\lam{S}$ but decrease as $\ms$ increases. On the other hand the dimension-8
prediction is a better approximation to the full model
for almost all the allowed range of $\lam{S}$
for $\ms \gtrsim 1$~TeV.

For each mass, we can see where the dimension-8 effects start to dominate and eventually diverge away from the full model prediction, indicating a lack of convergence of the SMEFT expansion.
When $\ms = 500$~GeV the dimension-8 constraint on $\lam{S}$
is closer to the full constraint at small $\lam{S}$, but is a
worse approximation than the dimension-6 constraint already for
$\lam{S} \gtrsim 0.5$.
In the lower panels  we see that the dimension-8
approximation to the full model as a function of
$\kappa_{S^3}/\ms$ is better than the dimension-6 constraint
for all positive values, but is worse than the dimension-6
approximation for negative values when $\ms \lesssim 1$~TeV.
In general, the agreement between dimension-8 and the full model spans a wider range of $\lam{S}$ and $\ksss$ with increasing $\ms$. Furthermore, the parts of the contours where the SMEFT convergence fails get pushed further into theoretically forbidden regions with increasing $\ms$, and by $\ms=2$ TeV the SMEFT at dimension-8 approximates the Higgs couplings quite well. 
In each panel, we can recognise the shapes of the constraints shown in Figs.~\ref{fig:singlet_kS_lS}--\ref{fig:singlet_kS_kS3_MS}, with the cancellation regions lying in between the solid and dashed contours at dimension-8. 

\begin{figure}[h!]
\centering
\includegraphics[width=0.8\textwidth]{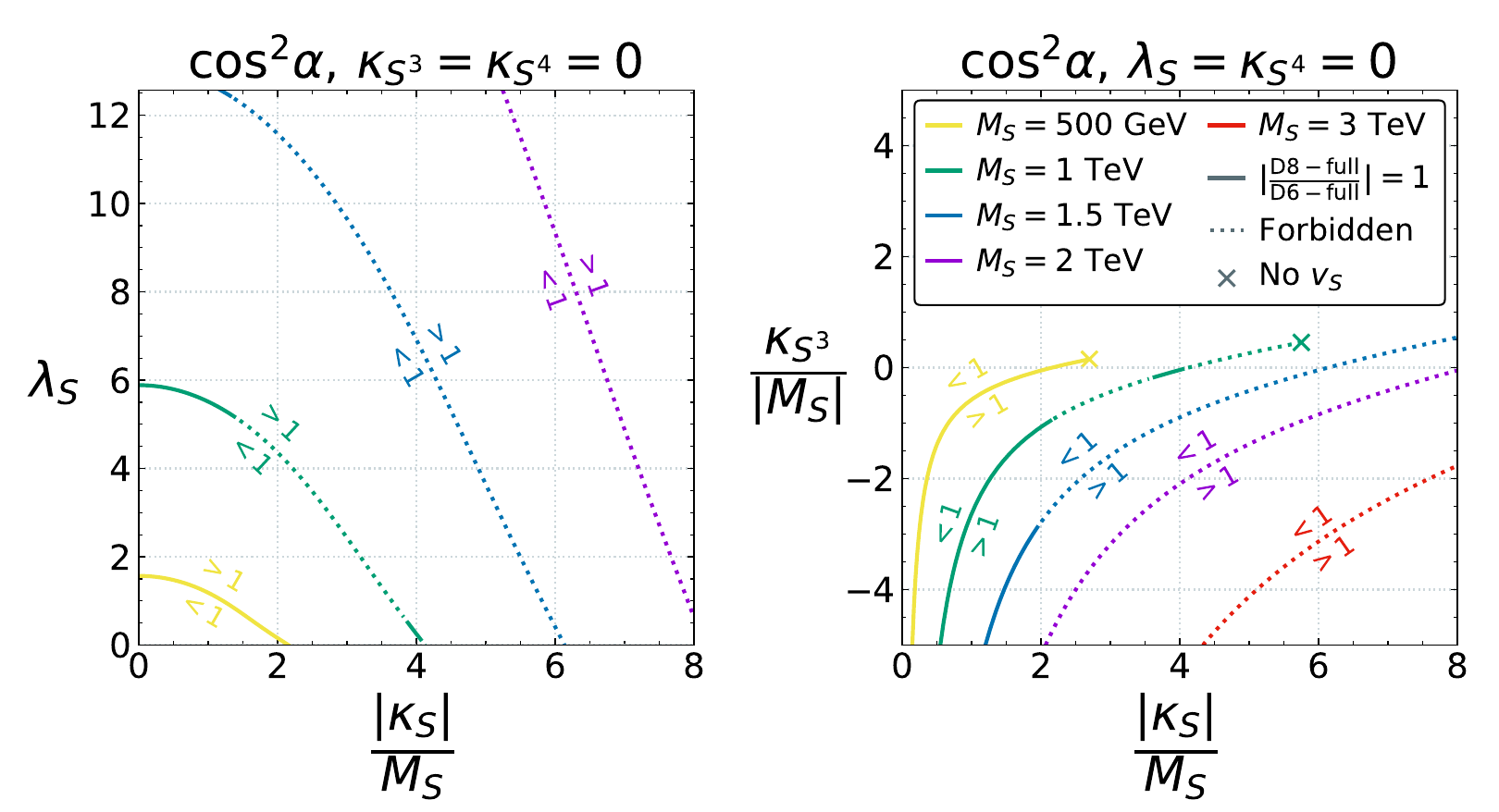}
\caption{\it
Isocontours where the dimension-6 and -8 predictions for the Higgs-singlet mixing parameter, $\cos^2\alpha$, differ from the full model prediction by the same amount, for $\ms = 500$~GeV--$3$ TeV. The regions where the dimension-8 prediction is closer to (further from) the full model prediction are indicated by the label ``$<1$'' (``$>1$''). The left panel depicts the $(\ks/\ms,\lam{S})$ plane, assuming $\ksss,\kssss=0$, while the right panel depicts the $(\ks/\ms,\ksss/\ms)$ plane, assuming $\lam{S},\kssss=0$. The dotted parts of the lines are excluded by theoretical constraints and the crosses indicate points beyond which no solution for $v_S$ exists in the full model.
\label{fig:singlet_comparison_2}
}
\end{figure}

In order to quantify further the comparison between dimension-6 and -8 predictions for the Higgs-singlet mixing angle, we can compare the differences over the parameter space between the EFT approximations at the two expansion orders from the full model. Specifically we take the absolute value of ratio of the differences between the two EFT approximations and the full model prediction for $\cos^2\alpha$:
\begin{align}
    \delta = \left| \frac{(\cos^2\alpha)_\mathrm{D8}-(\cos^2\alpha)_\mathrm{full}}{(\cos^2\alpha)_\mathrm{D6}-(\cos^2\alpha)_\mathrm{full}}\right| \, .
\end{align}
If this quantity is less than 1, dimension-8 provides a better approximation to the full model than dimension-6, while if it is greater than one, the converse is true, which indicates a non-converging SMEFT expansion. Fig.~\ref{fig:singlet_comparison_2} plots isocontours of the critical value of 1 for this ratio for values of $\ms$ between 500 GeV and 3 TeV in the $(\ks/\ms,\lam{S})$ plane, assuming $\ksss,\kssss=0$ (left panel) and the $(\ks/\ms,\ksss/\ms)$ plane, assuming $\lam{S},\kssss=0$ (right panel). The parameter plane for each mass is subject to the theoretical constraints from perturbativity and stability depicted, \emph{e.g.}, in Figs.~\ref{fig:singlet_kS_lS}--\ref{fig:singlet_kS_kS3_MS}. When a given line crosses into the forbidden region, it is drawn with a dotted style. Points marked with a cross indicate that the line has reached a point where no solution for $v_S$ exists in the full model, \emph{i.e.} the EW vacuum does not exist. 

In the regions labelled by ``$>1$'', the dimension-8 prediction is further from the full model than the dimension-6 one, meaning that the SMEFT expansion appears to break down and we likely cannot trust SMEFT approximations for Higgs coupling modifiers. The lines tend to lie between the normal and ``wrong-sign'' dimension-8 predictions in Fig.~\ref{fig:singlet_comparison}. For each $\ms$, we can draw similar conclusions to those drawn from Fig.~\ref{fig:singlet_comparison} in that, beyond a certain value of $\lam{S}$, the EFT validity is in question for a given value of $\ks$. We note that the validity criterion is trumped by the theoretical bounds for $\ms$ above 1.5 TeV, where the lines are completely within the forbidden regions. The 3 TeV line in the left panel is outside the plotted range, and in any case is rendered irrelevant by theoretical bounds.
We can now understand how the constraints we have derived do not accurately reflect the bounds on the singlet model over the whole parameter planes, highlighting, for each $\ms$, where the SMEFT approximation fails.

\subsection{Modifications of the quartic Higgs coupling}
We conclude our study of the singlet model by commenting briefly on the prospects for testing modifications to the quartic self-interaction of the Higgs boson, $d_4$, given in Eq.~(\ref{eq:d4_S}). Although a rather distant prospect, studies have show that this coupling could be constrained in the future by measuring triple Higgs production at a 
100 TeV proton-proton collider~\cite{Plehn:2005nk,Chen:2015gva,Papaefstathiou:2015paa,Fuks:2015hna,Fuks:2017zkg,Papaefstathiou:2019ofh} or a high-energy muon collider~\cite{Chiesa:2020awd}, or by precision measurements of double-Higgs production at through the loop-induced modifications due to $d_4$~\cite{Liu:2018peg,Maltoni:2018ttu,Borowka:2018pxx,Bizon:2018syu,DiMicco:2019ngk}. The projected sensitivities differ in optimism and also depend on the value of the trilinear Higgs self-coupling, $c_3$. They typically lie in the ballpark of $|d_4|\lesssim$ 2-20 at hadron colliders while more promising ${\cal O}(1)$ sensitivities are expected at high-energy muon colliders. 

To explore the scope for modified quartic Higgs interactions given the existing constraints on this model, we consider a simplified `post-LHC' scenario that represents the projected constraint set by the input dataset at that time on possible triple-Higgs rate measurements at future colliders. We assume that the current Higgs signal strengths are already systematics-dominated (which is true in many cases) and keep them fixed to their current values.  Di-Higgs measurements are, however, expected to improve significantly over the period of future LHC data taking. The ATLAS experiment has published a projected high-luminosity LHC sensitivity to the total rate based on extrapolations from their recent Run 2 analysis~\cite{ATLAS:2022okt}. Assuming a `baseline' scenario for the evolution of systematic uncertainties, the signal strength is expected to be measured with a relative uncertainty of -31\% and +34\%, which represents about a factor 3 improvement over the current measurement. We use this estimated measurement as our projected input data for the di-Higgs measurement, along with a corresponding projection for the CMS experiment obtained by improving the uncertainties of the existing CMS measurement by the same factor. 

\begin{figure}[h]
\centering
\includegraphics[width=\textwidth]{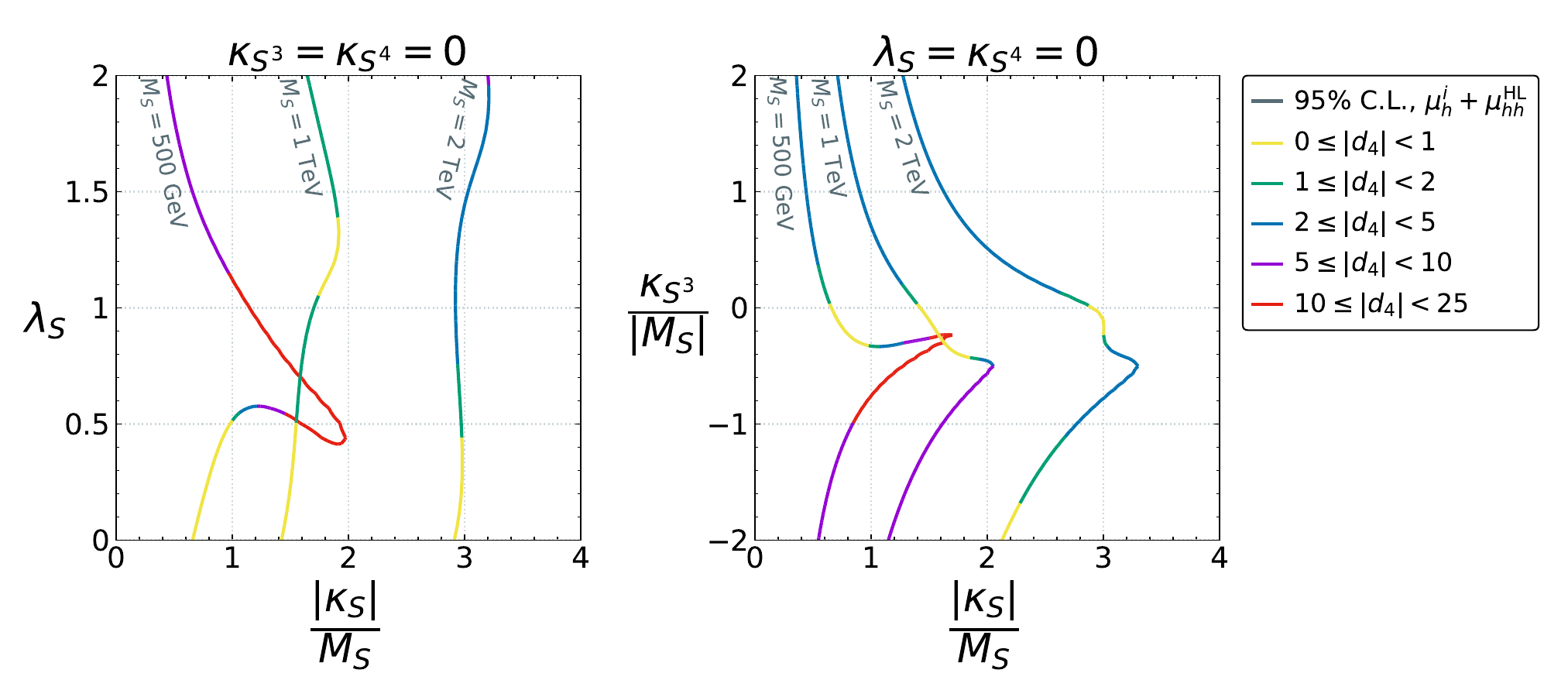}
\caption{\it Contours of the prospective 95\% C.L.
constraints from HL-LHC measurements of single- and di-Higgs production in the
$(\ks/\ms,\lam{S})$ plane assuming $\ksss,\kssss=0$ (left panel), and the $(\ks/\ms,\ksss/\ms)$ plane assuming $\lam{S},\kssss=0$ (right panel) for
specific values of $\ms = 500$~GeV, 1~TeV and 2~TeV, colour coded to
indicate the calculated values of $d_4$.
\label{fig:singlet_d4}
}
\end{figure}

Repeating the fit with the projected di-Higgs data yields moderate improvement in the prospective 95\% C.L. constraints on the parameter space, which are shown in Fig.~\ref{fig:singlet_d4}. These are given for specific values of $\ms = 500$~GeV, 1~TeV and 2~TeV in the $(\ks/\ms,\lam{S})$ plane assuming $\ksss,\kssss=0$ (left panel), and the $(\ks/\ms,\ksss/\ms)$ plane assuming $\lam{S},\kssss=0$ (right panel). The lines are  colour-coded to indicate the predicted values of $|d_4|$ along them, as per Eq.~\ref{eq:d4_S}, giving an idea of the possible modifications to the Higgs quartic coupling at the boundary of the allowed region of parameter space. We note that this is only an approximation of the true triple-Higgs production rate at dimension-8, since the singlet extension also predicts other, higher-derivative and higher-point interactions, such as $(\partial h)^4$, $t\bar{t}(\partial h)^2$ and $t\bar{t}h^3$ that would also affect this process.
Notwithstanding, we see for the singlet model that
the prospective HL-LHC measurements would not exclude the possibility that
$d_4$ may differ very significantly from the SM value, though we emphasise
that the largest deviations become possible only for relatively low values
of $\ms \lesssim 1$~TeV where the SMEFT approximation is generally less reliable.
Comparing to Fig.~\ref{fig:singlet_comparison_2}, we see that the very largest deviations occur in regions where the SMEFT expansion does not appear to converge, \emph{e.g.}, in the upper parts of the $\ms=500$ GeV line in the left plot or the lower parts of the 500 GeV and 1 TeV lines on the right plot. 

We do not go beyond $\ms=2$ TeV as the corresponding contours are largely excluded by theory constraints, although this does not preclude the possibility of significant $d_4$ modifications within the allowed parameter space for $\ms>2$ TeV. 
A more comprehensive exploration of the parameter space for $d_4$ may be interesting, as there may be points with large $d_4$ within the allowed region that we have not shown, and we have not considered the general case where all singlet model parameters are switched on. We have also neglected the singlet quartic coupling, $\kssss$, which is very poorly constrained by the existing data. It is, however, likely that the validity of the SMEFT approximation to triple Higgs production may be restricted to smaller regions than those we have determined using $\cos^2\alpha$, since the energy scale of the process is higher than single Higgs production and there is the possibility of resonant enhancements from on-shell singlet production.  Indeed, this process has been found to be interesting in the context of non-minimal scalar extensions of the SM~\cite{Papaefstathiou:2020lyp}.
Nevertheless, these exploratory results suggest that a measurement of the Higgs quartic coupling could provide an additional, complementary probe of the singlet model parameter space.

\section{Electroweak Triplet Scalar Model}
\label{sec:triplet}
We now move on to study the extension of the SM by a single real electroweak triplet 
scalar field $\Xi$ with hypercharge $Y_{\sss \Xi} = 0$. The tree-level SMEFT matching of this model to
dimension-6 has been calculated in Refs.~\cite{Henning:2014wua, Corbett:2017ieo, deBlas:2017xtg},
and partial dimension-8 results have been computed in Ref.~\cite{Corbett:2021eux}, focusing on operators that modify 3-point interactions between SM fields.
Similarly to the $S$ field, the $\Xi$ field interacts only with the SM Higgs
doublet at the renormalisable level.
Following again the conventions of Ref.~\cite{deBlas:2017xtg}, the full model Lagrangian is:
\begin{equation}
\begin{split}
    \mathcal{L}_{\sss{\Xi}} =& \frac{1}{2}(D_{\mu} \Xi^a)(D^{\mu} \Xi^a) - \frac{1}{2}\mx^2(\Xi^a \Xi^a) -  \kappa_{\sss{\Xi}}H^{\dagger} \Xi^a \sigma^a H +\\ &- \lambda_{\sss{\Xi}}(\Xi^a \Xi^a)(H^{\dagger}H) -  \frac{1}{4}\eta_{\sss{\Xi}}(\Xi^a \Xi^a)^2\,,
\end{split}
\label{eq:L_Xi}
\end{equation}
where $\sigma^a$ ($a = 1,2,3$) are the three Pauli matrices, and the couplings
have the following mass dimensions $[\kappa_{\sss{\Xi}}] = 1$ and
$[\lambda_{\sss{\Xi}}]= [\eta_{\sss{\Xi}}] = 0$. The potential parameters are subject to theoretical constraints~\cite{Khan:2016sxm,Krauss:2018orw} from boundedness:
\begin{align}
    \lambda,\ex \geq 0,\,\,|\lx| \geq -\sqrt{\lambda \ex} \, ,
\end{align}
and perturbative unitarity
\begin{align}
    |\lambda|\leq 4\pi,\,\,|\ex|\leq 4\pi,\,\,|6\lambda + 5 \ex \pm\sqrt{(6\lambda-5\ex)^2+48\lx^2}|\leq16\pi \, .
\end{align}
In general, the scalar potential can have more than one minimum, and we must ensure that the EW vacuum is the lowest energy state.

Writing Eq.~\eqref{eq:L_Xi} in a similar form to Eq.~\eqref{eq:L_S_B_and_U} defines the parameters:
\begin{align}
B^a = \kx \hsdh;\quad U = 2\lx \hdh \, .
\end{align}
The CDE procedure proceeds identically to that of the singlet, and is simplified thanks to the absence of a trilinear self-interaction for the triplet. The solution to the classical equation of motion is found to be
\begin{align}
  \label{eq:sol_triplet}
    \Xi_{\sss C}^a \approx \oo\left[
    B^a + \ex  \oo B^a(\oo B^b)^2
    \right]+\omm{6}\,,
\end{align}
including the higher-order term beyond the linearised solution.
Performing the inverse mass expansion and plugging it back into Eq.~\eqref{eq:L_Xi}, we find the following tree-level effective Lagrangian up to dimension-8:
\begin{align}
\begin{split}
    \label{eq:L_eff_Xi}
    \mathcal{L}^{(8)}_{\sss \text{eff.,tree},\Xi} =&    \frac{1}{2\mx^2}B^aB^a
         +\frac{1}{2\mx^4}B^a(P^2+U)B^a
         +\frac{1}{2\mx^6}B^a(P^2-U)(P^2-U)B^a \\
         &-\frac{\ex}{4\ms^8} (B^aB^a)^2+ \omm{5}\,.
\end{split}
\end{align}
This turns out to give the same result as using the linearised solution to the equation of motion.

The main difference with respect to the singlet matching to the dimension-6 Warsaw basis is that a redundant operator, $\Op{r}=\hdh (D^\mu H)^\dagger(D_\mu H)$, is generated that must be eliminated. This is done most efficiently using a Higgs field redefinition,
\begin{align}
    \label{eqn:hredef_eom}
    H\to& H\left(1-\frac{\kx^2}{\mx^4}\hdh\right),
\end{align}
in such a way that we keep track of the associated corrections out to dimension-8. Alternatively, one can make use of the Higgs equation of motion out to dimension-6, along with an additional correction stemming from the second-order variation of the action~\cite{Criado:2018sdb},
\begin{align}
    \frac{\delta S}{\delta \phi^A\delta\phi^B} f^A f^B;\quad \phi^{1(2)} = H^{(\dagger)}, f^1 = |H|^2 H, f^2=(f^1)^\dagger.
\end{align}
We verified that both methods yield the same dimension-8 correction to the low-energy effective action. For the dimension-8 terms generated by Eq.~\eqref{eq:L_eff_Xi}, the SM Higgs equation of motion~\eqref{eq:Higgs_EOM} can be used to reduce to the dimension-8 operators defined in Table~\ref{tab:dim8}. The SMEFT Wilson 
coefficients resulting from the matching of the triplet model at dimension-8 
are reported in Table~\ref{tab:triplet_results}. The relevant subset of our results agree with the computation of Ref.~\cite{Corbett:2021eux}.

The other detail that differs slightly from the singlet case is the appearance of the Higgs self-coupling parameter, $\lambda$, in the dimension-6 matching result for $\Cp{H}$. Because of the fact that a correction to this dimension-4 coupling is induced by this model, the $\lambda$ initially appearing in the matching result is not the same quantity that is expressed as a function of the input parameters in Eq.~\eqref{eq:dlam}. The appearance of SM couplings in tree-level matching conditions is always a consequence of basis reduction via field redefinitions or employing equations of motion. This means that any initial (\emph{i.e.} before basis reduction) contribution to the coefficient of $\Op{H^4}$ generated by the CDE can be absorbed into the definition of $\lambda$ without physical consequence. For the triplet, this corresponds to an initial redefintion of 
\begin{align}
         \lambda\to\lambda + \frac{\kx^2}{2\mx^2}.
\end{align}
However, subsequent contributions to $\Cp{H^4}$ generated by the basis reduction lead to the aforementioned mismatch, requiring a further redefinition of $\lambda$ that gets propagated to the Wilson coefficients of our basis. Taking into account the required shift:
\begin{align}
        \lambda\to\lambda - \frac{2\mu^2\kx^2}{\mx^4} + \frac{6\mu^4\kx^2}{\mx^6} \, ,
\end{align}
gives a correction to $\Cp{H}$ at the dimension-8 level. The results in Table~\ref{tab:triplet_results} are expressed only in terms of parameters appearing in the SMEFT Lagrangian, \emph{i.e.}, $\lambda$ and $\mu^2$ are defined as the low-energy coefficients of $\Op{H^4}$ and $\hdh$. As discussed in Section~\ref{sec:singlet}, this is not relevant for the singlet case, since $\lambda$ only appears in dimension-8 Wilson coefficients, meaning that all associated corrections occur beyond the order at which we are working. As in the case of the singlet scalar, we see that the dimension-8 matching also implies corrections to the dimension-6 coefficients. We detail in Appendix~\ref{app:hhhh} an analogous calculation of the $hh\to hh$ amplitude that validates a subset of our results at dimension-8, from which we can draw some confidence that the remainder of the matching results are reliable.

Finally, we can evaluate the dimension-6 and -8 contributions to the normalization of the physical Higgs field (\ref{eq:hredef})
in the triplet scalar extension of the SM:
\begin{equation}
\begin{split}
    \label{eq:Deltah_Xi}
    \Delta_h = & \frac{\kx^2 \vh^2}{\mx^4} \bigg[1 - \frac{1}{2\mx^2}(3\mh^2 +\lx \vh^2) + \frac{9\kx^2\vh^2}{8\mx^4} \bigg] 
    \approx  \frac{\kx^2 \vh^2}{\mx^4} \bigg( 1 - \frac{3\mh^2 \vh^2}{2\mx^2} + \frac{9\kx^2 \vh^2}{8\mx^4}\bigg) \, ,
\end{split}
\end{equation}
and also to the trilinear and quadrilinear Higgs self-couplings (\ref{eq:c3}, \ref{eq:d4}):
\begin{equation}
\begin{split}
    c_3 =& \, \frac{\kx^2\vh^2}{\mx^4} \bigg[-1 + \frac{2\lx\vh^2}{\mh^2} + \frac{\vh^2}{\mx^2}\bigg(-\frac{3\mh^2}{\vh^2} + 9\lx \bigg(1 - \frac{8\lx\vh^2}{9\mh^2}\bigg) \\ &\, + \frac{2\kx^2}{\mx^2}\bigg(1 - \frac{\vh^2}{\mh^2}\bigg( 3\lx - \frac{\ex}{2}\bigg)\bigg)\bigg) \bigg]\\
    \approx&\, -\frac{\kx^2\vh^2}{\mx^4} \bigg(1 + \frac{3\mh^2}{\mx^2} - \frac{2\kx^2\vh^2}{\mx^4}\bigg) \, ,
    \label{eq:c3_Xi}
\end{split}
\end{equation}
\begin{equation}
\begin{split}
    d_4 =& \, \frac{\kx^2\vh^2}{\mx^4} \bigg[-\frac{22}{3} + \frac{12\lx\vh^2}{\mh^2} + \frac{\vh^2}{\mx^2}\bigg(-\frac{82\mh^2}{3\vh^2} + 2\lx \bigg(41 - \frac{32\lx\vh^2}{\mh^2}\bigg) \\ &\, + \frac{\kx^2}{\mx^2}\bigg(21 - \frac{8\vh^2}{\mh^2}( 6\lx - \ex)\bigg)\bigg) \bigg]\\
    \approx&\, -\frac{\kx^2\vh^2}{\mx^4} \bigg(\frac{22}{3} + \frac{82\mh^2}{3\mx^2} - \frac{21\kx^2\vh^2}{\mx^4}\bigg) \, .
    \label{eq:d4_Xi}
\end{split}
\end{equation}
In each of the above equations, the third line, indicated by ``$\approx$'', corresponds to the case where all the couplings except $\kappa_\Xi$ are assumed to be negligible. 
\begin{table}[H]
    \begin{center}
    \renewcommand{\arraystretch}{1.4}
    \begin{tabular}{|c||c|c|}
    \hline 
     \multirow{4}{4em}{Dim - 6} & $C_H$ & $\frac{\kx^2}{\mx^2}\left( (4\lambda - \lx)\left(1 - \frac{4\mu^2}{\mx^2}\right)- \frac{5\mu^2\kx^2}{\mx^6}\right)$ \\[0.6ex]
     \cline{2-3}
     & $C_{HD}$ & $-\frac{2\kx^2}{\mx^2}\left(1 - \frac{4\mu^2}{\mx^2}\right)$\\[0.6ex]
     \cline{2-3}
     & $C_{H\Box}$ & $\frac{\kx^2}{2\mx^2}\left(1 - \frac{4\mu^2}{\mx^2}\right)$ \\[0.6ex]
     \cline{2-3}
     & $[C_{\psi H}]_{\text{\tiny wx}} $ & $ [y_{\psi}]_{\text{\tiny wx}} \frac{\kx^2}{\mx^2} \left(1 - \frac{4\mu^2}{\mx^2} \right)$; $\psi = u,d,e $\\[0.6ex]
     \hline \hline
     \multirow{20}{4em}[0cm]{Dim - 8} & $C_{H^8}$ & $\frac{2\kx^2}{\mx^2}\left((2\lambda - \lx)^2 + \frac{\kx^2}{\mx^2}(3\lx - 5\lambda - \frac{\ex}{8})\right)$ \\[0.6ex]
     \cline{2-3}
     & $C_{H^6}^{(1)}$ & $ - \frac{\kx^4}{\mx^4}$ \\[0.6ex]
     \cline{2-3}
     & $C_{H^6}^{(2)}$ & $\frac{4\kx^2}{\mx^2}\left(\lx - 2\lambda + \frac{\kx^2}{\mx^2}\right)$ \\[0.6ex]
     \cline{2-3}
     & $C_{H^4}^{(1)}$ & $\frac{4\kx^2}{\mx^2}$\\[0.6ex]
     \cline{2-3}
     & $C_{H^4}^{(3)}$ & $-\frac{2\kx^2}{\mx^2}$ \\[0.6ex]
     \cline{2-3}
     & $[C_{l\psi H^5}/C_{q\psi H^5}]_{\text{\tiny wx}}$ & $-[y_\psi]_{\text{\tiny wx}}\frac{2\kx^2}{\mx^2}\left(\lx - 2\lambda + \frac{\kx^2}{2\mx^2}\right)$; $\psi = u,d,e $ \\[0.6ex]
     \cline{2-3}
     & $[C_{l^2 \psi^2 H^2}^{(1)}/C_{q^2 \psi^2 H^2}^{(1)}]_{\text{\tiny wxyz}}$ & $-[y_\psi]_{\text{\tiny wz}}[y_\psi ^\dagger]_{\text{\tiny yx}}\frac{3\kx^2}{4\mx^2}$; $\psi=u,d,e$ \\[0.6ex]
     \cline{2-3}
     & $[C_{l^2e^2 H^2}^{(2)}/C_{q^2d^2 H^2}^{(2)}]_{\text{\tiny wxyz}}$ & $[y_\psi]_{\text{\tiny wz}}[y_\psi ^\dagger]_{\text{\tiny yx}}\frac{\kx^2}{4\mx^2}$; $\psi=d,e$ \\[0.6ex]
     \cline{2-3}
     & $[C_{q^2 u^2 H^2}^{(2)}]_{\text{\tiny wxyz}}$ & $-[y_u]_{\text{\tiny wz}}[y_u ^\dagger]_{\text{\tiny yx}}\frac{\kx^2}{4\mx^2}$\\[0.6ex]
     \cline{2-3}
     & $[C_{l^2\psi^2H^2}^{(3)}/C_{q^2\psi^2H^2}^{(5)}]_{\text{\tiny wxyz}}$ & $[y_\psi]_{\text{\tiny wx}}[y_\psi]_{\text{\tiny yz}}\frac{\kx^2}{2\mx^2}$; $\psi=u,d,e$ \\[0.6ex]
     \cline{2-3}
     & $[C_{lequH^2}^{(1)}]_{\text{\tiny wxyz}}$ & $ [y_e]_{\text{\tiny wx}}[y_u]_{\text{\tiny yz}}\frac{5\kx^2}{2\mx^2}$ \\[0.6ex]
     \cline{2-3} 
     & $[C_{leqdH^2}^{(1)}]_{\text{\tiny wxyz}}$ & $ [y_e]_{\text{\tiny wx}}[y_d^\dagger]_{\text{\tiny yz}}\frac{5\kx^2}{2\mx^2}$ \\[0.6ex]
     \cline{2-3}
     & $[C_{q^2udH^2}^{(1)}]_{\text{\tiny wxyz}}$ & $-[y_u]_{\text{\tiny wx}}[y_d]_{\text{\tiny yz}}\frac{5\kx^2}{2\mx^2}$ \\[0.6ex]
     \cline{2-3}
     & $[C_{lequH^2}^{(2)}]_{\text{\tiny wxyz}}$ & $ [y_e]_{\text{\tiny wx}}[y_u]_{\text{\tiny yz}}\frac{\kx^2}{2\mx^2}$ \\[0.6ex]
     \cline{2-3} 
     & $[C_{leqdH^2}^{(2)}]_{\text{\tiny wxyz}}$ & $ -[y_e]_{\text{\tiny wx}}[y_d^\dagger]_{\text{\tiny yz}}\frac{\kx^2}{2\mx^2}$ \\[0.6ex]
     \cline{2-3}
     & $[C_{q^2udH^2}^{(2)}]_{\text{\tiny wxyz}}$ & $[y_u]_{\text{\tiny wx}}[y_d]_{\text{\tiny yz}}\frac{\kx^2}{2\mx^2}$ \\[0.6ex]
     \cline{2-3}
     & $[C_{lequH^2}^{(5)}]_{\text{\tiny wxyz}}$ & $ [y_e]_{\text{\tiny wx}}[y_u^\dagger]_{\text{\tiny yz}}\frac{\kx^2}{\mx^2}$ \\[0.6ex]
     \cline{2-3} 
     & $[C_{leqdH^2}^{(3)}]_{\text{\tiny wxyz}}$ & $ [y_e]_{\text{\tiny wx}}[y_d]_{\text{\tiny yz}}\frac{\kx^2}{\mx^2}$ \\[0.6ex]
     \cline{2-3}
     & $[C_{q^2udH^2}^{(5)}]_{\text{\tiny wxyz}}$ & $[y_d]_{\text{\tiny wx}}[y_u^\dagger]_{\text{\tiny yz}}\frac{\kx^2}{\mx^2}$ \\[0.6ex]
     \cline{2-3}
     & $[C_{l \psi H^3D^2}^{(2)}/C_{q \psi H^3D^2}^{(2)}]_{\text{\tiny wx}}$ & $-[y_\psi]_{\text{\tiny wx}}\frac{4\kx^2}{\mx^2}$; $\psi=u,d,e$ \\[0.6ex]
     \hline
    \end{tabular}
    \renewcommand{\arraystretch}{1}
    \caption{\it Dimension-6 and -8 Wilson coefficients resulting from the tree-level matching of the triplet scalar model to the SMEFT. Flavour indices are denoted by Roman letters 
    $\{\mathrm{w,x,y,z}\}$. The parameters $\mu^2$ and $\lambda$ are the quadratic and quartic coefficients of the Higgs potential at the EW scale, respectively.}
    \label{tab:triplet_results}
\end{center}
\end{table}
\subsection{Constraints on Model Parameters}
Table~\ref{tab:triplet_results} shows that the triplet scalar model generates a few more Wilson coefficients at dimension-6 in addition to $\Cp{H\Box}$ and $\Cp{H}$, namely the Yukawa operators $\Cp{\psi H},\psi=u,d,e$ and most importantly the custodial symmetry-violating operator $\Cp{HD}$. The latter is commonly associated to the so-called $T$-parameter, since it contributes to the $Z$-boson mass term and therefore leads to a modified prediction for the $W$-boson mass as well as shifts in the $W$- and $Z$-boson couplings to the fermion currents with respect to the SM. As a result, the operator is strongly constrained by EWPOs, which represent by far the most important bounds on this model. For reference, we note that our dataset constrains the coefficient to lie within
\begin{align}
    \Cp{HD}\subset [-0.020,0.0040]
\end{align}
at 95\% CL, assuming $\Lambda=1$ TeV. This bound is completely dominated by the EWPOs, with the addition of single Higgs data modifying the lower and upper boundaries by 1 and 5\%, respectively. Even within the EWPO dataset, the bound is dominated by the $W$-mass measurement~\footnote{Pending scrutiny and confirmation of the recent CDF measurement of $m_W$~\cite{CDF:2022hxs}, we do not include it in our nominal analysis.}, since removing the $Z$-pole data only loosens the bounds by 10--20\%. This is 1--2 orders of magnitude more sensitive that the bounds on the $\Cp{H\Box}$ implied by the single Higgs data, see Eq.~\eqref{eq:CHbox_CH_bounds}. The bias towards negative values occurs because the inverse-variance weighted average of the pre-2022 $W$-mass measurements lies about 1.2 standard deviations above the SM prediction of 80.361 GeV used in our analysis, and $\Cp{HD}$ contributes linearly to $\mw$ with a negative coefficient. 

As in the singlet model, most of the dimension-6 Wilson coefficients only depend on $\kx$.
From the tight EWPO constraints discussed above, we can expect this quantity to be much more tightly constrained than the singlet analogue, $\ks$, and consequently for Higgs data to only have a modest impact on the allowed parameter space. The quartic portal coupling, $\lx$, only appears in $\Cp{H}$ at the dimension-6 level but, given the strong bounds on $\kx$, the deviations in di-Higgs production are expected to be below the current sensitivity. Overall, the interplay between the data and the model parameters in the triplet model is somewhat analogous to that of the singlet with the role of Higgs data being played by the EWPO and that of di-Higgs being played by single Higgs data. However, judging from the single operator analyses of $\Cp{HD}, \,\Cp{H\Box}$ and $\Cp{H}$, in this and the previous section, we can estimate that the relative sensitivity of EWPO to single Higgs is about 10 times better than that of single to di-Higgs data. This should result in a much milder impact of the less sensitive dataset (single Higgs in the triplet case) on the overall results. 

At the dimension-8 level, several more operators are generated than in the singlet case, including the custodial symmetry-violating operator $\Opp{H^6}{(2)}$ and an additional four-derivative quartic Higgs operator $\Opp{H^4}{(1)}$. We note that new $\kx^4/\mx^4$-dependent terms are now present in the matching results, due to the dimension-8 corrections from the dimension-6 reduction to the Warsaw basis operators. Most importantly, $\lx$ dependence is introduced into the EWPOs via $\Cpp{H^6}{(2)}$. This parameter also appears in coefficients relevant for single-Higgs data. Finally, as in the singlet model, the quartic self-coupling of the triplet, $\ex$, appears for the first time in the octic Higgs operator, $\Op{H^8}$. Since we are not even sensitive to this parameter in the singlet case, we can be sure that this will also be true for the triplet. We therefore do not discuss this parameter any further.

\begin{figure}[h!]
\centering
\includegraphics[width=0.46\textwidth,valign=t]{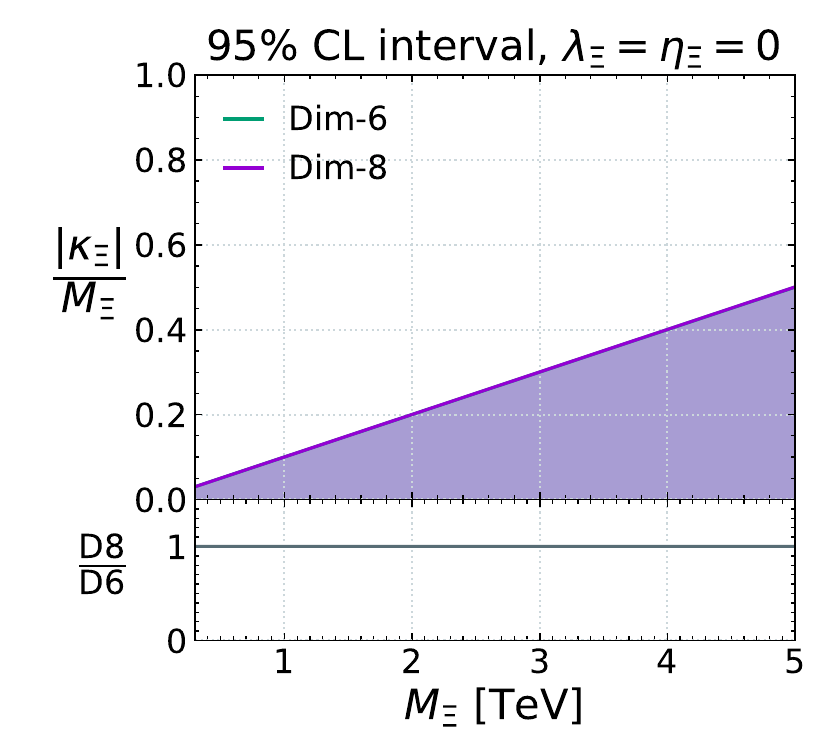}
\caption{\label{fig:Triplet_kxi}
\it The range of $|\kx|/\mx$ allowed as a function of $\mx$ by the current experimental data.
The green shaded area represents the allowed interval at 95\% Confidence Level from a dimension-6 analysis including only linear (interference) effects, and the purple shaded area is from an analysis including also linear dimension-8 contributions and quadratic dimension-6 contributions. The subplot shows the ratio of the dimension-8 and dimension-6 determinations of the upper bound.}
\end{figure}

The triplet matching results are insensitive to the sign of $\kx$, so we report subsequently results in the positive half of the parameter space. Fig.~\ref{fig:Triplet_kxi} shows the bounds implied by the data on the combination $|\kx|/\mx$, assuming the other parameters, $\lx$ and $\ex$, are set to 0. Bounds derived at the linear dimension-6 level are shown in green, while bounds obtained at the quadratic dimension-6 and linear dimension-8 level are shown in purple. 
The bounds obtained in the dimension-6 and -8 approximations coincide almost completely, differing only at the per-mille level. This is also shown in the subplot, where the ratio of the two upper bounds is exactly 1. Compared to the analogous parameter in the singlet model, $\ks$, $|\kx|/\mx$ is much better constrained thanks to the better precision of the EWPOs compared to Higgs data.
As discussed in the SMEFT case above, the bounds are dominated by the $W$-mass measurement. 

\begin{figure}[h!]
\centering
\includegraphics[width=\textwidth]{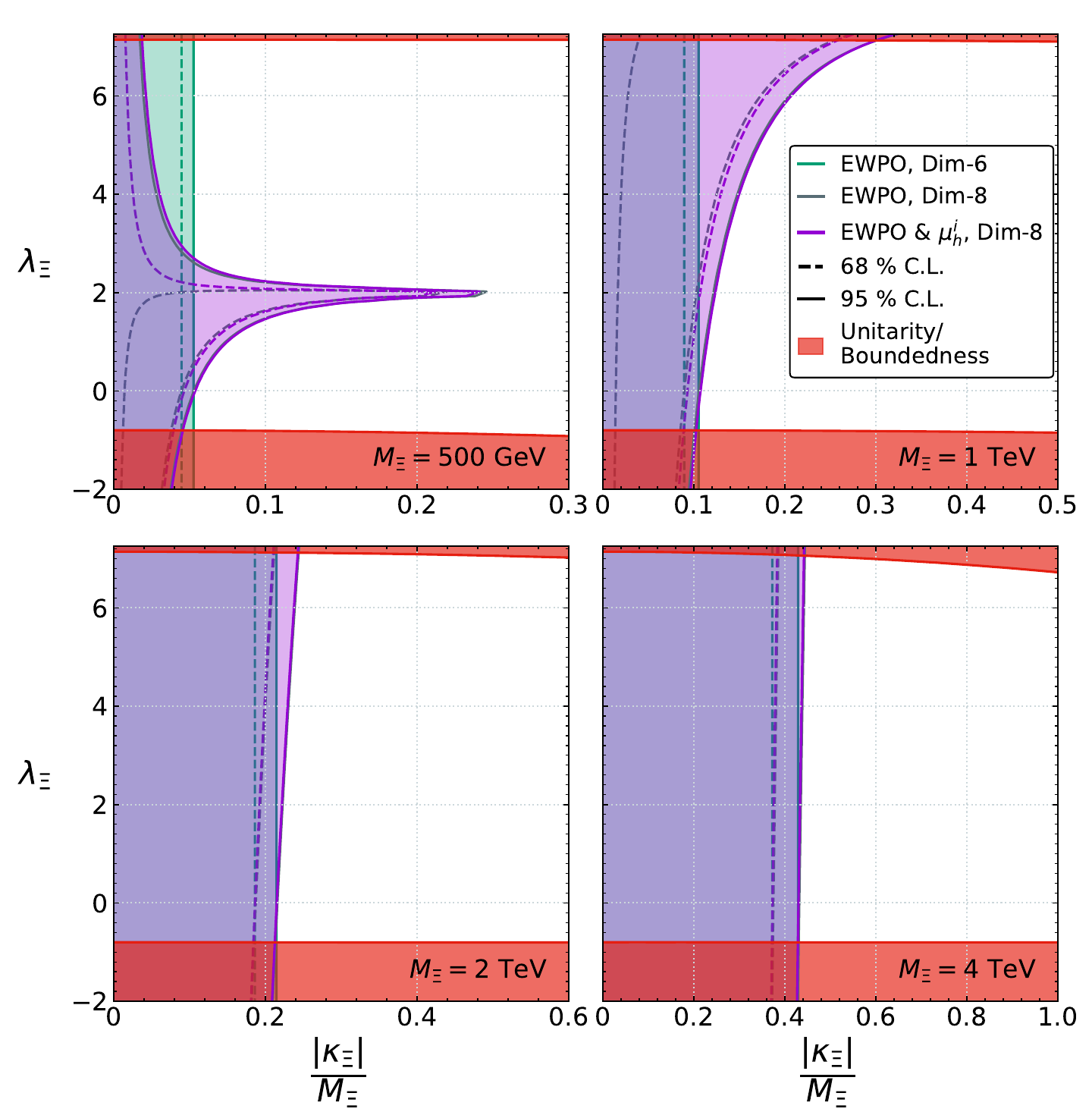}
\caption{\it
 Values of $\lx$ for $\mx = 1$~TeV and $\eta_\Xi = 0$
that are allowed at the 68\% CL (dashed lines) and 95\% CL (solid lines)
by the present electroweak precision observables including dimension-6 operators in the
linear approximation (green shading and lines),
including both the quadratic effects of dimension-6 operators and the linear effects of dimension-8
operators (grey shading and lines), and also including the present Higgs coupling measurements (purple 
shading and lines) as functions of
$|\kappa_\Xi|/\Lambda$. The red shaded regions at large $\lx$ are excluded by the 
perturbative unitarity constraint $\lambda \le 8 \pi/3$, and the boundedness of the scalar potential. These are determined by allowing for all possible values of $\ex$ allowed by perturbative unitarity and boundedness. }
\label{fig:Triplet_kxi_lxi_Mxi}
\end{figure}

We now allow for non-zero values of the quartic portal coupling, $\lx$ and examine the impact of the data on the 2-dimensional parameter space. 
In the four panels of Fig.~\ref{fig:Triplet_kxi_lxi_Mxi} we compare the values of 
$|\kx|/\mx$ that are allowed as functions of $\lx$
by the EWPOs for $\mx = 0.5,1,2$ and $4$~TeV  assuming $\ex = 0$,
at the 68\% CL (dashed lines) and 95\$ CL (solid lines),
including dimension-6 operators in the linear approximation (green shading and lines) and including both 
the quadratic effects of dimension-6 operators and the linear effects of dimension-8
operators (grey lines). We also show the effects of including the present 
Higgs coupling measurements (purple shading and lines). Regions that do not satisfy partial-wave unitarity bounds or where the potential is unbounded are shaded in red. These bounds are obtained in a conservative way, varying over all allowed values of $\ex$, such that they represent the absolute maximum available region of parameter space. In the specific case of $\ex=0$, for example, $\lx$ would be constrained to be strictly positive, while non-zero $\ex$ can permit small negative $\lx$.
We see that at the linear dimension-6 level the constraint on $|\kappa_\Xi|/\Lambda$ is independent
of $\lx$, which only enters in di-Higgs production at this level. We confirm that the di-Higgs data is not sensitive to this parameter, given the EWPO constraints. Instead, there is a strong dependence on $\lx$ when the quadratic effects of 
dimension-6 operators and the linear effects of dimension-8 operators are also included,
which arises from a cancellation between the dimension-6 and dimension-8 contributions.
We also note that the effects of including the Higgs coupling measurements at the dimension-8
level are minor, contributing mainly to rule out a small part of the tip of the cancellation region for $\mx=0.5$ TeV. Between $\mx=0.5$ and 1 TeV, the tip of this region disappears above the perturbative limit for $\lx\gtrsim4\pi/\sqrt{3}$. As $\mx$ increases, the dimension-8 effects gradually decouple. 

\subsection{Comparison between the Dimension-6 and -8 Results and the Full Model}
\label{sec:comparison}
\begin{figure}[h!]
\centering
\includegraphics[width=\textwidth]{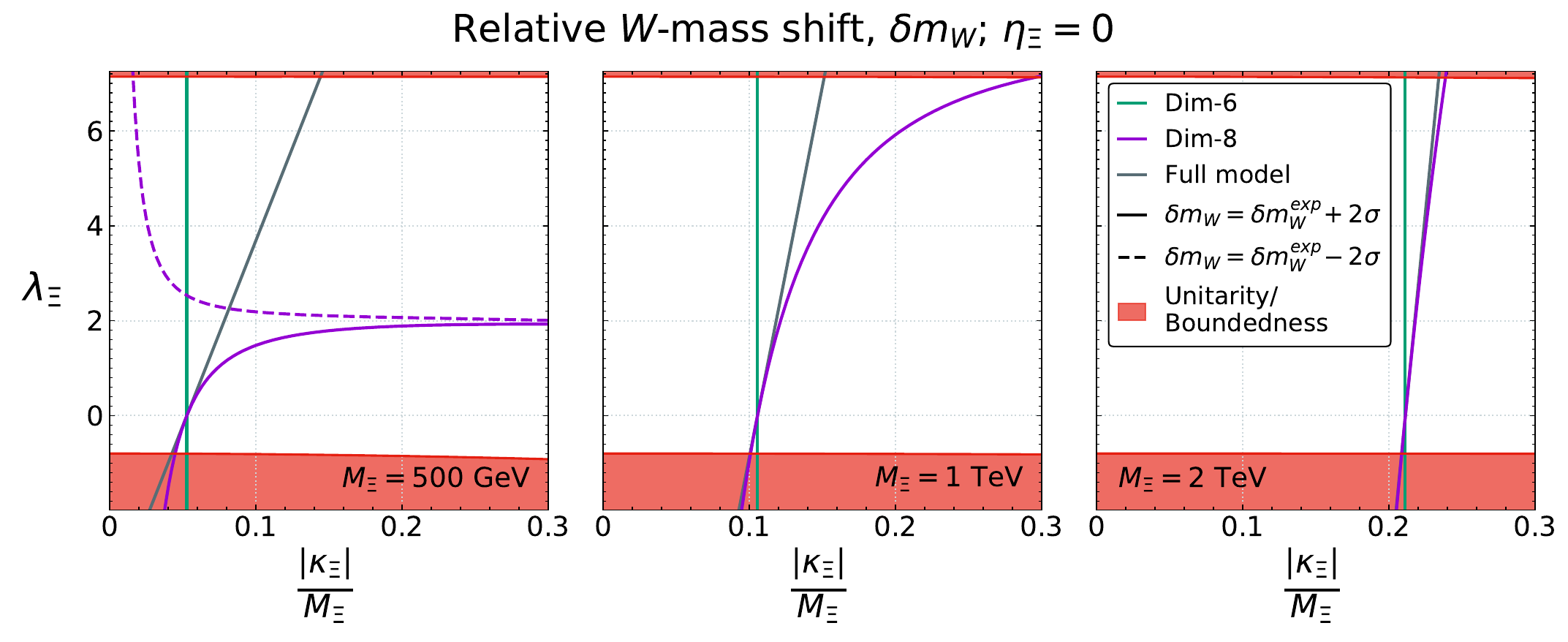}
\caption{\it Contours of constant $W$-boson mass as a function of $|\kx|/\mx$ and  $\lx$ for $\mx = 500$~GeV (left panel) $1$~TeV (middle panel) and $2$~TeV (right panel), all for $\eta_\Xi = 0$. Predictions at the linear dimension-6 level (green lines) are compared to the linear dimension-8/quadratic dimension-6 level (purple lines) and those from the full model (grey lines). The contours are plotted to match the $2\sigma$ upper (solid lines) and lower (dashed line in the left panel) bounds from pre-2022 $\mw$ measurements.}
\label{fig:Triplet_kxi_lxi_pre-CDF}
\end{figure}

The large differences between the dimension-6 and -8 results  signal that the  EFT description may be breaking down in some regions of the parameter space. In order to quantify this, we compare the two approximations to the full model description, focusing on the $\mw$ prediction that dominates the experimental constraints on the triplet scalar model. As with the singlet model, the fact that our matched, dimension-8 SMEFT prediction agrees with the Taylor-expanded full model prediction:
\begin{align}
    \frac{\mw}{\mw^{\sss \mathrm{SM}}} = 
    1+ \frac{\cwh^2}{\ctwh}\frac{\vh^2\kx^2}{\mx^4}\left(
     \frac{1}{2}
    - \frac{\vh^2}{\mx^2}\left(\lx 
    +  \frac{22\cwh^4-25\cwh^2+8}{8\ctwh^2}\frac{\kx^2}{\mx^2}
    \right)\right) \, ,
\end{align} 
provides an additional validation of our matching calculation.
Fig.~\ref{fig:Triplet_kxi_lxi_pre-CDF} depicts the constraint from the combined, pre-2022 $\mw$ measurements in the $(|\kx|/\mx,\lx)$ plane, for masses of $0.5,1$ and $2$ TeV and $\ex=0$. The constraint is determined by requiring that the $\mw$ prediction lie no more than 2 standard deviations above (solid lines) or below (dashed lines) the inverse-variance weighted average of the measurements. Constraints obtained with predictions at the dimension-6 and -8 levels, shown in green and purple, respectively, are contrasted with the full model prediction in grey. 

The shapes of the constraints in the EFT approximations are very similar to those of the full fits in Fig.~\ref{fig:Triplet_kxi_lxi_Mxi}. The dimension-6 bound does not depend on $\lx$, unlike the full model prediction, whereas a strong dependence is introduced at dimension-8, leading to a cancellation region that moves up with increasing $\mx$. In contrast, while the full model does display a $\lx$ dependence, the extreme cancellation displayed in the dimension-8 result is not present. The dimension-8 approximation very closely reproduces the full model at moderate $\lx$ but starts to diverge significantly at larger $\lx$ and smaller $\mx$, namely around $\lx\sim 1$ for $\mx=500$ GeV and around $\lx\sim 3$ for $\mx=1$~TeV. On the other hand, when $\mx = 2$~TeV the dimension-8 approximation is very accurate for all values of $\lx$ between the perturbative upper bound and the stability lower bound, shown as the red lines at $\lx \simeq 7$ and -1, respectively. We see that, in the dimension-8 approximation, the $\mw$ shift changes sign for large $\lx$, which can be traced back to the negative $\Opp{H^6}{(2)}$ contribution in Eqs.~\eqref{eq:dgwsq} and~\eqref{eq:mw} that depends linearly on $\lx$ (see Table~\ref{tab:triplet_results}). The full model dependence is such that this never occurs in reality, so this region is not well described by an EFT expansion to dimension-8. Fortunately the dimension-8 and full model descriptions agree better with increasing mass, and at $\mx=2$ TeV, the model is faithfully reproduced in this approximation. We also checked the regions of parameter space over which the dimension-8 predictions for $\mw$ are closer to the full model, finding that for $\mx> 850$ GeV, the 95\% C.L. contours derived from the global analysis lie completely within these regions.
Hence, although the dimension-8 analysis correctly introduces a $\lx$ dependence, it is not reliable for large values of this quartic coupling. For example, for $\mx>500$ GeV, the dimension-8 prediction is worse than the dimension-6 one above $\lx\sim2.5$.

\begin{figure}[h!]
\centering
\includegraphics[width=0.49\textwidth]{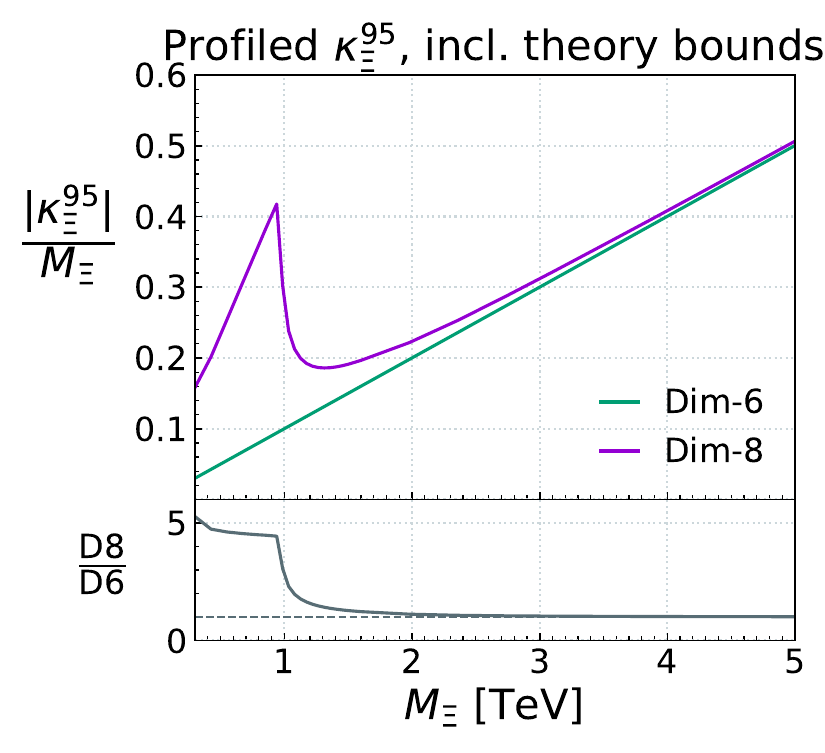}
\caption{\it
Profiled 95\% CL upper limit on $|\kx|/\mx$ as a function of $\mx$, where $|\lx|$ is varied within the region constrained by perturbative unitarity and boundedness, at the linear dimension-6 level (green) and at the 
dimension-8 level (purple). The subplot shows the ratio between the limits.}
\label{fig:Triplet_kxi_lxi_profiled}
\end{figure}

Fig.~\ref{fig:Triplet_kxi_lxi_profiled} shows the profiled 95\% CL upper limits 
on $|\kx|/\mx$ as functions of $\mx$ at the linear dimension-6 level (green) 
and at the dimension-8 level (purple), showing also the ratio between the limits, imposing the aforementioned
perturbativity and boundedness restrictions on $\lx$ and $\ex$.
Compared to the individual constraint shown in Fig.~\ref{fig:Triplet_kxi}, the profiled result highlights the impact of upgrading the analysis to dimension-8.  The spike in the dimension-8 limit,  which
reaches $\kx/\mx \simeq 0.42$ for $\mx \simeq 900$~GeV, arises in the cancellation region in the upper left panel of Fig.~\ref{fig:Triplet_kxi_lxi_Mxi}, and the dip at larger $|\kx|/\mx$ begins when the tip of this region starts to disappear beyond the perturbativity bound. However, as we have just discussed, the emergence of such a region is an artefact of the EFT approximation exaggerating the $\lx$ dependence of the $\mw$ prediction. This suggests that the EFT interpretation used to obtain this result is  not reliable below $\mx\sim1$ TeV. Above this mass, dimension-8 effects appreciably loosen the profiled bound, while the linear dimension-6
approximation gradually approaches the dimension-8 value at large $\mx$, indicating a nice convergence of the expansion.

\subsection{Interpreting the new CDF $\mw$ measurement}
\begin{figure}[h]
\centering
\includegraphics[width=\textwidth]{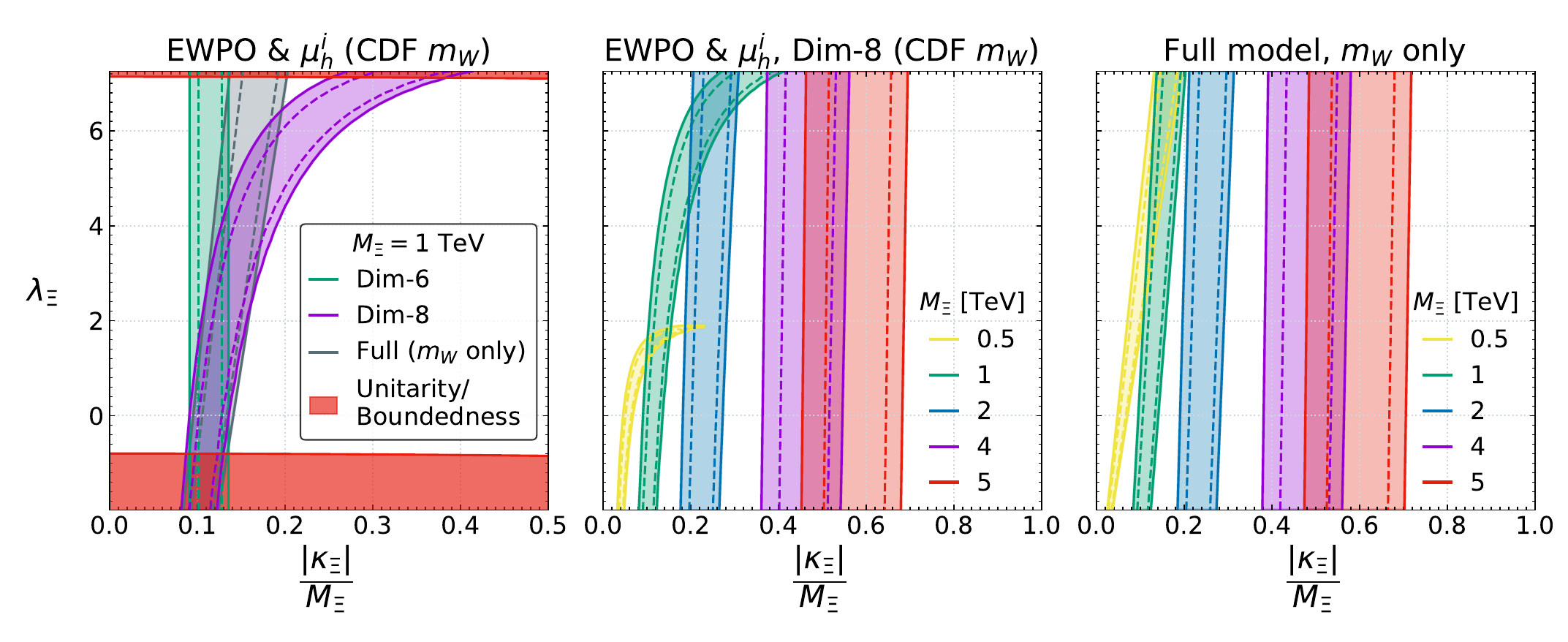}
\caption{\it Left panel: Values of $|\kx|/\mx$ that are allowed
as functions of $\lx$ for $\mx = 1$~TeV and $\eta_\Xi = 0$
at the 68\% CL (dashed lines) and 95\% CL (solid lines) 
by the present electroweak precision observables including the CDF measurement of $m_W$.
We compare results including dimension-6 operators in the linear approximation (green shading and lines)
, including both the quadratic effects of dimension-6 operators and the linear effects of dimension-8
operators (purple shading and lines) and the full model constraints using only the $\mw$ measurements (grey shading and lines). Middle panel: Constraints on  $|\kx|/\mx$
as functions of $\lx$ from the present electroweak precision observables 
including the CDF measurement of $m_W$, for the indicated values of $\mx$ and $\eta_\Xi = 0$
at the 68\% CL (dashed lines) and 95\% CL (solid lines) 
as analyzed including both the quadratic effects of dimension-6 operators and the 
linear effects of dimension-8 operators. Right panel: Same as middle panel, but for the full model and using only the $\mw$ measurements as input. }
\label{fig:Triplet_kxi_lxi_CDF}
\end{figure}

Finally, we comment on the recent measurement of $\mw$ by the CDF Collaboration~\cite{CDF:2022hxs}, which is in significant
tension with previous measurements and the SM prediction. The scalar triplet model is one of 
the extensions of the SM that is favoured as a possible scenario for accommodating the CDF
measurement, or at least mitigating this tension. Accordingly, we have explored the constraints
on this model that are imposed at the linear dimension-6 level and including quadratic
dimension-6 effects as well as dimension-8 operators in Fig.~\ref{fig:Triplet_kxi_lxi_CDF}. The left panel shows the constraints imposed on the parameter space for $\mx=1$ TeV in the two SMEFT approximations in green and purple, respectively. For reference, the constraints on the full model when considering only $\mw$ measurements are shown in grey. The middle panel shows how the preferred region at the dimension-8 level evolves for $\mx$ ranging between 0.5 and 5~TeV. The right panel shows the analogous preferred region for the full model, using only the $\mw$ measurements as input.

The left panel shows that at the linear dimension-6 level, a non-zero value of 
$|\kappa_\Xi|/\mx \sim 0.1$ is preferred, whereas at the dimension-8 level the allowed range of
$|\kx|/\mx$ increases as $\lx$ increases from negative values,
approaching 0.32 for $\lx \sim 4\pi/\sqrt{3}$. The dimension-8 bounds provide a much better approximation to the full model bounds for $\lx$ up to around 4, after which they start to diverge and become incompatible with the the true bound above $\lx=6$. Again, this emphasises the dominance of the $\mw$ measurements in the dataset, since the full model bounds only take those into account.
All of the allowed regions coincide around $\lx\sim 0$, showing that dimension-8 effects are negligible there and only matter for nonzero $\lx$. We see from the middle panel that the preferred value of $|\kx|/\mx$ scales linearly with $\mx$.
The band of allowed  $|\kx|/\mx$ and $\lx$ values for $\mx = 0.5$~TeV
has a similar shape which is cut off above $\lx\sim 2$, resulting in a spike around $|\kx|/\mx\sim 0.2$. For values of $\mx \ge 2$~TeV
the allowed band is tilted without a spike in the perturbative range for $|\lx|$.  The right panel confirms that the extreme dependence on $\lx$ introduced at dimension-8 for $\mx<2$ TeV is exaggerated with respect to the full model prediction, which predicts a linear relationship between $|\kx|/\mx$ and $\lx$ for a fixed modification to $\mw$. Nevertheless, the dimension-8 conclusions for lower $\lx$ are likely to be a good approximation of the constraints on the full model. 

Interpreting the new CDF $\mw$ measurement at face value could then suggest the presence of an EW triplet state with non-zero values of $\kx$ ranging from around 20 GeV for $\mx=$ 500 GeV to 2900 GeV for $\mx=5$ TeV. Our analysis has shown that an EFT analysis of the EWPO data presents a more faithful interpretation in terms of the triplet model when including dimension-8 effects. It has also quantified the regions of extreme $\lx$ in which the convergence of SMEFT expansion appears to break down and a HEFT description is likely to be more appropriate. 

\section{Summary\label{sec:summary}}
We have explored in this paper the present and prospective experimental constraints on singlet and zero-hypercharge triplet
scalar field extensions of the Standard Model using the SMEFT formalism with dimension-6 and -8  operators. As a first step, we computed the effect on gauge and Higgs interactions of the coefficients of the relevant SMEFT operators,
including both the linear and quadratic contributions of dimension-6 operators and the linear contributions of
dimension-8 operators.
Using the CDE method at tree level, we derived matching expressions for the coefficients 
of the relevant operators in the singlet and triplet models, using the Warsaw basis~\cite{Grzadkowski:2010es} at dimension-6 and the basis of Ref.~\cite{Murphy:2020rsh} at dimension~8. These generically involve corrections for dimension-6 operator coefficients as well as contributions to dimension-8 Wilson coefficients.  We found that, in order to perform a consistent matching to dimension-8 order using the CDE, it was necessary to solve the classical equation of motion for the heavy field beyond linear order. Although linear order is typically sufficient for dimension-6 and most dimension-8 matching results, in the presence of trilinear scalar self-interactions, higher orders are needed, which result in an additional contribution to the octic Higgs operator $\Op{H^8}=(\hdh)^4$.
These results enabled us to include consistently all the ${\cal O}(1/\Lambda^2)$ 
and ${\cal O}(1/\Lambda^4)$ contributions to rates for physical processes,
where $\Lambda$ is a heavy new physics scale that can be identified with the scalar mass. 
We also derived expressions for the model contributions to the deviations of the Higgs trilinear and
quartic couplings from their values in the Standard Model. We cross-checked our matching results by comparing them
with exact expressions for Higgs-Higgs scattering in both models, as well as expressions for the Higgs-singlet mixing angle and the $W$-boson mass in the singlet and triplet models, respectively.

In general, we observed that tree-level dimension-6 matching captures a subset of the potential parameter dependence of physical observables whose measurements can constrain such models. Tree-level matching relies on the presence of interaction terms that are linear in the heavy field. These parameters appear at the lowest-order, dimension-6 matching, and contribute in a leading way to the Higgs signal strengths and EWPOs for the singlet and triplet models, respectively. In both models, only modifications to the Higgs self-coupling via $\Op{H}=(\hdh)^3$ yield any sensitivity to the other parameters in the scalar potential at the dimension-6 level. At dimension-8, many more operators are generated and the parametric dependence is enriched. In particular, matching at this order introduces a dependence on the additional parameters of the scalar potential beyond the Higgs self-coupling, allowing for a more interesting interplay between Higgs signal strengths, EWPO and di-Higgs data.

In order to quantify this, we performed a global statistical analysis over the model parameter space, using the SMEFT framework as a bridge to the data. This allowed us to confront the two scalar extensions of the SM with a dataset comprising EWPO, Higgs signal strengths and di-Higgs cross section measurements. Performing this analysis to ${\cal O}(1/\Lambda^2)$ and ${\cal O}(1/\Lambda^4)$ allowed us to quantify the importance of dimension-8 effects and assess the validity and convergence of the SMEFT approximations. In addition, we compared the experimentally allowed regions with those allowed by theoretical constraints on the underlying model coming from boundedness, perturbative unitarity, and the requirement that the EW vacuum be the global minimum of the scalar potential. 

We showed how the expectations derived from our matching calculations played out in the data, demonstrating that the dimension-8 analysis leads to a non-trivial interplay driven by the balance between two effects. On one side, the dimension-6 effects are the leading ones, with dimension-8 contributions expected to contribute corrections suppressed by $\sim v^2/\Lambda^2$. On the other, the additional parametric dependence is limited to Higgs self-coupling modifications at dimension-6 order, which is only probed by the relatively less precise di-Higgs cross section measurements. Moving to dimension-8 order brings a dependence on these additional parameters into the Higgs signal strengths and EWPO, which are comparatively much better determined. There is hence a balance between the relative smallness of dimension-8 effects and the relatively poor experimental precision of di-Higgs rates with respect to other data. 

That said, we did find that di-Higgs data provide important additional information in the case of the singlet scalar model. At dimension-8 order the Higgs signal strengths alone yielded flat directions in the parameter space that were efficiently closed by the addition of di-Higgs rates. In particular, even when considering the linear coupling parameter $\ks$ alone, di-Higgs data served to exclude a region of large $\ks$ that would have otherwise been allowed by Higgs data alone, due to the appearance of a second minimum in the log-likelihood. Overall, we found that the allowed regions when including di-Higgs data were much reduced and overlapped significantly less with the theoretically forbidden regions. 
In contrast, we found that the EWPO data completely dominated the bounds on the triplet model parameter space, with Higgs data helping to exclude a very small additional portion and di-Higgs rates having no impact. Our results would certainly improve with the use of differential di-Higgs information, at the cost of a more delicate balance with the EFT validity criterion.

Because of the limited sensitivity of the di-Higgs data, we were unable to place any meaningful bounds on the quartic self-couplings of the scalar models, which only modify Higgs self-interactions at tree-level and dimension-8. 
Nevertheless, since our dimension-8 results lead to a more constraining likelihood over the parameter space, we were able to obtain profiled bounds on the linear coupling parameters, $\ks$ and $\kx$, allowing the other parameters of the scalar potential to vary within the theoretically allowed ranges. As expected, the significantly different interplay with data at dimension-8 order led to {\it prima facie} very different conclusions on the allowed values, particularly for the singlet model. This raised the question of the convergence of the SMEFT expansion, given that going from dimension-6 to -8 led to $O(1)$ or greater changes in the allowed parameter space. 
However, we found that much of the parameter space that was still allowed by the data and theoretical bounds corresponded to regions where the singlet acquired a large vev, $\vs \gg v$, in contrast with the SMEFT expectation that this be a suppressed quantity. Repeating the profiling analysis with the additional requirement that $|\vs|<\ms$ led to less drastic, but still significant, differences between the dimension-6 and -8 analyses. In the case of the triplet field, for $\mx<900$ GeV, we found that the profiled bounds on $\kx$ were exaggerated due to an unphysical cancellation in the dimension-8 predictions indicating an unreliable SMEFT expansion at large $\lx$. 

In order to explore further the validity of the SMEFT approximations, we also compared our SMEFT predictions at different orders with those of the full model. For the singlet, we used the Higgs-singlet mixing parameter $\cos^2\alpha$, which controls Higgs signal strengths, while for the triplet we used the $\mw$ prediction. We chose these because they are the observables that best constrain each model. In each case we uncovered the regions of parameter space in which the dimension-8 predictions represented improvements over dimension-6 ones as well as regions where they diverged significantly from the full model prediction. In the latter case, such regions of parameter space are not likely to admit a SMEFT approximation. The breakdown was found to occur for large values of the quartic portal couplings, $\lam{S}$ or $\lx$, and the singlet trilinear coupling, $\ksss$. For lower cutoff scales, the breakdown occurred earlier and above cutoff scales of around 2 TeV the dimension-8 predictions were found to give a much better approximation of the full model within the theoretically allowed parameter space.

Given the tight constraints from the EWPOs, there is no scope for observable modifications to Higgs self-couplings in the triplet model. Indeed, even the impact of the single Higgs data was found to be practically negligible in constraining this scenario. However for the singlet model, we found an interplay between single and di-Higgs data, and the trilinear self-coupling measurement already provides useful information in probing this extension of the SM. We then explored the scope for a modified Higgs quartic interaction in this model, which could potentially be probed by triple-Higgs production as well as indirectly through precision di-Higgs measurements at future colliders. By considering a simple `post-LHC' scenario using the projected HL-LHC sensitivities to di-Higgs cross section measurements, we found that significant deviations to the higgs quartic coupling may be possible within the allowed parameter space, although the most significant deviations were (unsurprisingly) observed in the regions where the SMEFT approximation is likely to break down. Within a restricted region where SMEFT predictions for single Higgs rates can be trusted, deviations with respect to the SM prediction by a factor of a few seem possible. This suggests that a more comprehensive analysis over the full parameter space may be interesting, complemented by explicit calculations of di-Higgs and triple-Higgs production rates in the SMEFT at dimension-8.

Finally we considered the impact of the recent CDF $W$-mass measurement on the parameter space of the triplet model, which is known to be one of the possible BSM candidates that could account for such an anomaly. By comparing to the full model predictions, we showed how the dimension-8 analysis allowed us to determine more accurately the preferred regions of parameter space. In particular, the fact that the preferred value of $\kx$ depends on $\lx$ is not captured by the dimension-6 analysis. This occurred up to a point where the SMEFT expansion became unreliable, in line with our previous conclusions about its validity. 

The extension of phenomenological applications of the SMEFT approach to include dimension-8 operators is less developed than
the corresponding dimension-6 analyses. However, explorations of the type discussed here may be interesting in other
SMEFT sectors, e.g., in top physics where the constraints on dimension-6 operator coefficients are weaker and hence
the possible contributions of dimension-8 operators may be more accessible. It will also be interesting to explore
other examples of the possible interplay between present and future SMEFT explorations of possible new physics
beyond the Standard Model.

\section*{Acknowledgements}

The work of JE was supported by the United Kingdom STFC via grants ST/P000258/1 and ST/T000759/1, and
also by the Estonian Research Council via a Mobilitas Pluss grant. The work of KM was supported by STFC grant
ST/T000759/1. KM would like to thank Ilaria Brivio for useful discussions concerning input schemes in the SMEFT and Tyler Corbett for providing material that helped us to validate our dimension-8 matching calculations.

\appendix
\section{Matching validation: $\mathbf{hh\to hh}$ scattering\label{app:hhhh}}
As a partial validation of the matching results reported in Table~\ref{tab:singlet_results}, we calculate the
amplitude for $hh\to hh$ scattering in both the full theory and in the SMEFT framework. 
\subsection{SMEFT \label{sec:hh2hh_SMEFT}}
In the SMEFT framework, the relevant Feynman rules for the cubic and quartic Higgs interactions in terms of the input parameters, the relevant dimension-6 and -8 operators and with all momenta incoming, are as follows:

\noindent
\begin{tabular}{p{3cm}p{0.1cm}p{11cm}}
\parbox{3cm}{
\includegraphics[width=3cm]{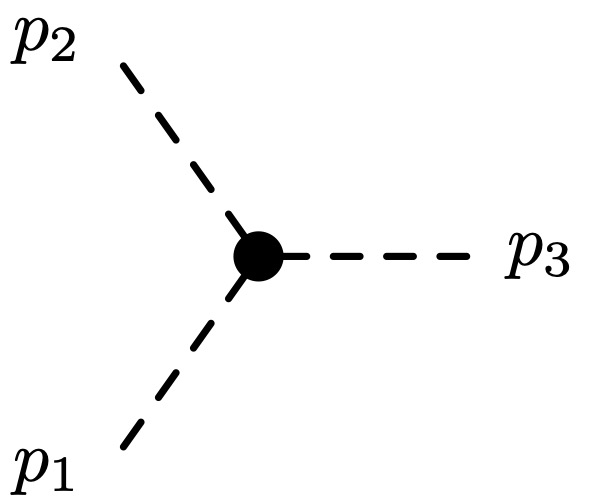}
}
&:&
\parbox{11cm}{{\small
\begin{equation}
\begin{split}
    -&\frac{3i\mh^2}{\hat{v}}\bigg[1 + \frac{\hat{v}^2}{\Lambda^2}\bigg(- \frac{2\hat{v}^2C_H}{\mh^2} +  C_{H\Box} - \frac{C_{HD}}{4}\bigg)\\&+
    \frac{\hat{v}^4}{\Lambda^4}\bigg(-\frac{4\hat{v}^2C_{H^8}}{\mh^2} - \frac{C_{H^6}^{(2)}}{4} + \frac{3C_{H\Box}^2}{2}  \\&+ \frac{3C_{HD}^2}{32} - \frac{6\hat{v}^2C_H C_{H\Box}}{\mh^2} + \frac{3\hat{v}^2 C_H C_{HD}}{\mh^2} - \frac{3C_{H\Box}C_{HD}}{4}\bigg)\bigg] \\+& i(p_1\cdot p_2 + p_1\cdot p_3 + p_2\cdot p_3)\bigg[\frac{\hat{v}}{\Lambda^2}\bigg(4C_{H\Box} - C_{HD}\bigg)  \\&+ \frac{\hat{v}^3}{\Lambda^4}\bigg(-C_{H^6}^{(1)} - C_{H^6}^{(2)} + 12C_{H\Box}^2+ \frac{3C_{HD}^2}{4} - 6C_{H\Box}C_{HD}\bigg)\bigg]\,,
    \label{eq:hhh}
\end{split}
\end{equation}
}}\\
\end{tabular}
\begin{tabular}{p{3cm}p{0.1cm}p{11cm}}
\parbox{3cm}{
\includegraphics[width=3cm]{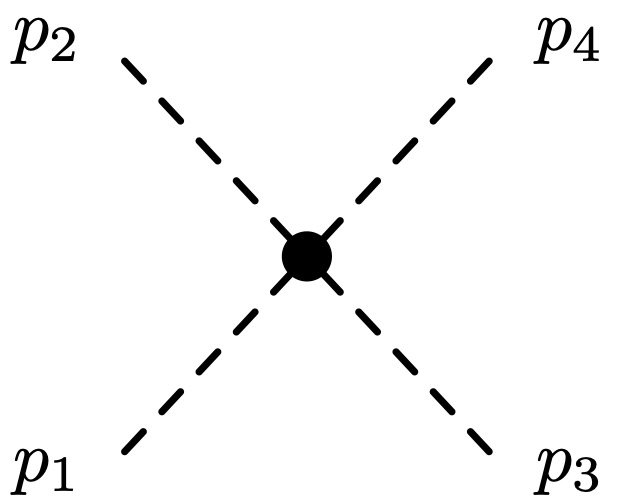}
}
&:&
\parbox{11cm}{{\small
\begin{equation}
\begin{split}
    -&\frac{3i\mh^2}{\hat{v}^2}\bigg[1 + \frac{\hat{v}^2}{\Lambda^2}\bigg(-\frac{12\hat{v}^2C_H}{\mh^2}+2C_{H\Box} - \frac{C_{HD}}{2}  \bigg) \\&+ \frac{\hat{v}^4}{\Lambda^4}\bigg(-\frac{32\hat{v}^2C_{H^8}}{\mh^2}  - \frac{C_{H^6}^{(2)}}{2} + 4C_{H\Box}^2 + \frac{C_{HD}^2}{4} \\&- \frac{48\hat{v}^2C_H C_{H\Box}}{\mh^2} + \frac{12\hat{v}^2C_H C_{HD}}{\mh^2} - 2C_{H\Box}C_{HD} \bigg)\bigg] \\+& i(p_1\cdot p_2 + p_1\cdot p_3 + p_1\cdot p_4 + p_2\cdot p_3 + p_2\cdot p_4 + p_3\cdot p_4) \\&\times\bigg[\frac{1}{\Lambda^2}\bigg(4C_{H\Box} - C_{HD}\bigg) \\&+ \frac{\hat{v}^2}{\Lambda^4}\bigg(-3C_{H^6}^{(1)} - 3C_{H^6}^{(2)} + 16C_{H\Box}^2 + C_{HD}^2 - 8C_{H\Box}C_{HD}\bigg)\bigg]  \\+& \frac{i}{\Lambda^4}( \,p_1\cdot p_2\,\,p_3\cdot p_4 + p_1\cdot p_3\,\,p_2\cdot p_4 + p_1\cdot p_4\,\,p_2\cdot p_3\, ) \\&\times\bigg(2C_{H^4}^{(1)} + 2C_{H^4}^{(3)}\bigg)\,.
    \label{eq:hhhh}
\end{split}
\end{equation}
}}
\end{tabular}
We use (\ref{eq:hhh}) and (\ref{eq:hhhh}) to find the following general expression for the 
$hh\to hh$ scattering amplitude in terms of the SMEFT coefficients to ${\cal O}(1/\Lambda^4)$, i.e., including contributions that are quadratic
in the dimension-6 operator coefficients and linear in the dimension-8 operator coefficients:
\begin{equation}
\label{eq:SMEFT_hhhh}
\begin{split}
    i\mathcal{M} = &D(s,t,u)
    \Bigg[
    \frac{9\mh^4}{\hat{v}^2} 
    + \frac{1}{\Lambda^2}\bigg(
       54\mh^4C_{H\Box} 
    - \frac{27\mh^4 C_{HD}}{2} 
    -36\mh^2\hat{v}^2C_H \bigg) \\&
    + \frac{1}{\Lambda^4}\bigg(
    - 72\mh^2\hat{v}^4C_{H^8} 
    - 9\mh^4\hat{v}^2C^{(1)}_{H^6} 
    -\frac{27\mh^4\hat{v}^2 C_{H^6}^{(2)}}{2} 
    + 36\hat{v}^6C_H^2 
    + 216\mh^4\hat{v}^2C_{H\Box}^2 \\&
    + \frac{27\mh^4\hat{v}^2C_{HD}^2}{2} 
    - 216\mh^2v_{\sss T}^4C_H C_{H\Box} 
    - 54\mh^2\hat{v}^4C_H C_{HD} 
    + 108\mh^4\hat{v^2}C_{H\Box}C_{HD} \bigg)\Bigg] \\& 
    + \frac{3i\mh^2}{\hat{v}^2} 
    + \frac{1}{\Lambda^2}\bigg(
    - 36\hat{v}^2C_H 
    + 50\mh^2C_{H\Box} 
    - \frac{25\mh^2C_{HD}}{2}
    \bigg) \\&
    + \frac{1}{\Lambda^4}\bigg(
    - 96\hat{v}^4C_{H^8} 
    - 15\mh^2\hat{v}^2C_{H^6}^{(1)} 
    - \frac{33\mh^2\hat{v}^2 C_{H^6}^{(2)}}{2} 
    + 264\mh^2 \hat{v}^2C_{H\Box}^2 \\&
    + \frac{33\mh^2\hat{v}^2C_{HD}^2}{2}
    - 216\hat{v}^4C_H C_{H\Box} 
    + 54\hat{v}^4C_H C_{HD} 
    - 132\mh^2\hat{v}^2C_{H\Box}C_{HD} \\&
    - \frac{1}{2}\left( s^2+t^2+u^2
    - 4\mh^4 \right)
    \left(C_{H^4}^{(1)} + C_{H^4}^{(3)}\right)
    \bigg)\,,
\end{split}
\end{equation}
Where $s,t,u$ are the usual Mandelstam variables and $D(s,t,u)$ is the $(s,t,u)$-symmetric Higgs boson propagator combination:
\begin{align}
   D(s,t,u)= \bigg[
    \frac{1}{s-\mh^2}+\frac{1}{t-\mh^2}+\frac{1}{u-\mh^2}
    \bigg].
\end{align}
We use this expression to validate the non-linear field redefinitions used to obtain Eqs.~\eqref{eq:c3} and~\eqref{eq:d4}, for the modified trilinear and quartic Higgs self couplings in a basis without any three or four point, two-derivative Higgs self-couplings.

\subsection{Singlet scalar \label{sec:hh2hh_singlet}}
Combining the general expression~\eqref{eq:SMEFT_hhhh} with the matching results of Table~\ref{tab:singlet_results} to
obtain the $hh\to hh$ amplitude in terms of the singlet model parameters, and the SM input parameters $\mh$ and $\vh$, we find:
\begin{equation}
\begin{split}
    i\mathcal{M} = &D(s,t,u)
    \bigg[
     \frac{9\mh^4}{\hat{v}^2} 
    - \frac{27\mh^4\ks^2}{\ms^4} 
    + \frac{36\mh^2\hat{v}^2\ks^2\lam{S}}{\ms^4}
    - \frac{54\mh^6\ks^2}{\ms^6} 
    + \frac{180\mh^4\hat{v}^2\ks^2\lam{S}}{\ms^6}\\&  
    - \frac{144\mh^2\hat{v}^4\ks^2\lam{S}^2}{\ms^6} 
    - \frac{36\mh^2\hat{v}^2\ks^3\ksss}{\ms^6} 
    + \frac{54\mh^4\hat{v}^2\ks^4}{\ms^8} 
    - \frac{108\mh^2\hat{v}^4\ks^4\lam{S}}{\ms^8} \\&
    + \frac{36\hat{v}^6\ks^4\lam{S}^2}{\ms^8} 
    - \frac{270\mh^4\hat{v}^2\ks^3\ksss}{\ms^8} 
    + \frac{432\mh^2\hat{v}^4\ks^3\lam{S}\ksss}{\ms^8}
    + \frac{72\mh^2\hat{v}^4\ks^4(\kssss)}{\ms^8} \\&
    - \frac{72\hat{v}^6\ks^5\lam{S}\ksss}{\ms^{10}}  
    - \frac{324\mh^2\hat{v}^4\ks^4\ksss^2}{\ms^{10}} 
    + \frac{108\mh^2\hat{v}^4\ks^5\ksss}{\ms^{10}} 
    + \frac{36\hat{v}^6\ks^6\ksss^2}{\ms^{12}}\bigg] \\&
    + \frac{3\mh^2}{\hat{v}^2} 
    - \frac{25\mh^2\ks^2}{\ms^4} 
    + \frac{36\hat{v}^2\ks^2\lam{S}}{\ms^4} 
    - \frac{78\mh^4\ks^2}{\ms^6} 
    + \frac{252\mh^2\hat{v}^2\ks^2\lam{S}}{\ms^6} 
    - \frac{192\hat{v}^4\ks^2\lam{S}^2}{\ms^6} \\&
    - \frac{\ks^2(s^2+t^2+u^2)}{\ms^6} 
    - \frac{36\hat{v}^2\ks^3\ksss}{\ms^6} 
    + \frac{66\mh^2\hat{v}^2(\ks^4)}{\ms^8} 
    - \frac{378\mh^2\hat{v}^2\ks^3\ksss}{\ms^8} \\&
    - \frac{108\hat{v}^4\ks^4\lam{S}}{\ms^8} 
    + \frac{576\hat{v}^4\ks^3\lam{S}\ksss}{\ms^8} 
    + \frac{96\hat{v}^4\ks^4(\kssss)}{\ms^8} 
    - \frac{432\hat{v}^4\ks^4\ksss^2}{\ms^{10}} 
    + \frac{108\hat{v}^4\ks^5\ksss}{\ms^{10}}
    \label{eq:Singlet_hhhh}
\end{split}
\end{equation}
in the SMEFT. 

To calculate the amplitude in the full theory, we start from the singlet scalar Lagrangian of Eq.~\eqref{eq:L_S}, repeated here for clarity:
\begin{align}
    \begin{split}
    \label{eq:Lag}
    \mathcal{L}_{s} &= 
    \frac{1}{2} \partial_\mu S\partial_\mu S 
  - \frac{1}{2} \ms^2 S^2 
  - \ks H^\dagger H S
  - \lam{S} H^\dagger H S^2  
  - \ksss S^3- \kssss S^4 \, . \\
    \end{split}
\end{align}
In order to verify our SMEFT matching calculation, we calculate $hh\to hh$ scattering in this theory taking the EFT limit, \emph{i.e.}, assuming $\ms,\ks,\ksss\gg v,s,|t|,|u|$. 
Since S has no quantum numbers, a tadpole term (linear in the field), $\mathcal{L}_{\textrm{tad.}}= \bs S$, is also permitted. 
Under a shift of the field by a constant $S\to S+c$, the parameters of the scalar potential are redefined according to:
\begin{align}
    \begin{split}
    \mu^2 &\to \mu^2 - c\, \ks - c^2 \lam{S} \, ,\\
    \bs &\to \bs - c\, \ms^2 - 3c^2 \ksss - 4c^3 \kssss \, ,\\
    \ms^2 &\to \ms^2 + 6 c\, \ksss + 12 c^2 \kssss \, ,\\
    \ks^2 &\to \ks + 2 c\, \lam{S} \, ,\\
    \ksss^2 &\to \ksss + 4 c\, \kssss \, ,
    \end{split}
\end{align}
where all unspecified parameters are unchanged and $\mu^2$ is the mass parameter of the SM Higgs doublet. Since physical observables must remain unchanged under such a field redefinition, there is an infinite freedom to choose the representation of the scalar singlet Lagrangian that yields the same physics. This freedom can be used, {e.g.}, to fix the vev of S to zero, or to eliminate the tadpole term. The latter choice has implicitly been made in Eq.~\eqref{eq:Lag}, meaning that S generically obtains a vev,  $\langle S \rangle = \vs \ne 0$, in this representation. If the former choice is made, the minimisation conditions to obtain $\vs = 0$ and $\langle H\rangle=v/\sqrt{2}$, fix the value of the tadpole coupling to be $\bs=-v^2\ks/2$. 

In the general case where S obtains a vev, the minimisation conditions read
\begin{align}
\begin{split}
    \label{eq:mincondS}
    \mu^2&=\lambda v^2+\vs(\ks + \vs\lam{S}) \, ,\\
    2\vs \ms^2&= - v^2(\ks + 2\lam{S}\vs) -6\vs^2\ksss -8\vs^3\kssss \, ,
\end{split}
\end{align}
which can be solved in the EFT limit for
\begin{align}
\label{eqn:vs_EFT}
    \vs=-\frac{\ks v^2}{2\ms^2}\left(
    1 - \frac{v^2\lam{S}}{\ms^2} + \frac{3}{2}\frac{ v^2\ks \ksss}{\ms^4} + \cdots
    \right) \, .
\end{align}
After the fields obtain their vevs, 
there is mass mixing between the excitations about the ground state, $h_0$ and $s_0$, that is diagonalised by a rotation through a mixing angle $\alpha$, thereby defining our two physical scalars:
\begin{align}
\begin{split}
    h&\equiv h_0\cos\alpha + s_0\sin\alpha \, , \quad 
    s\equiv s_0\cos\alpha- h_0\sin\alpha: \quad
    \tan 2\alpha = \frac{2 x y}{y^2-x^2} \, ,\\
    {\rm where} \quad x&=v(\ks + 2 \vs\lam{S}) \, ,\quad y=\ms^2 -  \mh^2 + 6 \vs(\ksss + 2 \vs \kssss) + v^2\lam{S} \, ,    
\end{split}
\end{align}
with masses $\mh$ and 
\begin{align}
    m_{\sss S,\mathrm{phys}}^2 = \mh^2 +\frac{x^2+y^2}{y} \, .
\end{align}
In the EFT limit, $\tan2\alpha\sim\frac{2 \ks v}{\ms^2}$, so the state $h$ is mostly comprised of the SM Higgs field, $h_0$.
The Higgs quartic coupling parameter can be written in terms of the other model parameters as
\begin{align}\label{eqn:lambda_singlet}
    \lambda = \frac{1}{2v^2}\left(\mh^2+\frac{x^2}{y}\right) \, .
\end{align}
For the purposes of validating our matching calculation, we stick to a parametrisation in terms of the original Lagrangian parameters rather than a set of physics inputs.

 In addition to a new scattering diagram mediated by the heavy state, $s$, the mixing between the fields leads to a modification of the trilinear and quartic Higgs self-interactions. We then expand our expression for the amplitude in inverse powers of $\ms$, assuming $\ks$ and $\ksss$ to be of order $\ms$, and obtain the same result as the SMEFT calculation in Eq.~\eqref{eq:Singlet_hhhh}. 
As an additional validation, we performed the calculation in both representations of the singlet Lagrangian mentioned above, where $\vs \neq 0$ and $\vs=0$, respectively. Computing the required $S$-field shift to translate from one to the other, we confirmed that the two give the same result up to dimension 8.

\subsection{Triplet Scalar \label{sec:hh2hh_triplet}}
Repeating the analysis of the previous Section for the triplet scalar case,
we combine the general expression (\ref{eq:SMEFT_hhhh}) with the matching results of 
Table~\ref{tab:triplet_results} to obtain the corresponding
$hh \to hh$ scattering amplitude in terms of the triplet model parameters and the SM input parameters $\mh$ and $\vh$:
\begin{equation}
\begin{split}\label{eq:Atriplet}
    i\mathcal{M} = &D(s,t,u)
    \bigg[
    \frac{9\mh^4}{\hat{v}^2} 
    - \frac{18\mh^4\kx^2}{\mx^4} 
    + \frac{36\mh^2\hat{v}^2\kx^2\lx}{\mx^4} 
    - \frac{54\mh^6\kx^2}{\mx^6} 
    + \frac{162\mh^4\hat{v}^2\kx^2\lx}{\mx^6} \\&
    - \frac{144\mh^2\hat{v}^4\kx^2\lx^2}{\mx^6} 
    + \frac{45\mh^4\hat{v}^2\kx^4}{\mx^8} 
    - \frac{144\mh^2\hat{v}^4\kx^4\lx}{\mx^8} 
    + \frac{36\hat{v}^6\kx^4\lx^2}{\mx^8} 
    + \frac{18\mh^2\hat{v}^4\kx^4\ex}{\mx^8}\bigg]\\& 
    + \frac{3\mh^2}{\hat{v}^2} 
    - \frac{22\mh^2\kx^2}{\mx^4} 
    + \frac{36\hat{v}^2\kx^2\lx}{\mx^4} 
    - \frac{78\mh^4\kx^2}{\mx^6} 
    + \frac{246\mh^2\hat{v}^2\kx^2\lx}{\mx^6} 
    - \frac{192\hat{v}^4\kx^2\lx^2}{\mx^6} \\&
    - \frac{(s^2+t^2+u^2)\kx^2}{\mx^6} 
    + \frac{63\mh^2\hat{v}^2\kx^4}{\mx^8} 
    - \frac{144\hat{v}^4\kx^4\lx}{\mx^8} 
    + \frac{24\hat{v}^4\kx^4\ex}{\mx^8} \, .
\end{split}
\end{equation}

To calculate the amplitude in the full theory, we extend the SM Lagrangian with the scalar triplet interactions of Eq.~\eqref{eq:L_Xi}:
\begin{equation}
\begin{split}
    \mathcal{L}_{\sss{\Xi}} =& \frac{1}{2}(D_{\mu} \Xi^a)(D^{\mu} \Xi^a) - \frac{1}{2}\mx^2(\Xi^a \Xi^a) -  \kappa_{\sss{\Xi}}H^{\dagger} \Xi^a \sigma^a H +\\ &- \lambda_{\sss{\Xi}}(\Xi^a \Xi^a)(H^{\dagger}H) -  \frac{1}{4}\eta_{\sss{\Xi}}(\Xi^a \Xi^a)^2\,,
\end{split}
\end{equation}
where the fields are defined as
\begin{equation}
    H = \frac{1}{\sqrt{2}} \begin{pmatrix}
    0 \\ h_0 + v
    \end{pmatrix} \,, \hspace{1cm} \Xi = \begin{pmatrix}
    0 \\ \xi_0 + v_t \\ 0
    \end{pmatrix}\,,
\end{equation}
since we are only interested in the neutral sector for the purpose of this exercise. The triplet vev contributes to the $W$-boson mass term:
\begin{align}
    \mw^2 = \frac{g^2}{4}(v^2+4v_t^2).
\end{align}
This, in turn, leads to a relation between the input parameter, $G_F$, which is determined from the muon decay lifetime, and the scalar vevs. 
\begin{align}
    v^2+4v_t^2=\frac{1}{\sqrt{2}G_F}=\vh^2\approx(246\,\mathrm{GeV})^2.
\end{align}
Given this constrain from the input data, the requirement that $v^2>0$ restricts the absolute value of $v_t$ to be less than $\vh/2$. The precisely measured EWPO constrain this quantity to be much smaller, on the order of a few GeV. 
The potential has the following minimisation conditions:
\begin{equation}
    \begin{split}
    \mu^2 &= v^2 \lambda - v_t \kx + v_t^2 \lambda_{\sss \Xi}\,, \\
    m_{\sss \Xi}^2 &= \frac{v^2 \kx}{2v_t} - v^2 \lambda_{\sss \Xi} - v_t^2 
    \eta_{\sss \Xi}\,,  
    \end{split}
\end{equation}
which can be solved for the triplet vacuum expectation value in the EFT limit:
\begin{align}
    v_t =\frac{\kx v^2}{2\mx^2}\left(1-\frac{\lx v^2}{\mx^2}+ \cdots\right) 
    =\frac{\kx \vh^2}{2\mx^2}\left(1-\frac{v^2}{\mx^2}\left(
    \lx +\frac{\kx^2}{\mx^2}
    \right)+ \cdots\right)\, .
\end{align}
After EWSB, there is mass mixing between the excitations about the ground state, $h_0$ and $\xi_0$, that can be diagonalised by a rotation through a mixing angle $\beta$:
\begin{equation}
    \tan 2\beta = \frac{2 v \kx - 4 v v_t \lambda_{\sss \Xi}}{3 v_t^2 \eta_{\sss \Xi} - 2 v^2 \lambda + v^2 \lambda_{\sss \Xi} + \mx^2}\,,
\end{equation}
thereby defining the two physical scalars:
\begin{equation}
    h \equiv h_0 \cos \beta + \xi_0 \sin \beta\,, \hspace{1cm} \xi \equiv \xi_0 \cos \beta - h_0 \sin \beta\,,
\end{equation}
with masses $\mh$ and
\begin{align}
    m^2_\xi = \mx^2+\lx v^2+3\ex v_t^2 + \frac{v^2(\kx - 2\lx v_t^2)^2}{ \mx^2+\lx v^2+3\ex v_t^2-\mh^2} \, .
\end{align}
The Higgs boson quartic coupling, $\lambda$, gets corrected with respect to the SM as follows:
\begin{align}
    \lambda = \frac{\mh^2}{2 v^2}
    +\frac{1}{2}\frac{(\kx - 2\lx v_t^2)^2}{ \mx^2+\lx v^2+3\ex v_t^2-\mh^2}
    \,.
\end{align}
We refer the reader to Refs.~\cite{Forshaw:2001xq,Khan:2016sxm} for more details on the phenomenology of the Triplet scalar model. In the same way as we did for the singlet model, we computed the $hh \to hh$ scattering amplitude in the EFT limit out to dimension-8, finding agreement with the matched amplitude of Eq.~\eqref{eq:Atriplet}.

\section{\{$\alpha_{\sss EM}, m_{\sss Z}, G_F$\} Input Scheme \label{app:input scheme}}

The SM Lagrangian parameters $g^2$, $g^{\prime 2}$, $v_{\sss T}^2$ and $\lambda$ appearing in this paper have been expressed in terms of measurable quantities using the \{$\alpha_{\sss EM}, m_{\sss Z}, G_F$\} scheme, following the formalism defined in Ref.~\cite{Brivio:2020onw}. For completeness, we review here the general formalism and results up to dimension~8.

We define the vector of independent SM parameters as $g \equiv \{g^{\prime 2}, g^2, v_{\sss T}^2, \lambda \}$ and the input observables to be ${\cal{O}} \equiv \{\alpha_{\sss EM}, m_{\sss Z}^2, G_F, \mh^2\}$, where $\mh$ is needed only to fix the value of $\lambda$.
Working order by order in the EFT expansion, each ${\cal{O}}_n$ can be expressed as:
\begin{equation}
    \mathcal{O}_n = F_n^{(0)}(g) + \frac{1}{\Lambda^2}F_n^{(2)}(g,C)+ \frac{1}{\Lambda^4}F_n^{(4)}(g,C)+...\,,
    \label{eq:O_n}
\end{equation}
where $F_n^{(0)}(g)$ depends only  on the parameters $g$ and represents the SM expression, while the $F_n^{(2,4)}(g,C)$ are SMEFT corrections that depend in addition on the Wilson coefficients $C$. The general solution to the system (\ref{eq:O_n}) is of the form:
\begin{equation}
    g_i = K_i^{(0)}(\mathcal{O})+\frac{1}{\Lambda^2}K_i^{(2)}(\mathcal{O},C)+\frac{1}{\Lambda^4}K_i^{(4)}(\mathcal{O},C)+...\,,
    \label{eq:g_i}
\end{equation}
where the leading term is the SM solution, defined by imposing 
\begin{equation}
    \mathcal{O}_n = F_n^{(0)}(K^{(0)}(\mathcal{O}))\,,
\end{equation}
while the following terms are SMEFT corrections defined, up to dimension 8, as:
\begin{equation}
    K_i^{(2)} = - (J^{-1})_{in} F_n^{(2)}\,,
    \label{eq:K2}
\end{equation}
\begin{equation}
    K_i^{(4)} = - (J^{-1})_{in}\bigg[ F_n^{(4)} + \frac{\partial F_n^{(2)}}{\partial g_k} K_k^{(2)} + \frac{1}{2}\frac{\partial^2 F_n^{(0)}}{\partial g_k \partial g_j}K_k^{(2)}K_j^{(2)}\bigg]\,,
    \label{eq:K4}
\end{equation}
where all the terms are evaluated using SM expressions and $J$ is the Jacobian matrix 
\begin{equation}
    J_{ni} = \frac{\partial F_n^{(0)}}{\partial g_i}\,.
    \label{eq:J}
\end{equation}
Eq. (\ref{eq:g_i}) allows us to express the parameters $g$ as 
\begin{equation}
    g_i = \hat{g}_i \bigg[1 + \frac{\delta g_i}{\hat{g}_i} \bigg]\,,
    \label{eq:g_shift}
\end{equation}
where $\hat{g}_i = K_i^{(0)}(\mathcal{O})$ is the SM expression in terms of the input observables and $\delta g_i$ represents the corrections depending on the Wilson coefficients. 

At tree level, the input observables $\mathcal{O}_n$ are defined as:

\begin{align}
\label{eqn:D_aEM}
    \alpha_{\sss EM}& = \frac{1}{4\pi}\frac{g^2 g^{\prime 2}}{g^2 + g^{\prime 2}}[1+\Delta^{(6)} \alpha_{\sss EM}+\Delta^{(8)} \alpha_{\sss EM}]\,,\\
\label{eqn:D_mz}
    m_{\sss Z}^2 &= \frac{(g^2 + g^{\prime 2})v_{\sss T}^2}{4}[1+\Delta^{(6)} m_{\sss Z}^2+\Delta^{(8)} m_{\sss Z}^2]\,,\\
\label{eqn:D_GF}
    G_F& = \frac{1}{\sqrt{2}v_{\sss T}^2}[1+\Delta^{(6)}G_F + \Delta^{(8)}G_F]\,,\\
\label{eqn:D_mh}
    \mh^2 &= 2\lambda v_{\sss T}^2[1+\Delta^{(6)}\mh^2 + \Delta^{(8)}\mh^2]\,,
\end{align}
where $\Delta^{(6)}$ and $\Delta^{(8)}$ represent the dimension-6 and -8 SMEFT corrections, respectively. Thus, in the notation of Eq. (\ref{eq:O_n}) one has:
\begin{equation}
    F_{\alpha_{\sss EM}}^{(0)} = \frac{1}{4\pi}\frac{g^2 g^{\prime 2}}{g^2 + g^{\prime 2}}\,, \hspace{0.3cm} \frac{1}{\Lambda^2}F_{\alpha_{\sss EM}}^{(2)} = F_{\alpha_{\sss EM}}^{(0)}\Delta^{(6)} \alpha_{\sss EM}\,, \hspace{0.3cm} \frac{1}{\Lambda^4}F_{\alpha_{\sss EM}}^{(4)} = F_{\alpha_{\sss EM}}^{(0)}\Delta^{(8)} \alpha_{\sss EM}\,,
\end{equation}
and similarly for the other quantities.

To express the parameters $g$ as in Eq. (\ref{eq:g_shift}
), we first determine the SM solutions $\hat{g}_i$:
\begin{equation}
\label{eq:SMsol}
    \hat{g}^{\prime 2} = \frac{4\pi \alpha_{\sss EM}}{c^2_w}\,, \hspace{0.5cm} \hat{g}^2 = \frac{4\pi \alpha_{\sss EM}}{s^2_w}\,, \hspace{0.5cm} \hat{v}^2 = \frac{1}{\sqrt{2}G_F}\,, \hspace{0.5cm} \hat{\lambda} = \frac{\mh^2 G_F}{\sqrt{2}}\,,
\end{equation}
where the weak angle $\hat{\theta}_w$ is defined as
\begin{equation}
    s^2_w = \sin^2 \hat{\theta}_w \equiv \frac{\hat{g}^{\prime 2}}{\hat{g}^2+\hat{g}^{\prime 2}} = \frac{1}{2}\bigg[ 1 - \sqrt{1-\frac{2\sqrt{2}\pi \alpha_{\sss EM}}{G_F m^2_{\sss Z}}}\bigg]\,.
\end{equation}
We can then determine the shifts $\delta g_i$ by computing the Jacobian (\ref{eq:J}), 
\begin{equation}
    J = \begin{pmatrix}
    c^4_{\hat{w}}/4\pi & s^4_{\hat{w}}/4\pi & & \\
    \hat{v}^2/4 & \hat{v}^2/4 & \hat{g}^2/4c^2_{\hat{w}} &\\
    & & -1/\sqrt{2}\hat{v}^4 & \\
    & & 2\hat{\lambda} & 2\hat{v}^2
    \end{pmatrix}\,,
\end{equation}
and the remaining $K^{(2,4)}$ terms (\ref{eq:K2}, \ref{eq:K4}) in Eq.~(\ref{eq:g_i}).

\bibliography{main}

\providecommand{\href}[2]{#2}\begingroup\raggedright\begin{thebibliography}{10}

\bibitem{Buchmuller:1985jz}
W.~Buchmuller and D.~Wyler, \emph{{Effective Lagrangian Analysis of New
  Interactions and Flavor Conservation}},
  \href{https://doi.org/10.1016/0550-3213(86)90262-2}{\emph{Nucl. Phys. B}
  {\bfseries 268} (1986) 621--653}.

\bibitem{Grzadkowski:2010es}
B.~Grzadkowski, M.~Iskrzynski, M.~Misiak and J.~Rosiek, \emph{{Dimension-Six
  Terms in the Standard Model Lagrangian}},
  \href{https://doi.org/10.1007/JHEP10(2010)085}{\emph{JHEP} {\bfseries 10}
  (2010) 085}, [\href{https://arxiv.org/abs/1008.4884}{{\ttfamily 1008.4884}}].

\bibitem{Giudice:2007fh}
G.~F. Giudice, C.~Grojean, A.~Pomarol and R.~Rattazzi, \emph{{The
  Strongly-Interacting Light Higgs}},
  \href{https://doi.org/10.1088/1126-6708/2007/06/045}{\emph{JHEP} {\bfseries
  06} (2007) 045}, [\href{https://arxiv.org/abs/hep-ph/0703164}{{\ttfamily
  hep-ph/0703164}}].

\bibitem{Pomarol:2013zra}
A.~Pomarol and F.~Riva, \emph{{Towards the Ultimate SM Fit to Close in on Higgs
  Physics}}, \href{https://doi.org/10.1007/JHEP01(2014)151}{\emph{JHEP}
  {\bfseries 01} (2014) 151},
  [\href{https://arxiv.org/abs/1308.2803}{{\ttfamily 1308.2803}}].

\bibitem{Berthier:2015oma}
L.~Berthier and M.~Trott, \emph{{Towards consistent Electroweak Precision Data
  constraints in the SMEFT}},
  \href{https://doi.org/10.1007/JHEP05(2015)024}{\emph{JHEP} {\bfseries 05}
  (2015) 024}, [\href{https://arxiv.org/abs/1502.02570}{{\ttfamily
  1502.02570}}].

\bibitem{Berthier:2015gja}
L.~Berthier and M.~Trott, \emph{{Consistent constraints on the Standard Model
  Effective Field Theory}},
  \href{https://doi.org/10.1007/JHEP02(2016)069}{\emph{JHEP} {\bfseries 02}
  (2016) 069}, [\href{https://arxiv.org/abs/1508.05060}{{\ttfamily
  1508.05060}}].

\bibitem{Berthier:2016tkq}
L.~Berthier, M.~Bj\o{}rn and M.~Trott, \emph{{Incorporating doubly resonant
  $W^\pm$ data in a global fit of SMEFT parameters to lift flat directions}},
  \href{https://doi.org/10.1007/JHEP09(2016)157}{\emph{JHEP} {\bfseries 09}
  (2016) 157}, [\href{https://arxiv.org/abs/1606.06693}{{\ttfamily
  1606.06693}}].

\bibitem{Brivio:2017bnu}
I.~Brivio and M.~Trott, \emph{{Scheming in the SMEFT... and a
  reparameterization invariance!}},
  \href{https://doi.org/10.1007/JHEP07(2017)148}{\emph{JHEP} {\bfseries 07}
  (2017) 148}, [\href{https://arxiv.org/abs/1701.06424}{{\ttfamily
  1701.06424}}].

\bibitem{Biekoetter:2018ypq}
A.~Biekoetter, T.~Corbett and T.~Plehn, \emph{{The Gauge-Higgs Legacy of the
  LHC Run II}},
  \href{https://doi.org/10.21468/SciPostPhys.6.6.064}{\emph{SciPost Phys.}
  {\bfseries 6} (2019) 064},
  [\href{https://arxiv.org/abs/1812.07587}{{\ttfamily 1812.07587}}].

\bibitem{deBlas:2019rxi}
J.~de~Blas et~al., \emph{{Higgs Boson Studies at Future Particle Colliders}},
  \href{https://doi.org/10.1007/JHEP01(2020)139}{\emph{JHEP} {\bfseries 01}
  (2020) 139}, [\href{https://arxiv.org/abs/1905.03764}{{\ttfamily
  1905.03764}}].

\bibitem{Ellis:2020unq}
J.~Ellis, M.~Madigan, K.~Mimasu, V.~Sanz and T.~You, \emph{{Top, Higgs, Diboson
  and Electroweak Fit to the Standard Model Effective Field Theory}},
  \href{https://doi.org/10.1007/JHEP04(2021)279}{\emph{JHEP} {\bfseries 04}
  (2021) 279}, [\href{https://arxiv.org/abs/2012.02779}{{\ttfamily
  2012.02779}}].

\bibitem{Corbett:2015ksa}
T.~Corbett, O.~J.~P. Eboli, D.~Goncalves, J.~Gonzalez-Fraile, T.~Plehn and
  M.~Rauch, \emph{{The Higgs Legacy of the LHC Run I}},
  \href{https://doi.org/10.1007/JHEP08(2015)156}{\emph{JHEP} {\bfseries 08}
  (2015) 156}, [\href{https://arxiv.org/abs/1505.05516}{{\ttfamily
  1505.05516}}].

\bibitem{Brivio:2019ius}
I.~Brivio, S.~Bruggisser, F.~Maltoni, R.~Moutafis, T.~Plehn, E.~Vryonidou
  et~al., \emph{{O new physics, where art thou? A global search in the top
  sector}}, \href{https://doi.org/10.1007/JHEP02(2020)131}{\emph{JHEP}
  {\bfseries 02} (2020) 131},
  [\href{https://arxiv.org/abs/1910.03606}{{\ttfamily 1910.03606}}].

\bibitem{Durieux:2019rbz}
G.~Durieux, A.~Irles, V.~Miralles, A.~Pe\~nuelas, R.~P\"oschl, M.~Perell\'o
  et~al., \emph{{The electro-weak couplings of the top and bottom quarks
  \textemdash{} Global fit and future prospects}},
  \href{https://doi.org/10.1007/JHEP12(2019)098}{\emph{JHEP} {\bfseries 12}
  (2019) 98}, [\href{https://arxiv.org/abs/1907.10619}{{\ttfamily
  1907.10619}}].

\bibitem{Ethier:2021bye}
{\scshape SMEFiT} collaboration, J.~J. Ethier, G.~Magni, F.~Maltoni,
  L.~Mantani, E.~R. Nocera, J.~Rojo et~al., \emph{{Combined SMEFT
  interpretation of Higgs, diboson, and top quark data from the LHC}},
  \href{https://doi.org/10.1007/JHEP11(2021)089}{\emph{JHEP} {\bfseries 11}
  (2021) 089}, [\href{https://arxiv.org/abs/2105.00006}{{\ttfamily
  2105.00006}}].

\bibitem{Murphy:2020rsh}
C.~W. Murphy, \emph{{Dimension-8 operators in the Standard Model Eective Field
  Theory}}, \href{https://doi.org/10.1007/JHEP10(2020)174}{\emph{JHEP}
  {\bfseries 10} (2020) 174},
  [\href{https://arxiv.org/abs/2005.00059}{{\ttfamily 2005.00059}}].

\bibitem{Li:2020gnx}
H.-L. Li, Z.~Ren, J.~Shu, M.-L. Xiao, J.-H. Yu and Y.-H. Zheng, \emph{{Complete
  set of dimension-eight operators in the standard model effective field
  theory}}, \href{https://doi.org/10.1103/PhysRevD.104.015026}{\emph{Phys. Rev.
  D} {\bfseries 104} (2021) 015026},
  [\href{https://arxiv.org/abs/2005.00008}{{\ttfamily 2005.00008}}].

\bibitem{Hays:2018zze}
C.~Hays, A.~Martin, V.~Sanz and J.~Setford, \emph{{On the impact of
  dimension-eight SMEFT operators on Higgs measurements}},
  \href{https://doi.org/10.1007/JHEP02(2019)123}{\emph{JHEP} {\bfseries 02}
  (2019) 123}, [\href{https://arxiv.org/abs/1808.00442}{{\ttfamily
  1808.00442}}].

\bibitem{Corbett:2021iob}
T.~Corbett and T.~Rasmussen, \emph{{Higgs decays to two leptons and a photon
  beyond leading order in the SMEFT}},
  \href{https://doi.org/10.21468/SciPostPhys.13.5.112}{\emph{SciPost Phys.}
  {\bfseries 13} (2022) 112},
  [\href{https://arxiv.org/abs/2110.03694}{{\ttfamily 2110.03694}}].

\bibitem{Asteriadis:2022ras}
K.~Asteriadis, S.~Dawson and D.~Fontes, \emph{{Double insertions of SMEFT
  operators in gluon fusion Higgs boson production}},
  \href{https://doi.org/10.1103/PhysRevD.107.055038}{\emph{Phys. Rev. D}
  {\bfseries 107} (2023) 055038},
  [\href{https://arxiv.org/abs/2212.03258}{{\ttfamily 2212.03258}}].

\bibitem{Corbett:2021eux}
T.~Corbett, A.~Helset, A.~Martin and M.~Trott, \emph{{EWPD in the SMEFT to
  dimension eight}}, \href{https://doi.org/10.1007/JHEP06(2021)076}{\emph{JHEP}
  {\bfseries 06} (2021) 076},
  [\href{https://arxiv.org/abs/2102.02819}{{\ttfamily 2102.02819}}].

\bibitem{Alioli:2020kez}
S.~Alioli, R.~Boughezal, E.~Mereghetti and F.~Petriello, \emph{{Novel angular
  dependence in Drell-Yan lepton production via dimension-8 operators}},
  \href{https://doi.org/10.1016/j.physletb.2020.135703}{\emph{Phys. Lett. B}
  {\bfseries 809} (2020) 135703},
  [\href{https://arxiv.org/abs/2003.11615}{{\ttfamily 2003.11615}}].

\bibitem{Boughezal:2021tih}
R.~Boughezal, E.~Mereghetti and F.~Petriello, \emph{{Dilepton production in the
  SMEFT at $ \mathcal{O} $(1/\ensuremath{\Lambda}$^{4}$)}},
  \href{https://doi.org/10.1103/PhysRevD.104.095022}{\emph{Phys. Rev. D}
  {\bfseries 104} (2021) 095022},
  [\href{https://arxiv.org/abs/2106.05337}{{\ttfamily 2106.05337}}].

\bibitem{Kim:2022amu}
T.~Kim and A.~Martin, \emph{{Monolepton production in SMEFT to $ \mathcal{O}
  $(1/\ensuremath{\Lambda}$^{4}$) and beyond}},
  \href{https://doi.org/10.1007/JHEP09(2022)124}{\emph{JHEP} {\bfseries 09}
  (2022) 124}, [\href{https://arxiv.org/abs/2203.11976}{{\ttfamily
  2203.11976}}].

\bibitem{Degrande:2023iob}
C.~Degrande and H.-L. Li, \emph{{Impact of Dimension-8 SMEFT operators on
  Diboson Productions}},  \href{https://arxiv.org/abs/2303.10493}{{\ttfamily
  2303.10493}}.

\bibitem{Corbett:2023qtg}
T.~Corbett, J.~Desai, O.~J.~P. Eboli, M.~C. Gonzalez-Garcia, M.~Martines and
  P.~Reimitz, \emph{{Impact of dimension-eight SMEFT operators in the EWPO and
  Triple Gauge Couplings analysis in Universal SMEFT}},
  \href{https://arxiv.org/abs/2304.03305}{{\ttfamily 2304.03305}}.

\bibitem{Ellis:2022uxv}
J.~Ellis, N.~E. Mavromatos, P.~Roloff and T.~You, \emph{{Light-by-light
  scattering at future $e^+e^-$ colliders}},
  \href{https://doi.org/10.1140/epjc/s10052-022-10565-w}{\emph{Eur. Phys. J. C}
  {\bfseries 82} (2022) 634},
  [\href{https://arxiv.org/abs/2203.17111}{{\ttfamily 2203.17111}}].

\bibitem{Ellis:2018cos}
J.~Ellis and S.-F. Ge, \emph{{Constraining Gluonic Quartic Gauge Coupling
  Operators with gg\textrightarrow{}\ensuremath{\gamma}\ensuremath{\gamma}}},
  \href{https://doi.org/10.1103/PhysRevLett.121.041801}{\emph{Phys. Rev. Lett.}
  {\bfseries 121} (2018) 041801},
  [\href{https://arxiv.org/abs/1802.02416}{{\ttfamily 1802.02416}}].

\bibitem{Ellis:2021dfa}
J.~Ellis, S.-F. Ge and K.~Ma, \emph{{Hadron collider probes of the quartic
  couplings of gluons to the photon and Z boson}},
  \href{https://doi.org/10.1007/JHEP04(2022)123}{\emph{JHEP} {\bfseries 04}
  (2022) 123}, [\href{https://arxiv.org/abs/2112.06729}{{\ttfamily
  2112.06729}}].

\bibitem{Ellis:2019zex}
J.~Ellis, S.-F. Ge, H.-J. He and R.-Q. Xiao, \emph{{Probing the scale of new
  physics in the $ZZ\gamma$ coupling at $e^+e^-$ colliders}},
  \href{https://doi.org/10.1088/1674-1137/44/6/063106}{\emph{Chin. Phys. C}
  {\bfseries 44} (2020) 063106},
  [\href{https://arxiv.org/abs/1902.06631}{{\ttfamily 1902.06631}}].

\bibitem{Ellis:2020ljj}
J.~Ellis, H.-J. He and R.-Q. Xiao, \emph{{Probing new physics in dimension-8
  neutral gauge couplings at e$^{+}$e$^{-}$ colliders}},
  \href{https://doi.org/10.1007/s11433-020-1617-3}{\emph{Sci. China Phys. Mech.
  Astron.} {\bfseries 64} (2021) 221062},
  [\href{https://arxiv.org/abs/2008.04298}{{\ttfamily 2008.04298}}].

\bibitem{Ellis:2022zdw}
J.~Ellis, H.-J. He and R.-Q. Xiao, \emph{{Probing neutral triple gauge
  couplings at the LHC and future hadron colliders}},
  \href{https://doi.org/10.1103/PhysRevD.107.035005}{\emph{Phys. Rev. D}
  {\bfseries 107} (2023) 035005},
  [\href{https://arxiv.org/abs/2206.11676}{{\ttfamily 2206.11676}}].

\bibitem{Corbett:2015lfa}
T.~Corbett, O.~J.~P. \'Eboli and M.~C. Gonzalez-Garcia, \emph{{Inverse
  amplitude method for the perturbative electroweak symmetry breaking sector:
  The singlet Higgs portal as a study case}},
  \href{https://doi.org/10.1103/PhysRevD.93.015005}{\emph{Phys. Rev. D}
  {\bfseries 93} (2016) 015005},
  [\href{https://arxiv.org/abs/1509.01585}{{\ttfamily 1509.01585}}].

\bibitem{Dawson:2021xei}
S.~Dawson, S.~Homiller and M.~Sullivan, \emph{{Impact of dimension-eight SMEFT
  contributions: A case study}},
  \href{https://doi.org/10.1103/PhysRevD.104.115013}{\emph{Phys. Rev. D}
  {\bfseries 104} (2021) 115013},
  [\href{https://arxiv.org/abs/2110.06929}{{\ttfamily 2110.06929}}].

\bibitem{Dawson:2022cmu}
S.~Dawson, D.~Fontes, S.~Homiller and M.~Sullivan, \emph{{Role of
  dimension-eight operators in an EFT for the 2HDM}},
  \href{https://doi.org/10.1103/PhysRevD.106.055012}{\emph{Phys. Rev. D}
  {\bfseries 106} (2022) 055012},
  [\href{https://arxiv.org/abs/2205.01561}{{\ttfamily 2205.01561}}].

\bibitem{Banerjee:2022thk}
U.~Banerjee, J.~Chakrabortty, C.~Englert, S.~U. Rahaman and M.~Spannowsky,
  \emph{{Integrating out heavy scalars with modified equations of motion:
  Matching computation of dimension-eight SMEFT coefficients}},
  \href{https://doi.org/10.1103/PhysRevD.107.055007}{\emph{Phys. Rev. D}
  {\bfseries 107} (2023) 055007},
  [\href{https://arxiv.org/abs/2210.14761}{{\ttfamily 2210.14761}}].

\bibitem{Banerjee:2023bzl}
U.~Banerjee, J.~Chakrabortty, C.~Englert, W.~Naskar, S.~U. Rahaman and
  M.~Spannowsky, \emph{{EFT, Decoupling, Higgs Mixing and All That Jazz}},
  \href{https://arxiv.org/abs/2303.05224}{{\ttfamily 2303.05224}}.

\bibitem{ATLAS:2022vkf}
{\scshape ATLAS} collaboration, \emph{{A detailed map of Higgs boson
  interactions by the ATLAS experiment ten years after the discovery}},
  \href{https://doi.org/10.1038/s41586-022-04893-w}{\emph{Nature} {\bfseries
  607} (2022) 52--59}, [\href{https://arxiv.org/abs/2207.00092}{{\ttfamily
  2207.00092}}].

\bibitem{CMS:2022dwd}
{\scshape CMS} collaboration, A.~Tumasyan et~al., \emph{{A portrait of the
  Higgs boson by the CMS experiment ten years after the discovery}},
  \href{https://doi.org/10.1038/s41586-022-04892-x}{\emph{Nature} {\bfseries
  607} (2022) 60--68}, [\href{https://arxiv.org/abs/2207.00043}{{\ttfamily
  2207.00043}}].

\bibitem{CDF:2022hxs}
{\scshape CDF} collaboration, T.~Aaltonen et~al., \emph{{High-precision
  measurement of the $W$ boson mass with the CDF II detector}},
  \href{https://doi.org/10.1126/science.abk1781}{\emph{Science} {\bfseries 376}
  (2022) 170--176}.

\bibitem{Bagnaschi:2022whn}
E.~Bagnaschi, J.~Ellis, M.~Madigan, K.~Mimasu, V.~Sanz and T.~You, \emph{{SMEFT
  analysis of m$_{W}$}},
  \href{https://doi.org/10.1007/JHEP08(2022)308}{\emph{JHEP} {\bfseries 08}
  (2022) 308}, [\href{https://arxiv.org/abs/2204.05260}{{\ttfamily
  2204.05260}}].

\bibitem{Papaefstathiou:2015paa}
A.~Papaefstathiou and K.~Sakurai, \emph{{Triple Higgs boson production at a 100
  TeV proton-proton collider}},
  \href{https://doi.org/10.1007/JHEP02(2016)006}{\emph{JHEP} {\bfseries 02}
  (2016) 006}, [\href{https://arxiv.org/abs/1508.06524}{{\ttfamily
  1508.06524}}].

\bibitem{Roloff:2019crr}
{\scshape CLICdp} collaboration, P.~Roloff, U.~Schnoor, R.~Simoniello and
  B.~Xu, \emph{{Double Higgs boson production and Higgs self-coupling
  extraction at CLIC}},
  \href{https://doi.org/10.1140/epjc/s10052-020-08567-7}{\emph{Eur. Phys. J. C}
  {\bfseries 80} (2020) 1010},
  [\href{https://arxiv.org/abs/1901.05897}{{\ttfamily 1901.05897}}].

\bibitem{Kamionkowski:1993fg}
M.~Kamionkowski, A.~Kosowsky and M.~S. Turner, \emph{{Gravitational radiation
  from first order phase transitions}},
  \href{https://doi.org/10.1103/PhysRevD.49.2837}{\emph{Phys. Rev. D}
  {\bfseries 49} (1994) 2837--2851},
  [\href{https://arxiv.org/abs/astro-ph/9310044}{{\ttfamily
  astro-ph/9310044}}].

\bibitem{Barger:2003rs}
V.~Barger, T.~Han, P.~Langacker, B.~McElrath and P.~Zerwas, \emph{{Effects of
  genuine dimension-six Higgs operators}},
  \href{https://doi.org/10.1103/PhysRevD.67.115001}{\emph{Phys. Rev. D}
  {\bfseries 67} (2003) 115001},
  [\href{https://arxiv.org/abs/hep-ph/0301097}{{\ttfamily hep-ph/0301097}}].

\bibitem{Helset:2020yio}
A.~Helset, A.~Martin and M.~Trott, \emph{{The Geometric Standard Model
  Effective Field Theory}},
  \href{https://doi.org/10.1007/JHEP03(2020)163}{\emph{JHEP} {\bfseries 03}
  (2020) 163}, [\href{https://arxiv.org/abs/2001.01453}{{\ttfamily
  2001.01453}}].

\bibitem{Brivio:2020onw}
I.~Brivio, \emph{{SMEFTsim 3.0 \textemdash{} a practical guide}},
  \href{https://doi.org/10.1007/JHEP04(2021)073}{\emph{JHEP} {\bfseries 04}
  (2021) 073}, [\href{https://arxiv.org/abs/2012.11343}{{\ttfamily
  2012.11343}}].

\bibitem{Goertz:2014qta}
F.~Goertz, A.~Papaefstathiou, L.~L. Yang and J.~Zurita, \emph{{Higgs boson pair
  production in the D=6 extension of the SM}},
  \href{https://doi.org/10.1007/JHEP04(2015)167}{\emph{JHEP} {\bfseries 04}
  (2015) 167}, [\href{https://arxiv.org/abs/1410.3471}{{\ttfamily 1410.3471}}].

\bibitem{Azatov:2015oxa}
A.~Azatov, R.~Contino, G.~Panico and M.~Son, \emph{{Effective field theory
  analysis of double Higgs boson production via gluon fusion}},
  \href{https://doi.org/10.1103/PhysRevD.92.035001}{\emph{Phys. Rev. D}
  {\bfseries 92} (2015) 035001},
  [\href{https://arxiv.org/abs/1502.00539}{{\ttfamily 1502.00539}}].

\bibitem{CMS:2019qfk}
{\scshape CMS} collaboration, A.~M. Sirunyan et~al., \emph{{Search for
  anomalous electroweak production of vector boson pairs in association with
  two jets in proton-proton collisions at 13 TeV}},
  \href{https://doi.org/10.1016/j.physletb.2019.134985}{\emph{Phys. Lett. B}
  {\bfseries 798} (2019) 134985},
  [\href{https://arxiv.org/abs/1905.07445}{{\ttfamily 1905.07445}}].

\bibitem{Zhang:2018shp}
C.~Zhang and S.-Y. Zhou, \emph{{Positivity bounds on vector boson scattering at
  the LHC}}, \href{https://doi.org/10.1103/PhysRevD.100.095003}{\emph{Phys.
  Rev. D} {\bfseries 100} (2019) 095003},
  [\href{https://arxiv.org/abs/1808.00010}{{\ttfamily 1808.00010}}].

\bibitem{ALEPH:2005ab}
{\scshape ALEPH, DELPHI, L3, OPAL, SLD, LEP Electroweak Working Group, SLD
  Electroweak \& Heavy Flavour Groups} collaboration, S.~Schael et~al.,
  \emph{{Precision electroweak measurements on the $Z$ resonance}},
  \href{https://doi.org/10.1016/j.physrep.2005.12.006}{\emph{Phys. Rept.}
  {\bfseries 427} (2006) 257--454},
  [\href{https://arxiv.org/abs/hep-ex/0509008}{{\ttfamily hep-ex/0509008}}].

\bibitem{CDF:2013dpa}
{\scshape CDF, D0} collaboration, T.~A. Aaltonen et~al., \emph{{Combination of
  CDF and D0 $W$-Boson Mass Measurements}},
  \href{https://doi.org/10.1103/PhysRevD.88.052018}{\emph{Phys. Rev. D}
  {\bfseries 88} (2013) 052018},
  [\href{https://arxiv.org/abs/1307.7627}{{\ttfamily 1307.7627}}].

\bibitem{LHCb:2021bjt}
{\scshape LHCb} collaboration, R.~Aaij et~al., \emph{{Measurement of the W
  boson mass}}, \href{https://doi.org/10.1007/JHEP01(2022)036}{\emph{JHEP}
  {\bfseries 01} (2022) 036},
  [\href{https://arxiv.org/abs/2109.01113}{{\ttfamily 2109.01113}}].

\bibitem{hepdata.130266}
{ATLAS Collaboration}, ``{\it A detailed map of Higgs boson interactions by the
  ATLAS experiment ten years after the discovery}.'' {HEPData (collection)},
  2022, \url{https://doi.org/10.17182/hepdata.130266}.

\bibitem{hepdata.127765}
{CMS Collaboration}, ``{\it A portrait of the Higgs boson by the CMS experiment
  ten years after the discovery}.'' {HEPData (collection)}, 2022,
  \url{https://doi.org/10.17182/hepdata.127765}.

\bibitem{ATLAS:2022kbf}
{\scshape ATLAS} collaboration, \emph{{Constraining the Higgs boson
  self-coupling from single- and double-Higgs production with the ATLAS
  detector using $pp$ collisions at $\sqrt{s}=13$ TeV}},  ATLAS-CONF-2022-050,
  2022, \url{http://cds.cern.ch/record/2816332}.

\bibitem{Brivio:2017vri}
I.~Brivio and M.~Trott, \emph{{The Standard Model as an Effective Field
  Theory}}, \href{https://doi.org/10.1016/j.physrep.2018.11.002}{\emph{Phys.
  Rept.} {\bfseries 793} (2019) 1--98},
  [\href{https://arxiv.org/abs/1706.08945}{{\ttfamily 1706.08945}}].

\bibitem{Alwall:2014hca}
J.~Alwall, R.~Frederix, S.~Frixione, V.~Hirschi, F.~Maltoni, O.~Mattelaer
  et~al., \emph{{The automated computation of tree-level and next-to-leading
  order differential cross sections, and their matching to parton shower
  simulations}}, \href{https://doi.org/10.1007/JHEP07(2014)079}{\emph{JHEP}
  {\bfseries 07} (2014) 079},
  [\href{https://arxiv.org/abs/1405.0301}{{\ttfamily 1405.0301}}].

\bibitem{NNPDF:2017mvq}
{\scshape NNPDF} collaboration, R.~D. Ball et~al., \emph{{Parton distributions
  from high-precision collider data}},
  \href{https://doi.org/10.1140/epjc/s10052-017-5199-5}{\emph{Eur. Phys. J. C}
  {\bfseries 77} (2017) 663},
  [\href{https://arxiv.org/abs/1706.00428}{{\ttfamily 1706.00428}}].

\bibitem{Rizzo:1980gz}
T.~G. Rizzo, \emph{{Decays of Heavy Higgs Bosons}},
  \href{https://doi.org/10.1103/PhysRevD.22.722}{\emph{Phys. Rev. D} {\bfseries
  22} (1980) 722}.

\bibitem{Carvalho:2015ttv}
A.~Carvalho, M.~Dall'Osso, T.~Dorigo, F.~Goertz, C.~A. Gottardo and M.~Tosi,
  \emph{{Higgs Pair Production: Choosing Benchmarks With Cluster Analysis}},
  \href{https://doi.org/10.1007/JHEP04(2016)126}{\emph{JHEP} {\bfseries 04}
  (2016) 126}, [\href{https://arxiv.org/abs/1507.02245}{{\ttfamily
  1507.02245}}].

\bibitem{Zuk:1985sw}
J.~A. Zuk, \emph{{A Functional Approach to Derivative Expansion of the
  Effective Lagrangian}},
  \href{https://doi.org/10.1103/PhysRevD.32.2653}{\emph{Phys. Rev. D}
  {\bfseries 32} (1985) 2653}.

\bibitem{Cheyette:1985ue}
O.~Cheyette, \emph{{Derivative Expansion of the Effective Action}},
  \href{https://doi.org/10.1103/PhysRevLett.55.2394}{\emph{Phys. Rev. Lett.}
  {\bfseries 55} (1985) 2394}.

\bibitem{Gaillard:1985uh}
M.~K. Gaillard, \emph{{The Effective One Loop Lagrangian With Derivative
  Couplings}}, \href{https://doi.org/10.1016/0550-3213(86)90264-6}{\emph{Nucl.
  Phys. B} {\bfseries 268} (1986) 669--692}.

\bibitem{Henning:2014wua}
B.~Henning, X.~Lu and H.~Murayama, \emph{{How to use the Standard Model
  effective field theory}},
  \href{https://doi.org/10.1007/JHEP01(2016)023}{\emph{JHEP} {\bfseries 01}
  (2016) 023}, [\href{https://arxiv.org/abs/1412.1837}{{\ttfamily 1412.1837}}].

\bibitem{Drozd:2015rsp}
A.~Drozd, J.~Ellis, J.~Quevillon and T.~You, \emph{{The Universal One-Loop
  Effective Action}},
  \href{https://doi.org/10.1007/JHEP03(2016)180}{\emph{JHEP} {\bfseries 03}
  (2016) 180}, [\href{https://arxiv.org/abs/1512.03003}{{\ttfamily
  1512.03003}}].

\bibitem{Haisch:2020ahr}
U.~Haisch, M.~Ruhdorfer, E.~Salvioni, E.~Venturini and A.~Weiler,
  \emph{{Singlet night in Feynman-ville: one-loop matching of a real scalar}},
  \href{https://doi.org/10.1007/JHEP04(2020)164}{\emph{JHEP} {\bfseries 04}
  (2020) 164}, [\href{https://arxiv.org/abs/2003.05936}{{\ttfamily
  2003.05936}}].

\bibitem{deBlas:2017xtg}
J.~de~Blas, J.~C. Criado, M.~Perez-Victoria and J.~Santiago, \emph{{Effective
  description of general extensions of the Standard Model: the complete
  tree-level dictionary}},
  \href{https://doi.org/10.1007/JHEP03(2018)109}{\emph{JHEP} {\bfseries 03}
  (2018) 109}, [\href{https://arxiv.org/abs/1711.10391}{{\ttfamily
  1711.10391}}].

\bibitem{Jenkins:2017jig}
E.~E. Jenkins, A.~V. Manohar and P.~Stoffer, \emph{{Low-Energy Effective Field
  Theory below the Electroweak Scale: Operators and Matching}},
  \href{https://doi.org/10.1007/JHEP03(2018)016}{\emph{JHEP} {\bfseries 03}
  (2018) 016}, [\href{https://arxiv.org/abs/1709.04486}{{\ttfamily
  1709.04486}}].

\bibitem{Corbett:2017ieo}
T.~Corbett, A.~Joglekar, H.-L. Li and J.-H. Yu, \emph{{Exploring Extended
  Scalar Sectors with Di-Higgs Signals: A Higgs EFT Perspective}},
  \href{https://doi.org/10.1007/JHEP05(2018)061}{\emph{JHEP} {\bfseries 05}
  (2018) 061}, [\href{https://arxiv.org/abs/1705.02551}{{\ttfamily
  1705.02551}}].

\bibitem{Jiang:2018pbd}
M.~Jiang, N.~Craig, Y.-Y. Li and D.~Sutherland, \emph{{Complete one-loop
  matching for a singlet scalar in the Standard Model EFT}},
  \href{https://doi.org/10.1007/JHEP02(2019)031}{\emph{JHEP} {\bfseries 02}
  (2019) 031}, [\href{https://arxiv.org/abs/1811.08878}{{\ttfamily
  1811.08878}}].

\bibitem{Dawson:2021jcl}
S.~Dawson, P.~P. Giardino and S.~Homiller, \emph{{Uncovering the High Scale
  Higgs Singlet Model}},
  \href{https://doi.org/10.1103/PhysRevD.103.075016}{\emph{Phys. Rev. D}
  {\bfseries 103} (2021) 075016},
  [\href{https://arxiv.org/abs/2102.02823}{{\ttfamily 2102.02823}}].

\bibitem{Chen:2014ask}
C.-Y. Chen, S.~Dawson and I.~M. Lewis, \emph{{Exploring resonant di-Higgs boson
  production in the Higgs singlet model}},
  \href{https://doi.org/10.1103/PhysRevD.91.035015}{\emph{Phys. Rev. D}
  {\bfseries 91} (2015) 035015},
  [\href{https://arxiv.org/abs/1410.5488}{{\ttfamily 1410.5488}}].

\bibitem{Espinosa:2011ax}
J.~R. Espinosa, T.~Konstandin and F.~Riva, \emph{{Strong Electroweak Phase
  Transitions in the Standard Model with a Singlet}},
  \href{https://doi.org/10.1016/j.nuclphysb.2011.09.010}{\emph{Nucl. Phys. B}
  {\bfseries 854} (2012) 592--630},
  [\href{https://arxiv.org/abs/1107.5441}{{\ttfamily 1107.5441}}].

\bibitem{Brehmer:2015rna}
J.~Brehmer, A.~Freitas, D.~Lopez-Val and T.~Plehn, \emph{{Pushing Higgs
  Effective Theory to its Limits}},
  \href{https://doi.org/10.1103/PhysRevD.93.075014}{\emph{Phys. Rev. D}
  {\bfseries 93} (2016) 075014},
  [\href{https://arxiv.org/abs/1510.03443}{{\ttfamily 1510.03443}}].

\bibitem{Goodsell:2018fex}
M.~D. Goodsell and F.~Staub, \emph{{Improved unitarity constraints in
  Two-Higgs-Doublet-Models}},
  \href{https://doi.org/10.1016/j.physletb.2018.11.030}{\emph{Phys. Lett. B}
  {\bfseries 788} (2019) 206--212},
  [\href{https://arxiv.org/abs/1805.07310}{{\ttfamily 1805.07310}}].

\bibitem{Krauss:2018orw}
M.~E. Krauss and F.~Staub, \emph{{Unitarity constraints in triplet extensions
  beyond the large s limit}},
  \href{https://doi.org/10.1103/PhysRevD.98.015041}{\emph{Phys. Rev. D}
  {\bfseries 98} (2018) 015041},
  [\href{https://arxiv.org/abs/1805.07309}{{\ttfamily 1805.07309}}].

\bibitem{Goodsell:2018tti}
M.~D. Goodsell and F.~Staub, \emph{{Unitarity constraints on general scalar
  couplings with SARAH}},
  \href{https://doi.org/10.1140/epjc/s10052-018-6127-z}{\emph{Eur. Phys. J. C}
  {\bfseries 78} (2018) 649},
  [\href{https://arxiv.org/abs/1805.07306}{{\ttfamily 1805.07306}}].

\bibitem{Buchalla:2016bse}
G.~Buchalla, O.~Cata, A.~Celis and C.~Krause, \emph{{Standard Model Extended by
  a Heavy Singlet: Linear vs. Nonlinear EFT}},
  \href{https://doi.org/10.1016/j.nuclphysb.2017.02.006}{\emph{Nucl. Phys. B}
  {\bfseries 917} (2017) 209--233},
  [\href{https://arxiv.org/abs/1608.03564}{{\ttfamily 1608.03564}}].

\bibitem{Falkowski:2019tft}
A.~Falkowski and R.~Rattazzi, \emph{{Which EFT}},
  \href{https://doi.org/10.1007/JHEP10(2019)255}{\emph{JHEP} {\bfseries 10}
  (2019) 255}, [\href{https://arxiv.org/abs/1902.05936}{{\ttfamily
  1902.05936}}].

\bibitem{Cohen:2020xca}
T.~Cohen, N.~Craig, X.~Lu and D.~Sutherland, \emph{{Is SMEFT Enough?}},
  \href{https://doi.org/10.1007/JHEP03(2021)237}{\emph{JHEP} {\bfseries 03}
  (2021) 237}, [\href{https://arxiv.org/abs/2008.08597}{{\ttfamily
  2008.08597}}].

\bibitem{Plehn:2005nk}
T.~Plehn and M.~Rauch, \emph{{The quartic higgs coupling at hadron colliders}},
  \href{https://doi.org/10.1103/PhysRevD.72.053008}{\emph{Phys. Rev. D}
  {\bfseries 72} (2005) 053008},
  [\href{https://arxiv.org/abs/hep-ph/0507321}{{\ttfamily hep-ph/0507321}}].

\bibitem{Chen:2015gva}
C.-Y. Chen, Q.-S. Yan, X.~Zhao, Y.-M. Zhong and Z.~Zhao, \emph{{Probing
  triple-Higgs productions via 4b2\ensuremath{\gamma} decay channel at a 100
  TeV hadron collider}},
  \href{https://doi.org/10.1103/PhysRevD.93.013007}{\emph{Phys. Rev. D}
  {\bfseries 93} (2016) 013007},
  [\href{https://arxiv.org/abs/1510.04013}{{\ttfamily 1510.04013}}].

\bibitem{Fuks:2015hna}
B.~Fuks, J.~H. Kim and S.~J. Lee, \emph{{Probing Higgs self-interactions in
  proton-proton collisions at a center-of-mass energy of 100 TeV}},
  \href{https://doi.org/10.1103/PhysRevD.93.035026}{\emph{Phys. Rev. D}
  {\bfseries 93} (2016) 035026},
  [\href{https://arxiv.org/abs/1510.07697}{{\ttfamily 1510.07697}}].

\bibitem{Fuks:2017zkg}
B.~Fuks, J.~H. Kim and S.~J. Lee, \emph{{Scrutinizing the Higgs quartic
  coupling at a future 100 TeV proton\textendash{}proton collider with taus and
  b-jets}}, \href{https://doi.org/10.1016/j.physletb.2017.05.075}{\emph{Phys.
  Lett. B} {\bfseries 771} (2017) 354--358},
  [\href{https://arxiv.org/abs/1704.04298}{{\ttfamily 1704.04298}}].

\bibitem{Papaefstathiou:2019ofh}
A.~Papaefstathiou, G.~Tetlalmatzi-Xolocotzi and M.~Zaro, \emph{{Triple Higgs
  boson production to six $b$-jets at a 100 TeV proton collider}},
  \href{https://doi.org/10.1140/epjc/s10052-019-7457-1}{\emph{Eur. Phys. J. C}
  {\bfseries 79} (2019) 947},
  [\href{https://arxiv.org/abs/1909.09166}{{\ttfamily 1909.09166}}].

\bibitem{Chiesa:2020awd}
M.~Chiesa, F.~Maltoni, L.~Mantani, B.~Mele, F.~Piccinini and X.~Zhao,
  \emph{{Measuring the quartic Higgs self-coupling at a multi-TeV muon
  collider}}, \href{https://doi.org/10.1007/JHEP09(2020)098}{\emph{JHEP}
  {\bfseries 09} (2020) 098},
  [\href{https://arxiv.org/abs/2003.13628}{{\ttfamily 2003.13628}}].

\bibitem{Liu:2018peg}
T.~Liu, K.-F. Lyu, J.~Ren and H.~X. Zhu, \emph{{Probing the quartic Higgs boson
  self-interaction}},
  \href{https://doi.org/10.1103/PhysRevD.98.093004}{\emph{Phys. Rev. D}
  {\bfseries 98} (2018) 093004},
  [\href{https://arxiv.org/abs/1803.04359}{{\ttfamily 1803.04359}}].

\bibitem{Maltoni:2018ttu}
F.~Maltoni, D.~Pagani and X.~Zhao, \emph{{Constraining the Higgs self-couplings
  at e$^{+}$e$^{-}$ colliders}},
  \href{https://doi.org/10.1007/JHEP07(2018)087}{\emph{JHEP} {\bfseries 07}
  (2018) 087}, [\href{https://arxiv.org/abs/1802.07616}{{\ttfamily
  1802.07616}}].

\bibitem{Borowka:2018pxx}
S.~Borowka, C.~Duhr, F.~Maltoni, D.~Pagani, A.~Shivaji and X.~Zhao,
  \emph{{Probing the scalar potential via double Higgs boson production at
  hadron colliders}},
  \href{https://doi.org/10.1007/JHEP04(2019)016}{\emph{JHEP} {\bfseries 04}
  (2019) 016}, [\href{https://arxiv.org/abs/1811.12366}{{\ttfamily
  1811.12366}}].

\bibitem{Bizon:2018syu}
W.~Bizo\'n, U.~Haisch and L.~Rottoli, \emph{{Constraints on the quartic Higgs
  self-coupling from double-Higgs production at future hadron colliders}},
  \href{https://doi.org/10.1007/JHEP10(2019)267}{\emph{JHEP} {\bfseries 10}
  (2019) 267}, [\href{https://arxiv.org/abs/1810.04665}{{\ttfamily
  1810.04665}}].

\bibitem{DiMicco:2019ngk}
J.~Alison et~al., \emph{{Higgs boson potential at colliders: Status and
  perspectives}}, \href{https://doi.org/10.1016/j.revip.2020.100045}{\emph{Rev.
  Phys.} {\bfseries 5} (2020) 100045},
  [\href{https://arxiv.org/abs/1910.00012}{{\ttfamily 1910.00012}}].

\bibitem{ATLAS:2022okt}
{\scshape ATLAS} collaboration, \emph{{Projected sensitivity of Higgs boson
  pair production combining the $b\bar{b}\gamma\gamma$ and
  $b\bar{b}\tau^{+}\tau^{-}$ final states with the ATLAS detector at the
  HL-LHC}},  ATL-PHYS-PUB-2022-005, 2022,
  \url{http://cds.cern.ch/record/2802127}.

\bibitem{Papaefstathiou:2020lyp}
A.~Papaefstathiou, T.~Robens and G.~Tetlalmatzi-Xolocotzi, \emph{{Triple Higgs
  Boson Production at the Large Hadron Collider with Two Real Singlet
  Scalars}}, \href{https://doi.org/10.1007/JHEP05(2021)193}{\emph{JHEP}
  {\bfseries 05} (2021) 193},
  [\href{https://arxiv.org/abs/2101.00037}{{\ttfamily 2101.00037}}].

\bibitem{Khan:2016sxm}
N.~Khan, \emph{{Exploring the hyperchargeless Higgs triplet model up to the
  Planck scale}},
  \href{https://doi.org/10.1140/epjc/s10052-018-5766-4}{\emph{Eur. Phys. J. C}
  {\bfseries 78} (2018) 341},
  [\href{https://arxiv.org/abs/1610.03178}{{\ttfamily 1610.03178}}].

\bibitem{Criado:2018sdb}
J.~C. Criado and M.~P\'erez-Victoria, \emph{{Field redefinitions in effective
  theories at higher orders}},
  \href{https://doi.org/10.1007/JHEP03(2019)038}{\emph{JHEP} {\bfseries 03}
  (2019) 038}, [\href{https://arxiv.org/abs/1811.09413}{{\ttfamily
  1811.09413}}].

\bibitem{Forshaw:2001xq}
J.~R. Forshaw, D.~A. Ross and B.~E. White, \emph{{Higgs mass bounds in a
  triplet model}},
  \href{https://doi.org/10.1088/1126-6708/2001/10/007}{\emph{JHEP} {\bfseries
  10} (2001) 007}, [\href{https://arxiv.org/abs/hep-ph/0107232}{{\ttfamily
  hep-ph/0107232}}].

\end{thebibliography}\endgroup

\end{document}